\documentclass[onecollarge]{svjour2}       
\smartqed  

\usepackage{graphicx,epsfig}
\usepackage{natbib}
\usepackage{amssymb}

\newcommand{\beq}{\begin{equation}} 
\newcommand{\eeq}{\end{equation}} 
\newcommand{\beqa}{\begin{eqnarray}} 
\newcommand{\eeqa}{\end{eqnarray}} 
\newcommand{\bea}{\begin{array}} 
\newcommand{\ea}{\end{array}} 

\newcommand{\dd}{{\rm d}}
\newcommand{\pl}{\partial}
\newcommand{\inta}{\int_{-i\infty}^{+i\infty}} 
\newcommand{\lag}{\langle} 
\newcommand{\rag}{\rangle}
\newcommand{\law}{\stackrel{\rm law}{=}}
\newcommand{\ii}{{\rm i}}

\newcommand{\cP}{{\cal P}}
\newcommand{\gam}{\gamma}
\newcommand{\om}{\omega}
\newcommand{\cJ}{{\cal J}}
\newcommand{\cH}{{\cal H}}
\newcommand{\bv}{\bar{v}}
\newcommand{\cFi}{{\cal F}_{\infty}}
\newcommand{\hcFi}{\hat{\cal F}_{\infty}}

\newcommand{\cFs}{{\cal F}_0}
\newcommand{\tcFs}{\tilde{\cal F}_0}
\newcommand{\bPsi}{\overline{\Psi}}

\newcommand{\Ax}{A_{s_1,s_2}}
\newcommand{\hx}{h_{s_1,s_2}}
\newcommand{\Phix}{\Phi_{s_1,s_2}}
\newcommand{\cGs}{{\cal G}_0}
\newcommand{\Pshock}{P^{\rm shock}}

\newcommand{\rhoh}{\hat{\rho}}
\newcommand{\bx}{\bf{x}}
\newcommand{\cD}{{\cal D}}
\newcommand{\bq}{\bf{q}}

\newcommand{\Ai}{\mbox{Ai}}
\newcommand{\Aip}{\mbox{Ai}\,'}

\journalname{Journal of Statistical Physics}

\begin{document}

\title{Some statistical properties of the Burgers equation with white-noise initial velocity}
\titlerunning{Burgers equation with white-noise initial velocity}        

\author{Patrick Valageas}

\institute{Institut de Physique Th{\'e}orique, CEA Saclay, 
91191 Gif-sur-Yvette, France\\
\email{patrick.valageas@cea.fr}}

\date{Received: date / Accepted: date}

\maketitle

\begin{abstract}
We revisit the one-dimensional Burgers equation in the inviscid limit for white-noise
initial velocity. We derive the probability distributions of velocity and Lagrangian
increments, measured on intervals of any length $x$. This also gives the velocity 
structure functions. Next, for the case where the initial density is uniform, 
we obtain the distribution of the density, over any scale $x$, and we derive the
density two-point correlation and power spectrum. Finally, we consider the Lagrangian
displacement field and we derive the distribution of increments of the Lagrangian
map. We check that this gives back the well-known mass function of shocks.
For all distributions we describe the limiting scaling functions that are obtained 
in the large-scale and small-scale limits. We also discuss how these results 
generalize to other initial conditions, or to higher dimensions, and make the
connection with a heuristic multifractal formalism.
We note that the formation of point-like masses generically leads to a universal
small-scale scaling for the density distribution, which is known as the 
``stable-clustering ansatz'' in the cosmological context (where the Burgers dynamics
is also known as the ``adhesion model'').
\keywords{Inviscid Burgers equation \and Turbulence \and Cosmology: large-scale structure of the universe}
\end{abstract}

\section{Introduction}
\label{sec:intro}

The Burgers equation \cite{Burgersbook}, which describes the advection of a velocity
field by itself, with a non-zero viscosity, is a very popular nonlinear evolution
equation that appears in many physical problems, see \cite{Bec2007}
for a recent review. It was first introduced 
as a simplified model of fluid turbulence, as it shares the same hydrodynamical
(advective) nonlinearity and several conservation laws with the Navier-Stokes
equation. Even though it was shown later on by \cite{Hopf1950} and \cite{Cole1951}
that it can be explicitly integrated and lacks the chaotic character associated
with actual turbulence, it still retains much interest for hydrodynamical studies,
particularly as a useful benchmark for approximation schemes \cite{Fournier1983}.
On the other hand, it has appeared in many other
physical situations, such as the propagation of nonlinear acoustic waves in
non-dispersive media \cite{Gurbatov1991}, the study of disordered systems and
pinned manifolds \cite{LeDoussal2008}, or the formation of large-scale
structures in cosmology \cite{Gurbatove1989,Vergassola1994}. There,
in the limit of vanishing viscosity, it is known as the ``adhesion model'' and it
provides a good description of the large-scale filamentary structure of the
cosmic web \cite{Melott1994}. In this context, one is interested in the
statistical properties of the dynamics, starting with random Gaussian initial
conditions \cite{Kida1979,Gurbatov1997} (i.e. ``decaying Burgers turbulence'' in 
the hydrodynamical context). 
Moreover, in addition to the velocity field, one is also interested in the
properties of the density field generated by this dynamics, starting with 
an initial uniform density. 

This problem has led to many studies in the inviscid limit, focusing 
on power-law initial spectra (fractional Brownian motion), $E_0(k) \propto k^n$,
especially for the two peculiar cases of white-noise initial velocity ($n=0$)
\cite{Burgersbook,Kida1979,She1992,Frachebourg2000} or Brownian motion initial
velocity ($n=-2$) \cite{She1992,Sinai1992,Bertoin1998,Valageas2008}. 
The initial velocity fluctuations are dominated by short wavelengths in the former 
case and by large wavelengths in the latter case. Therefore, they provide two
simple examples for two more general classes of random initial conditions,
associated with $-1<n<1$ and $-3<n<-1$ \cite{Gurbatov1991,Gurbatov1997},
which show both common and different significant behaviors. For instance, the
integral scale of turbulence, $L(t)$, and the tail of the shock mass function,
scale with $n$ as $L(t) \sim t^{2/(n+3)}$ and $\ln[n(>m)] \sim -m^{n+3}$
over the whole range $-3<n<1$ \cite{She1992,Molchan1997,Gurbatov1997,Noullez2005},
even though shocks are dense for $-3<n<-1$ but isolated for $-1<n<1$ 
\cite{She1992}. Then, the specific advantage of these two cases, $n=0$ and
$n=-2$, is that in both cases the initial velocity field is built from a white-noise 
stochastic field (either directly or through one integration), which gives rise
to Markovian processes and allows to derive many explicit analytical results. 

In parallel with a study of the Brownian case in \cite{Valageas2008}, we revisit in
this article the white-noise case, taking advantage of the results obtained
in \cite{Frachebourg2000}. In particular, we pay attention to issues that arise in 
the hydrodynamical context (velocity structure functions, Lagrangian 
displacement field) as well as in the cosmological context (statistics of the 
density field). Thus, the main goal of this article is to provide explicit results
for the distributions of velocity increments and density fluctuations. As explained
above, this complements the study \cite{Valageas2008} of the Brownian case,
so that we now have explicit exact results for these quantities for the two representative
cases $n=0$ and $n=-2$. This should prove useful to check the validity of
approximation schemes devised for generic initial conditions and higher
dimensions. as in \cite{Valageas2009} where the tails of these probability
distributions are studied in the general case.

We first describe in section~\ref{sec:Initial} the white-noise initial conditions 
and the standard geometrical interpretation in terms of parabolas of the
Hopf-Cole solution of the dynamics \cite{Burgersbook}. Then, we recall
in section~\ref{Known-Eulerian-distributions} the Eulerian one-point and
two-point distributions, $p_x(q)$ and $p_{x_1,x_2}(q_1,q_2)$, associated
with the inverse Lagrangian map $x\mapsto q$, that were obtained in
\cite{Frachebourg2000}.
This allows us to derive in section~\ref{Lagrangian and velocity increments}
the distributions of the inverse Lagrangian increment
and velocity increment, as well as the velocity
structure functions. We also describe the limiting large-scale and small-scale
distributions. Next, we consider in section~\ref{Density} the distribution of the 
density within intervals of size $x$, and the density two-point correlation and 
power spectrum. Then, turning to a Lagrangian point of view, we study the Lagrangian
displacement field in section~\ref{Lagrangian-displacement}.
Finally, we describe in section~\ref{Heuristic} how the small-scale scalings shown by
these exact results can be generalized to other initial conditions and higher dimensions
within a heuristic approach.

\section{Initial conditions and geometrical solution}
\label{sec:Initial}

\subsection{Equation of motion}
\label{Equation-of-motion}

We consider in this article the one-dimensional Burgers equation for the velocity 
field $v(x,t)$ in the limit of zero viscosity,
\beq
\frac{\pl v}{\pl t} + v \frac{\pl v}{\pl x} = \nu \frac{\pl^2 v}{\pl x^2}
\hspace{1cm} \mbox{with} \hspace{1cm} \nu \rightarrow 0^+ .
\label{Burg}
\eeq
Let us recall here that in the cosmological context, the time $t$ in the Burgers 
equation (\ref{Burg}) actually stands for the linear growing mode $D_+(t)$ of the 
density fluctuations, the spatial coordinate $x$ is a comoving coordinate (that 
follows the uniform Hubble expansion) and, up to a time-dependent factor, the velocity 
$v$ is the peculiar velocity (where the Hubble expansion has been subtracted), 
see \cite{Gurbatove1989,Vergassola1994}. 
In these coordinates, the evolution of the density field is still given by the 
continuity equation (\ref{continuity}) below, where the density $\rho$ is the comoving 
density. 
If we take $\nu=0$, that is we remove the right-hand side in Eq.(\ref{Burg}), this 
is the well-known Zeldovich approximation \cite{Zeldovich1970,Valageas2007}, where 
particles always keep their initial velocity and merely follow straight trajectories. 
The diffusive term of (\ref{Burg}) is then added as a phenomenological device to 
prevent particles from escaping to infinity after crossing each other and to mimic 
the gravitational trapping of particles within the potential wells formed by the 
overdensities \cite{Gurbatove1989}.
Of course, this cannot describe the inner structure of collapsed objects 
(e.g., galaxies) but it provides a good description of the large-scale structure
of the cosmic web \cite{Melott1994}.

As is well known \cite[]{Hopf1950,Cole1951}, introducing the velocity potential 
$\psi(x,t)$ and making the change of variable $\psi(x,t)=-2\nu\ln\theta(x,t)$ 
transforms the nonlinear Burgers equation into the linear heat equation. 
This gives the explicit solution 
\beq
v(x,t) = \frac{\pl\psi}{\pl x} \hspace{0.5cm} \mbox{with} \hspace{0.5cm} 
\psi(x,t)= -2\nu \ln \int_{-\infty}^{\infty} \frac{\dd q}{\sqrt{4\pi\nu t}} \;
\exp\left[-\frac{(x-q)^2}{4\nu t}-\frac{\psi_0(q)}{2\nu}\right] ,
\label{psinu}
\eeq
where we introduced the initial condition $\psi_0(q)=\psi(q,t=0)$. 
Then, in the limit $\nu \rightarrow 0^+$ the steepest-descent method
gives
\beq
\psi(x,t) = \min_q \left[ \psi_0(q) + \frac{(x-q)^2}{2t} \right]
\hspace{0.5cm} \mbox{and} \hspace{0.5cm} v(x,t) = \frac{x-q(x,t)}{t} ,
\label{psinu0}
\eeq
where we introduced the Lagrangian coordinate $q(x,t)$ defined by
\beq
\psi_0(q) + \frac{(x-q)^2}{2t} \hspace{0.5cm} \mbox{is minimum at the point} 
\hspace{0.5cm} q = q(x,t) .
\label{qmin}
\eeq
The Eulerian locations $x$ where there are two solutions, $q_-<q_+$, to the
minimization problem (\ref{qmin}) correspond to shocks (and all the matter
initially between $q_-$ and $q_+$ is gathered at $x$). The application
$q \mapsto x(q,t)$ is usually called the Lagrangian map, and
$x \mapsto q(x,t)$ the inverse Lagrangian map (which is discontinuous at
shock locations) \cite{Bec2007}. 
For the case of white-noise initial velocity that we consider
in this paper, it is known that there is only a finite number of shocks
per unit length \cite{She1992,AvellanedaE1995}.

\subsection{Initial conditions}
\label{Initial-conditions}

In this article, we consider a white-noise initial velocity field $v_0(q)$,
normalized by
\beq
\lag v_0(q)\rag=0 , \;\;\;\; \lag v_0(q_1) v_0(q_2)\rag = D \, \delta(q_1-q_2) ,
\label{v0def}
\eeq
where $\lag .. \rag$ is the average over all realizations of the initial velocity
field. The velocity potential is defined up to a constant, and we may choose to
normalize the initial potential $\psi_0(q)$ by $\psi_0(0)=0$, whence
\beq
\psi_0(q)= \int_0^q\dd q' \, v_0(q') , \;\; \lag \psi_0(q)\rag=0 , \;\;
\lag \psi_0(q_1) \psi_0(q_2)\rag = D \, q_1 , 
\;\;\; \mbox{for} \;\;\; 0 \leq q_1 \leq q_2 .
\label{psi0def}
\eeq
Thus, the initial velocity potential is a bilateral Brownian motion that starts 
from the origin. Then, thanks to the scale invariance of the
Brownian motion, the scaled initial potential $\psi_0(\lambda q)$ has the same
probability distribution as $\lambda^{1/2} \psi_0(q)$, for any $\lambda>0$.
Hence, using the explicit solution (\ref{psinu0}) we obtain the scaling laws
\beq
\psi(x,t) \law t^{1/3} \psi(x/t^{2/3},1) , \;\;\; v(x,t) \law t^{-1/3} 
v(x/t^{2/3},1) , \;\;\; q(x,t) \law t^{2/3} q(x/t^{2/3},1) ,
\label{scalings}
\eeq
where $\law$ means that both sides have the same probability distribution.
Thus, any equal-time statistics at a given time $t>0$ can be expressed in terms
of the same quantity at the time $t=1$ through appropriate rescalings.
In this article we only investigate equal-time statistics, so that $t$ can
be seen as a mere parameter in the explicit solution (\ref{psinu}).
Then, it is convenient to introduce the dimensionless coordinates,
\beq
Q= \frac{q}{\gam} , \;\; X= \frac{x}{\gam}, \;\; V=\frac{tv}{\gam} , \;\;
\Psi= \frac{t\psi}{\gam^2} , \;\; C= \frac{t c}{\gam^2} ,
\;\; \mbox{with} \;\; \gam = (2D t^2)^{1/3} ,
\label{QXdef}
\eeq
which express the scaling laws (\ref{scalings}) (here $c$ is the parabola height
that will be introduced below in Eq.(\ref{paraboladef})). Thus, probability distributions
written in terms of these variables no longer depend on time, and the scale $X=1$
is the characteristic length of the system, at any time. On large quasi-linear
scales, $X \gg 1$, density fluctuations are small and the distributions are
strongly peaked around their mean, with tails that are directly governed by
the initial conditions (but shocks cannot be neglected).
On small nonlinear scales, $X \ll 1$, density fluctuations are large (e.g., most
Eulerian intervals are empty) and probability distributions show broad
power-law regions. These behaviors will be clearly seen in the following
sections.

\subsection{Geometrical interpretation}
\label{Geometrical-interpretation}

As is well known \cite[]{Burgersbook}, the minimization problem (\ref{qmin})
has a nice geometrical solution. Indeed, let us consider the 
downward parabola $\cP_{x,c}(q)$ centered at $x$ and of maximum $c$, i.e. of 
vertex $(x,c)$, of equation
\beq
\cP_{x,c}(q) = - \frac{(q-x)^2}{2 t} + c .
\label{paraboladef}
\eeq
Then, starting from below with a large negative value of $c$, such that the
parabola is everywhere well below $\psi_0(q)$ (this is possible thanks to the
scaling $\psi_0(\lambda q) \law \lambda^{1/2} \psi_0(q)$ which shows 
that $\psi_0(q)$ only grows as $|q|^{1/2}$
at large $|q|$), we increase $c$ until the two curves touch one another.
Then, the abscissa of the point of contact is the Lagrangian coordinate
$q(x,t)$ and the potential is given by $\psi(x,t)=c$.
In order to use this geometrical construction, it will be more convenient
in the following to normalize the potential $\psi_0$ by 
$\psi_0(q_-)=0$, where we first restrict the system to the finite interval 
$[q_-,q_+]$, and to eventually take the limits $q_{\pm} \rightarrow \pm\infty$
\cite{Frachebourg2000}, instead of normalizing at the origin $q=0$ as in 
(\ref{psi0def}). Indeed, this avoids making the point $q=0$ artificially play 
a special role. With this choice, the initial potential $\psi_0(q)$ is a single
Brownian motion that starts from the left boundary $q_-$. 

For the white-noise initial conditions (\ref{v0def}), the process 
$q\mapsto\psi_0$ is Markovian. Then, following the approach of 
\cite{Frachebourg2000}, from the geometrical construction 
(\ref{paraboladef})
one can see that a key quantity is the conditional probability density
$K_{x,c}(q_1,\psi_1;q_2,\psi_2)$ for the Markov process 
$\psi_0(q)$, starting from $\psi_1$ at $q_1$, to end at
$\psi_2$ at $q_2 \geq q_1$, while staying above the parabolic barrier,
$\psi_0(q)>\cP_{x,c}(q)$, for $q_1\leq q\leq q_2$. This kernel was obtained
in \cite{Frachebourg2000} and we recall its expression in
Appendix~\ref{sec:transition} with our notations. We also derive the
closely related kernel $E_{x,c}(q_1,\psi_1;q_2,\psi_2;q)$,
defined in Eq.(\ref{E1}), which only counts among these initial conditions
the ones that have a last excursion below $\cP_{x,c+\dd c}$ in the range
$[q,q+\dd q]$.

\section{Known Eulerian distributions}
\label{Known-Eulerian-distributions}

We briefly recall in this section the expressions of the one-point distributions,
$p_x(q)$ and $p_x(v)$, of the Lagrangian coordinate $q(x,t)$ and velocity
$v(x,t)$ at the Eulerian point $x$. We also consider the two-point distributions
$p_{x_1,x_2}(q_1,q_2)$ and $p_{x_1,x_2}(q_1,q_2)$. These results
were already obtained in \cite{Frachebourg2000}, but they are the basis of our
computation in the following sections of the distributions of Lagrangian and
velocity increments, from which we obtain the distribution of the matter density,
and of the distribution of the Eulerian increment. We give more details and
explicit expressions in Appendix~\ref{Eulerian-distributions}.

\subsection{One-point Eulerian distributions $p_x(q)$ and $p_x(v)$}
\label{One-point-Eulerian}

To any Eulerian point $x$ we can associate the Lagrangian coordinate $q(x,t)$
defined as the location of the minimum in Eq.(\ref{qmin}), except at shock 
locations where there are two (or more) contact points
between the initial potential $\psi_0(q)$ and the first-contact parabola
$\cP_{x,c}$. Since shocks are in finite number per unit length 
\cite{AvellanedaE1995,She1992},
Eulerian points have a well-defined Lagrangian coordinate $q(x,t)$ with
probability one. However, note that the Eulerian position $x$ is usually not 
``occupied'' by the infinitesimal mass that was initially located at $q$, as 
all the matter is collected within shocks (thus a given Eulerian point has almost
surely a zero matter density) \cite{Vergassola1994}. Nevertheless, through 
Eq.(\ref{psinu0}) one can derive the properties of the velocity field from the 
Lagrangian coordinate $q(x,t)$.

The one-point distribution, $p_x(q)$, of the Lagrangian coordinate $q$ at point
$x$, can be readily obtained from the kernel $E_{x,c}$ given in Eq.(\ref{ExcG}),
or the kernel $K_{x,c}$, as shown in \cite{Frachebourg2000}.
For instance, from the definition of $E_{x,c}$ we can write
\beq
p_x(q) = \lim_{q_{\pm}\rightarrow\pm\infty} \int \dd c\dd\psi_+ 
\; E_{x,c}(q_-,0;q_+,\psi_+;q) ,
\label{pxq1}
\eeq
where we normalized the initial potential by $\psi_0(q_-)=0$ 
and we let $q_{\pm}\rightarrow\pm\infty$ as the size of the system goes to 
infinity, as discussed below (\ref{paraboladef}). 
Thus, in Eq.(\ref{pxq1}) we count all initial conditions $\psi_0(q)$ that
have a first-contact point of abscissa $q$ with a parabola $\cP_{x,c}$,
and we integrate over all possible heights $c$. We recall in
Appendix~\ref{One-point} the explicit expressions of $P_X(Q)$ and $P_X(V)$,
in terms of the scaling variables (\ref{QXdef}), see Eqs.(\ref{PXQJJ})-(\ref{Jdef})
and \cite{Frachebourg2000}.
Both distributions are related through the change of variable $X=Q+V$,
that expresses the second equation (\ref{psinu0}). 
Thanks to the homogeneity and isotropy of the system, the distribution $P_X(Q)$ 
only depends on the distance $|Q-X|$, whereas $P(V)$ is even and no longer
depends on $X$.
The asymptotic behavior of the distribution of the velocity $V$ 
(and of the Lagrangian coordinate $Q=X-V$),
\beq
|V| \gg 1 : \;\;\; P(V) \sim \frac{2\,|V|}{\Aip(-\om_1)} \, e^{-\om_1 |V|-|V|^3/3} ,
\label{PVinf}
\eeq
shows that $P(V)$ decreases faster than a Gaussian at large $V$
\cite{Avellaneda1995,Frachebourg2000}. Contrary to cases where the initial
velocity field has no ultraviolet divergence (i.e. the initial variance
$\sigma_{v_0}^2(0)=\lag v_0^2\rag$ is finite, as for the case of Brownian initial
velocity \cite{Valageas2008}), the large-$v$ tail cannot be directly understood
from the statistics of rare local peaks in the initial velocity field, 
Here, as we have recalled above, at any time $t>0$ all the matter has collapsed 
within a finite number of shocks per unit length, which merge in the course of time
to build increasingly massive shocks within larger voids 
\cite{She1992,AvellanedaE1995,Frachebourg2000}. Then, the typical velocities 
observed in the system are governed by this merging process, rather than by the
initial velocities of regular points that would not have collided yet.
Nevertheless, the cubic exponential tail (\ref{PVinf}) can be understood as follows.
A structure with a large velocity $v$ has traveled by time $t$ over a distance 
of order $x\sim v t$. On the other hand, the mean velocity $\bv_0(q)$ of the mass 
that was initially located in the Lagrangian interval $[q_1,q_2]$, of size 
$q=q_2-q_1$, is $\bv_0(q)=\int_{q_1}^{q_2}\dd q' \, v_0(q')/q=(\psi_2-\psi_1)/q$.
It is Gaussian with a variance $\sigma_{\bv_0}^2(q) = D/q$ from Eq.(\ref{psi0def}). 
Since momentum is conserved by the inviscid Burgers dynamics \cite{Burgersbook},
so that the momentum of a shock is equal to the sum of the initial momenta of
all the particles it contains,
we can associate to the velocity $v$ and the distance $x=v t$ the Gaussian weight
$\sim e^{-v^2/\sigma_{\bv_0}^2(vt)} \sim e^{-v^3t/D}$, where we did not write
numerical factors in the exponential. This gives back the cubic exponential tail 
(\ref{PVinf}). Even though we followed shocks in the previous argument, in spite
of the fact that they occupy a set of zero measure in Eulerian space, this still
sets the tail of the Eulerian velocity field as the velocity of a shock located
at position $x$ is related to the local velocity field as 
$v^{\rm shock}=(v(x^-)+v(x^+))/2$, and $v(x,t)$ has a constant slope of $1/t$
in-between shocks, see \cite{Burgersbook}.
Cubic exponential tails such as (\ref{PVinf}) are characteristic of probability
distributions obtained for these white-noise initial conditions 
\cite{Avellaneda1995,AvellanedaE1995,Frachebourg2000}.

\subsection{Two-point Eulerian distributions $p_{x_1,x_2}(q_1,q_2)$ and 
$p_{x_1,x_2}(v_1,v_2)$}
\label{Two-point-Eulerian}

We now consider the two-point distribution, $p_{x_1,x_2}(q_1,q_2)$, of the
Lagrangian coordinates $\{q_1,q_2\}$ associated with the Eulerian locations
$\{x_1,x_2\}$. We take $x_1<x_2$, which implies that $q_1\leq q_2$ since
particles do not cross each other and therefore remain well-ordered.
One needs to consider the two cases, i) $q_1\neq q_2$, and ii) $q_1=q_2$. 
The first case, associated with different first-contact points, gives
the contribution \cite{Frachebourg2000}
\beq
P_{X_1,X_2}^{\neq}(Q_1,Q_2) = \theta(Q_2-Q_1) \, \cJ(Q_1-X_1) \, \cJ(X_2-Q_2) \, 
\cH_{X_1,X_2}(Q_1,Q_2) ,
\label{PQ1Q2i}
\eeq
where $\theta(Q_2-Q_1)$ is the Heaviside function and we introduced the functions 
$\cJ$ and $\cH$ given by Eqs.(\ref{Jdef}) and (\ref{Hdef}), whereas the
second case, associated with a common first-contact point, gives the contribution
\beq
P_{X_1,X_2}^=(Q_1,Q_2) = \delta(Q_2-Q_1) \, \cJ(Q_1-X_1) \, \cJ(X_2-Q_2) \,
e^{-(Q_1-X_1)^3/3+(Q_2-X_2)^3/3} .
\label{PQ1Q2ii}
\eeq
One can check that the function $\cH_{X_1,X_2}(Q_1,Q_2)$, whence the contribution
$P_{X_1,X_2}^{\neq}(Q_1,Q_2)$, and the contribution $P_{X_1,X_2}^=(Q_1,Q_2)$
are invariant with respect to uniform translations of $X_i$ and $Q_i$, in agreement
with the statistical homogeneity of the system.
Then, the full distribution $P_{X_1,X_2}(Q_1,Q_2)$ is given by the sum of
both contributions (\ref{PQ1Q2i}) and (\ref{PQ1Q2ii}).
Next, the two-point velocity distribution, $P_{X_1,X_2}(V_1,V_2)$, is
obtained from Eqs.(\ref{PQ1Q2i}), (\ref{PQ1Q2ii}), by using $V_i=X_i-Q_i$.

We can note here that, thanks to the Markovian character of the process
$q\mapsto \psi_0(q)$, the $n$-point distributions of the velocities $v_i$,
and of the Lagrangian coordinates $q_i$, factorize as \cite{Frachebourg2000}
\beq
p_{x_1,..,x_n}(v_1,..,v_n) = p_{x_1}(v_1) \, p(x_2,v_2|x_1,v_1) .. 
p(x_n,v_n|x_{n-1},v_{n-1}) ,
\label{pnvn}
\eeq
and
\beq
p_{x_1,..,x_n}(q_1,..,q_n) = p_{x_1}(q_1) \, p(x_2,q_2|x_1,q_1) .. 
p(x_n,q_n|x_{n-1},q_{n-1}) ,
\label{pnqn}
\eeq
with the transition kernels
\beq
p(x_2,v_2|x_1,v_1) = \frac{p_{x_1,x_2}(v_1,v_2)}{p_{x_1}(v_1)} \;\;\;\;
\mbox{and} \;\;\;\; p(x_2,q_2|x_1,q_1) = \frac{p_{x_1,x_2}(q_1,q_2)}{p_{x_1}(q_1)} ,
\label{p2bar1}
\eeq
that can be obtained from the two-point and one-point distributions derived above.
Again, the kernels $p(x_2,v_2|x_1,v_1)$ and $p(x_2,q_2|x_1,q_1)$ are invariant 
with respect to uniform translations of the spatial coordinates $x_i$ and $q_i$.
However, contrary to the case of Brownian initial velocity
\cite{Bertoin1998,Valageas2008}, the transition kernel
does not only depend on the two relative distances $x_2-x_1$ and $q_2-q_1$
(thus it also depends on the third distance $q_1-x_1$). This means that the inverse
Lagrangian map, $x\mapsto q$, does not have independent increments.

\section{Probability distributions of the Lagrangian and velocity increments}
\label{Lagrangian and velocity increments}

We now consider the probability distributions, $p_x(q)$ and $p_x(v)$, of the Lagrangian
increment, $q=q_2-q_1$, and of the velocity increment, $v=v_2-v_1$, over the
Eulerian distance $x= x_2-x_1$. These distributions can be directly obtained from
the two-point distributions (\ref{PQ1Q2i}) and (\ref{PQ1Q2ii}), but they were not
studied in previous works (except for the singular part (\ref{PX0}) associated with
voids). In particular, as noticed in the conclusion
of \cite{Frachebourg2000}, the asymptotics of $p_x(v)$ at large $v$ cannot be obtained
in a straightforward manner from the estimations of their section~5, as the latter apply
to the limit of large distance $x$ at fixed $v_1$ and $v_2$.
Next, we shall need the distribution $p_x(q)$ to derive the distribution of the
overdensity at scale $x$ in section~\ref{Density}.

\subsection{Lagrangian increment, $q=q_2-q_1$, and velocity increment, $v=v_2-v_1$}
\label{Lagrangian-increment}

The probability distribution $P_X(Q)$ of the Lagrangian increment, $Q=Q_2-Q_1$,
can be obtained by integrating the sum of the bivariate 
distributions (\ref{PQ1Q2i}) and (\ref{PQ1Q2ii}) over the variable $Q_1$ at fixed
$Q=Q_2-Q_1$. This gives
\beq
P_X(Q) = \int_{-\infty}^{\infty} \dd Q_1 \, \cJ(Q_1) \, \cJ(X-Q-Q_1) 
\left[\theta(Q) \cH_{0,X}(Q_1,Q_1+Q) + \delta(Q) e^{-Q_1^3/3+(Q_1-X)^3/3} \right] ,
\label{PXQ1}
\eeq
where $\theta(Q)$ is the Heaviside function. The second term gives a contribution
of the form $P_X^=(Q) = \delta(Q) \, P_X^0$, given by Eq.(\ref{PX0}) in
Appendix~\ref{Void-distribution}. 
Note that Eulerian intervals with $Q=0$ also have a zero matter content so that
$P_X^0$ is also the probability for an interval of size $X$ to be empty
(see section~\ref{Density} below where we discuss the matter density field),
in agreement with the result obtained in \cite{Frachebourg2000} for this void
probability.
We recall the properties of this distribution of voids in Appendix~\ref{Void-distribution}
and Fig.~\ref{figPvoid}.

In this article we are mostly interested in the regular part, $P_X^{\neq}(Q)$, associated
with non-empty Eulerian intervals, which has not been studied in previous works. 
From the first term in expression (\ref{PXQ1}) it reads as
\beqa
P_X^{\neq}(Q) & = & \theta(Q) \, 2\sqrt{\pi X} e^{-X^3/12} \inta 
\frac{\dd s \dd s_1 \dd s_2}{(2\pi\ii)^3} \,
\frac{e^{s (Q-X)+(s_1+s_2)X/2+(s_1-s_2)^2/(4X)}}
{\Ai(s_1)\Ai(s_2)\Ai(s_1-s)\Ai(s_2-s)}
\nonumber \\
&& \times \int_0^{\infty} \dd r \, e^{Xr} \, \Ai(r+s_1) \Ai(r+s_2) .
\label{PXQs1}
\eeqa
We recall in Appendix~\ref{Laplace} an alternative expression for the integral
over $r$ that appears in Eq.(\ref{PXQs1}), obtained in \cite{Frachebourg2000},
which is useful to derive asymptotic behaviors.
Thus, at large distances, $X\gg 1$, Eq.(\ref{PXQs1}) yields the asymptotic 
behaviors
\beqa
X\gg 1 & : & \;\; P_X^{\neq}(Q) \sim \frac{\sqrt{X}}{\Aip(-\om_1)^2} \, Q^{-1/2}  
\, e^{-\om_1 X-X^3/12} \hspace{1.27cm} \mbox{for} \;\; 0 < Q \ll X^{-2} , 
\label{PXQasympXl1} \\
&& \;\; P_X^{\neq}(Q) \sim \frac{\sqrt{\pi}}{\Aip(-\om_1)^2} \, 
|V|^{3/2} \, e^{-\om_1|V|-|V|^3/12} \hspace{1cm} \mbox{for} \;\;\; |V|\gg 1 
\;\; \mbox{and} \;\; Q\gg X^{-2} ,
\label{PXQasympXl2}
\eeqa
where $V=X-Q$ is the dimensionless velocity as in (\ref{QXdef}).
At small distances, $X\ll 1$, Eq.(\ref{PXQs1}) leads to
\beqa
X\ll 1 & : & \;\; P_X^{\neq}(Q) \sim \frac{X}{\sqrt{\pi}} \, Q^{-1/2} 
\hspace{3.2cm} \mbox{for} \;\; 0 < Q \ll 1 , \label{PXQasympXs1} \\
&& \;\;  P_X^{\neq}(Q) \sim 2\sqrt{\pi} X \, Q^{5/2} \, e^{-\om_1 Q - Q^3/12}
\hspace{1cm} \mbox{for} \;\; 1 \ll Q \ll X^{-1/2} , \label{PXQasympXs2} \\
&& \;\;  P_X^{\neq}(Q) \sim 2 \pi \sqrt{X} \, Q^{3/2} \, e^{-\om_1 Q - Q^3/12}
\hspace{1cm} \mbox{for} \;\; Q \gg X^{-1/2} .
\label{PXQasympXs3}
\eeqa
Thus, at all scales $X$ the distribution $P_X^{\neq}(Q)$ displays an inverse
square root tail at low $Q$. At large $X$ this tail has an exponentially small 
weight, that scales as the weight $P_X^0$ of the empty cells, and it is restricted
to very low $Q$, whereas at small $X$ it describes the full low-$Q$ regime.
As expected, we can check from Eq.(\ref{PXQasympXl2}) that on large scales, 
$X\gg 1$, the Lagrangian increment is centered on $X$, with the usual cubic
exponential tails encountered for this white-noise initial velocity spectrum, 
whereas on small scales, $X \ll 1$, the distribution shows a monotonous decline.

\begin{figure}
\begin{center}
\epsfxsize=6.3 cm \epsfysize=5 cm {\epsfbox{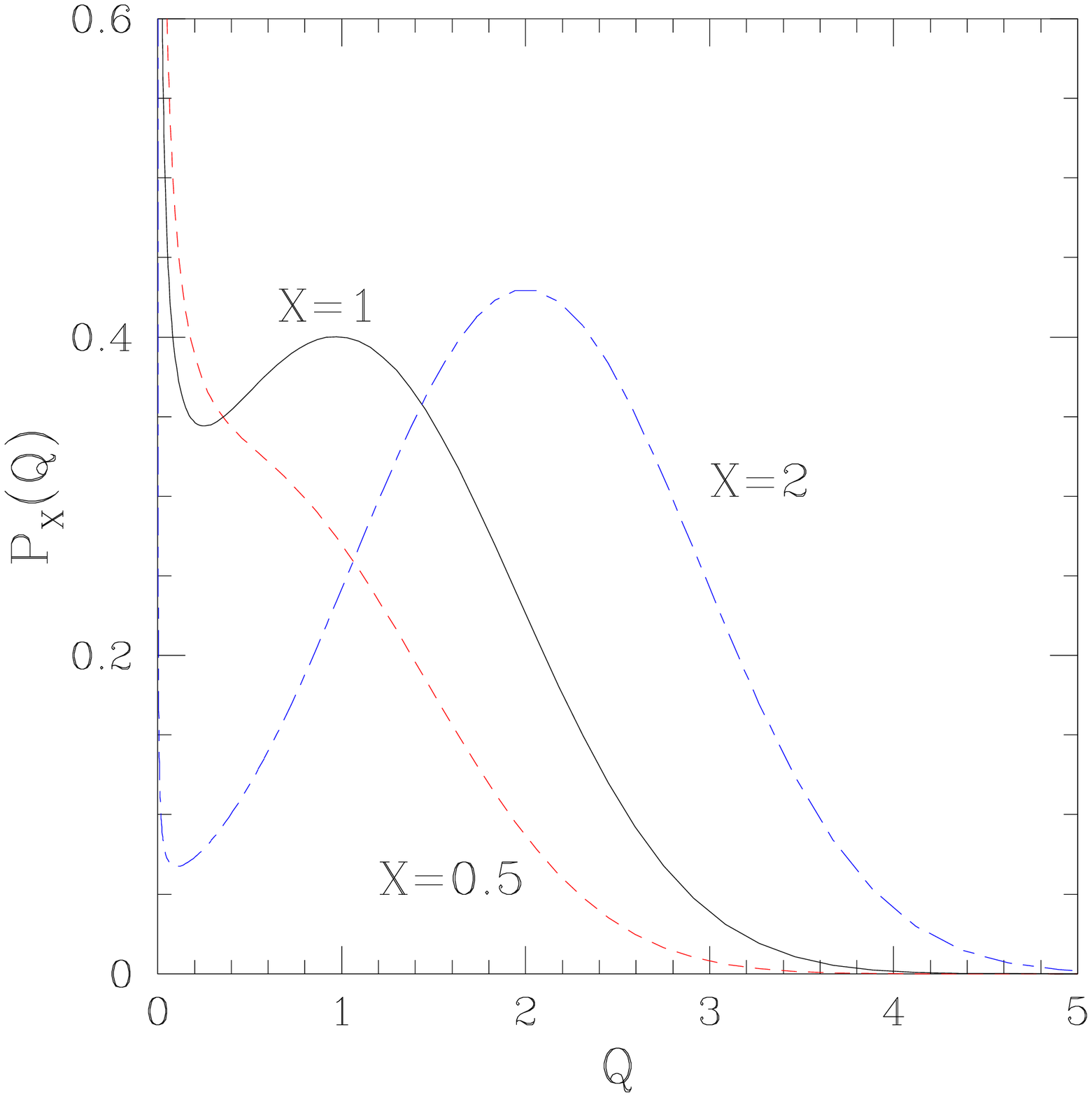}}
\epsfxsize=6.3 cm \epsfysize=5 cm {\epsfbox{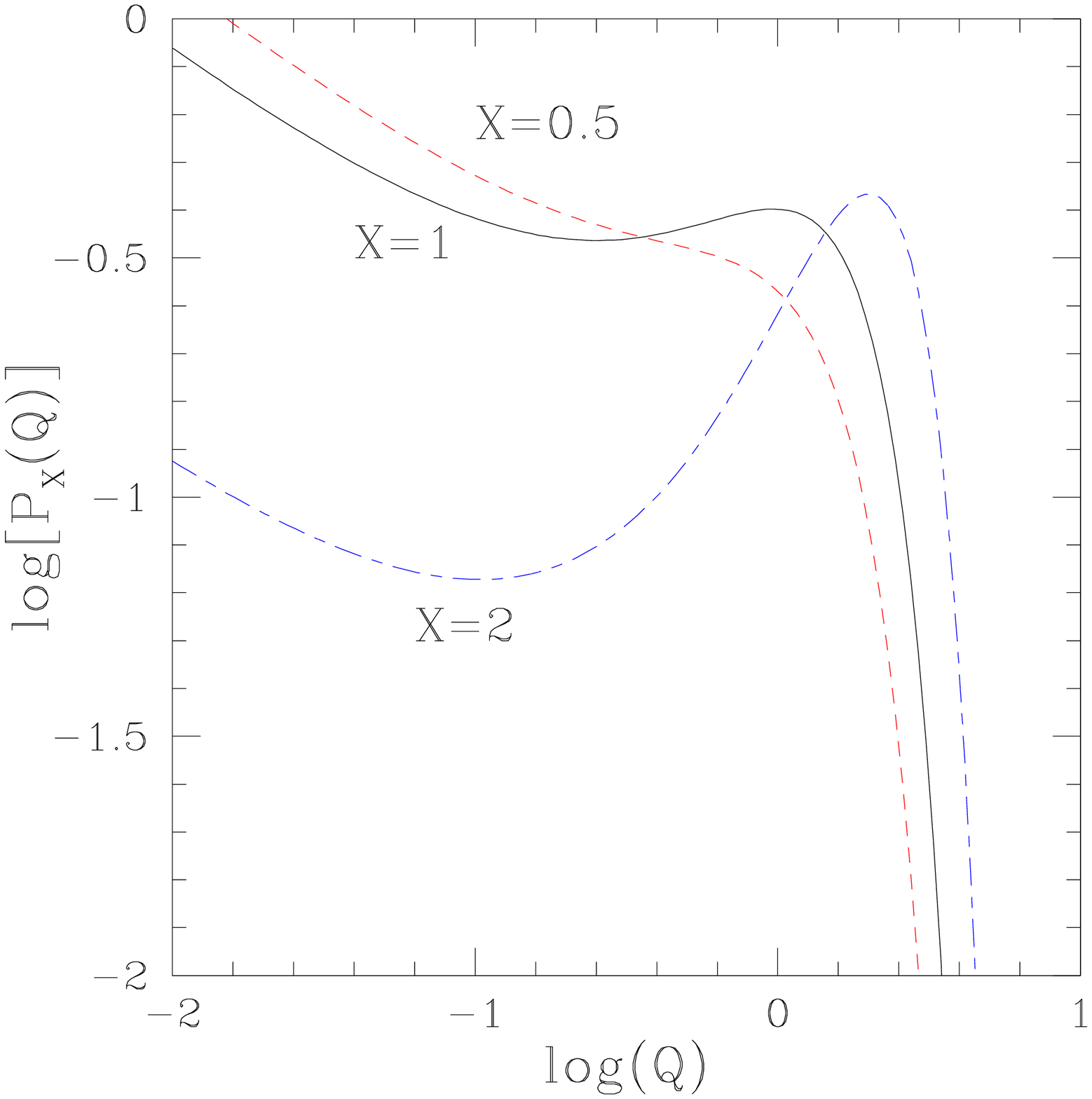}}
\end{center}
\caption{(Color online) {\it Left panel:} The probability distribution $P_X(Q)$
of the Lagrangian increment $Q$, for three Eulerian sizes, $X=0.5,1$ and $2$,
from Eq.(\ref{PXQs1}).
We have $Q\geq 0$ and all curves display an inverse square root singularity
$\propto 1/\sqrt{Q}$ at $Q\rightarrow 0^+$. 
In addition, there is a Dirac contribution, $P_X^0 \delta(Q)$, with the weight 
$P_X^0$ displayed in Fig.~\ref{figPvoid}.
{\it Right panel:} Same as left panel but on a logarithmic scale.}
\label{figPXQ}
\end{figure}

We display in Fig.~\ref{figPXQ} the probability distribution $P_X(Q)$ for three 
Eulerian sizes. This clearly shows the change of shape as we go from large
to small scales, as well as the translation of the mean $\lag Q\rag$, that follows
$X$ from the conservation of matter (see Eq.(\ref{PXeta}) below).
Note that for numerical purposes, in order to follow the evolution of $P_X(Q)$
with $X$, and its behavior over the different characteristic domains listed in
Eqs.(\ref{PXQasympXl1})-(\ref{PXQasympXs3}), it is useful to gradually move
the integration contours in the complex plane of Eq.(\ref{PXQs1}) as one
goes from one regime to another one (but making sure that one does not cross
singularities).
One interest of these results is to provide an explicit example that is representative
of initial conditions in the range $-1<n<1$, where $n$ is the slope of the initial
energy spectrum (i.e. $E_0(k) \propto k^{n+1-D}$ in $D$ dimensions), which show
significant power at high wavenumbers. Then, we can see that the distribution
$P_X^{\neq}(Q)$ always diverges at low $Q$ (i.e. at low density) as $1/\sqrt{Q}$.
This implies in particular that, contrary to the cases $-3<n<-1$, the very low-$Q$
part of the distribution cannot be estimated through steepest-descent approaches
that apply to rare events, as discussed in \cite{Valageas2009}.
Nevertheless, these approaches can give an estimate of $P_X^{\neq}(Q)$ in the
quasi-linear limit $X\rightarrow\infty$ at fixed $Q$, studied in section~\ref{Large-scales}
below.

\begin{figure}
\begin{center}
\epsfxsize=6.3 cm \epsfysize=5 cm {\epsfbox{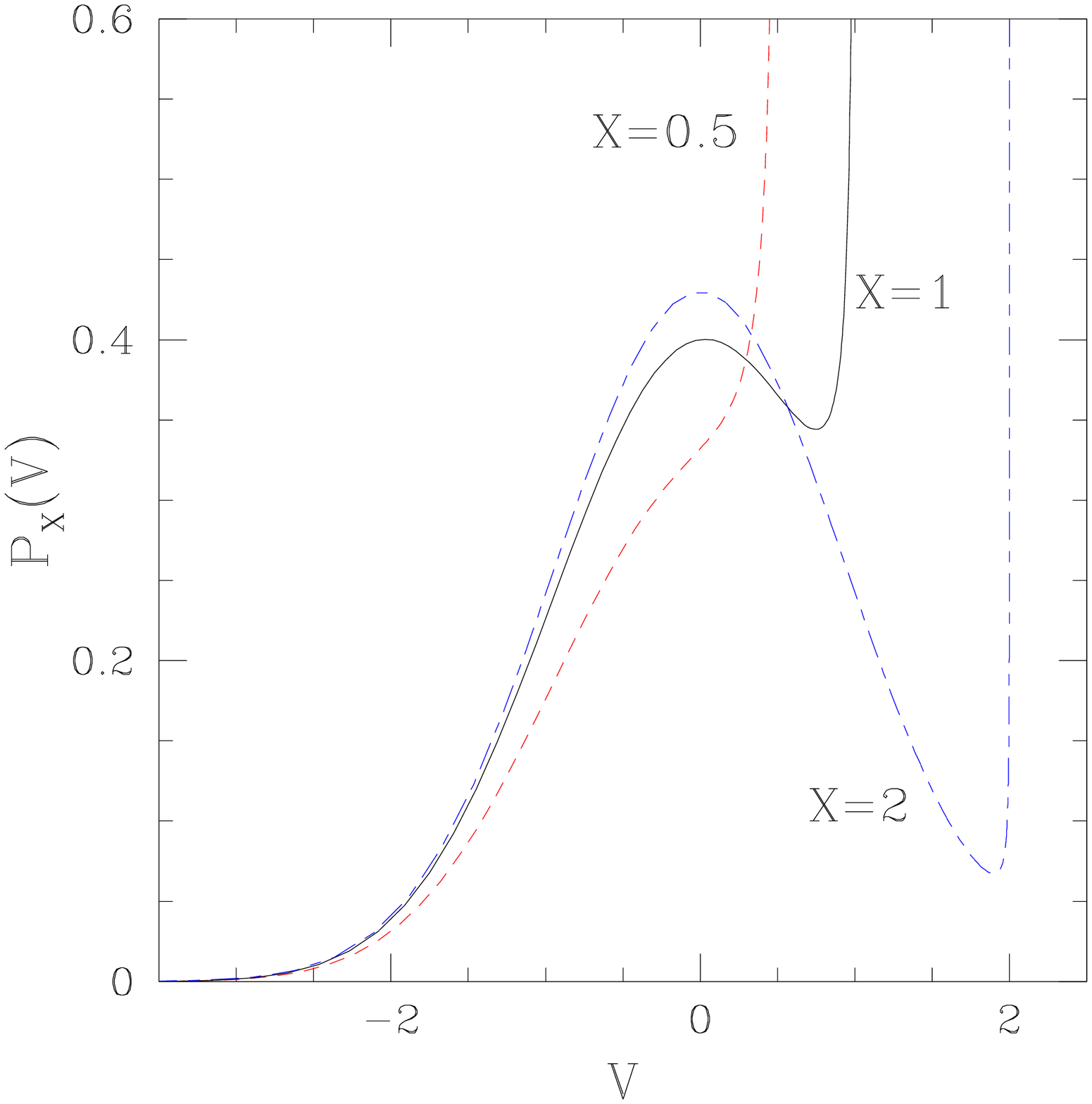}}
\epsfxsize=6.3 cm \epsfysize=5 cm {\epsfbox{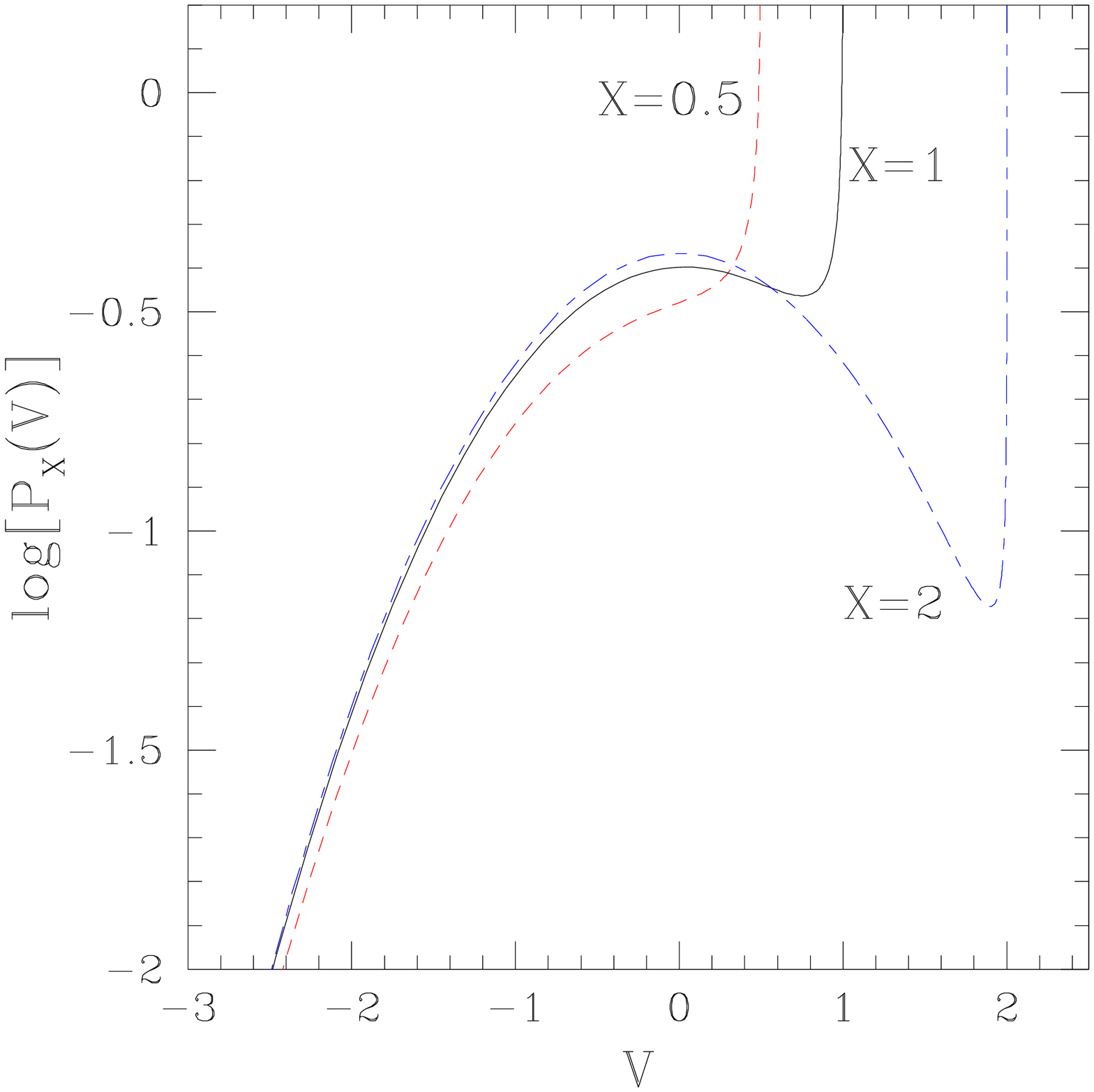}}
\end{center}
\caption{(Color online) {\it Left panel:} The probability distribution $P_X(V)$
of the velocity increment $V$, for three Eulerian sizes, $X=0.5,1$ and $2$,
from Eq.(\ref{PXQs1}).
We have $V\leq X$ and all curves display an inverse square root singularity
$\propto 1/\sqrt{X-V}$ at $V\rightarrow X^-$. 
In addition, there is a Dirac contribution, $P_X^0 \delta(X-V)$, with the weight 
$P_X^0$ displayed in Fig.~\ref{figPvoid}.
{\it Right panel:} Same as left panel but on a semi-logarithmic scale.}
\label{figPXV}
\end{figure}

The probability distribution, $P_X(V)$, of the velocity increment, 
$V=V_2-V_1$, is obtained from the distribution $P_X(Q)$ by using the relation
$V=X-Q$. We show our results in Fig.~\ref{figPXV}, for the same three 
Eulerian scales as for $P_X(Q)$ displayed in Fig.~\ref{figPXQ}. 
There is no longer a translation of the typical velocity, since $\lag V\rag=0$
for any scale $X$, but we clearly see the translation of the inverse-square
root tail, $\sim 1/\sqrt{X-V}$, that follows the upper bound $V\leq X$ associated
with empty cells (as seen from the relation $X=Q+V$ and the constraint $Q\geq 0$).

\subsection{Asymptotic distribution on large scales}
\label{Large-scales}

On large scales, $X\gg 1$, the distributions $P_X(Q)$ and $P_X(V)$, except for 
the exponentially small contributions associated with $Q=0$ and $Q\ll X^{-2}$
in (\ref{PX0}) and (\ref{PXQasympXl1}), can be described by the symmetric 
distribution $\cFi$,
\beq
X \gg 1: \;\;  P_X(Q) \sim \cFi(X-Q) \hspace{0.7cm} \mbox{and} 
\hspace{0.7cm} P_X(V) \sim \cFi(V) ,
\label{PxFi}
\eeq
with a Fourier transform $\hcFi$ given by:
\beq
\cFi(V) = \int_{-\infty}^{\infty} \frac{\dd k}{2\pi} \, e^{ikV} \hcFi(k)
\;\;\;\; \mbox{with} \;\;\;\;  
\hcFi(k) = \left( \inta\frac{\dd s'}{2\pi\ii} \, \frac{1}{\Ai(s') \Ai(s'+\ii k)}
\right)^2 .
\label{hFidef}
\eeq
Equation (\ref{hFidef}) is obtained from Eq.(\ref{PXQs1}) by using the 
asymptotic behavior of the integral over $r$, as given by the first term
in Eq.(\ref{gPhih}),
and next making the change of variable $s=\ii k$.
The scaling function $\cFi(V)$ no longer depends on $X$: the distribution
of the velocity increment $V$ converges to the finite distribution $\cFi$
on large scales $X \rightarrow\infty$.
Note that in this limit the upper boundary on $V$, $V\leq X$, associated 
with the positivity of $Q=X-V$, goes to $+\infty$ so that the limiting function
$\cFi(V)$ is defined over the whole real axis. This is also why the Laplace
transform (\ref{PXQs1}) naturally gives rise to the Fourier transform 
(\ref{hFidef}) in this limit. Moreover, we can see from Eq.(\ref{hFidef})
that $\cFi(V)$ is even (the change of integration variable
$s'\rightarrow s'-\ii k/2$ in Eq.(\ref{hFidef}) readily shows that $\hcFi(k)$ 
is even).
The asymptotic behaviors of $\cFi(V)$ at large $V$ can be read from 
Eq.(\ref{PXQasympXl2}):
\beq
|V| \gg 1 : \;\;\; \cFi(V) \sim \frac{\sqrt{\pi}}{\Aip(-\om_1)^2} \, 
|V|^{3/2} \, e^{-\om_1|V|-|V|^3/12} .
\label{Fiasymp}
\eeq
Thus, we recover the cubic exponential tails that are characteristic of
white-noise initial conditions
\cite{Avellaneda1995,AvellanedaE1995,Frachebourg2000} and
can be understood at a qualitative level following the discussion
below Eq.(\ref{PVinf}). In fact, as shown in \cite{Valageas2009},
on a quantitative level it is possible to obtain the factor $1/12$
in the exponential (\ref{Fiasymp}) through a simple steepest-descent
approach, that identifies the initial conditions (i.e. the relevant saddle-points)
that give the main contribution to these tails. Then,
the exact result (\ref{Fiasymp}) provides a useful non-trivial test of such
general approaches that rest on some additional assumptions (for instance,
one only looks for symmetric saddle-points).

\begin{figure}
\begin{center}
\epsfxsize=6.3 cm \epsfysize=5 cm {\epsfbox{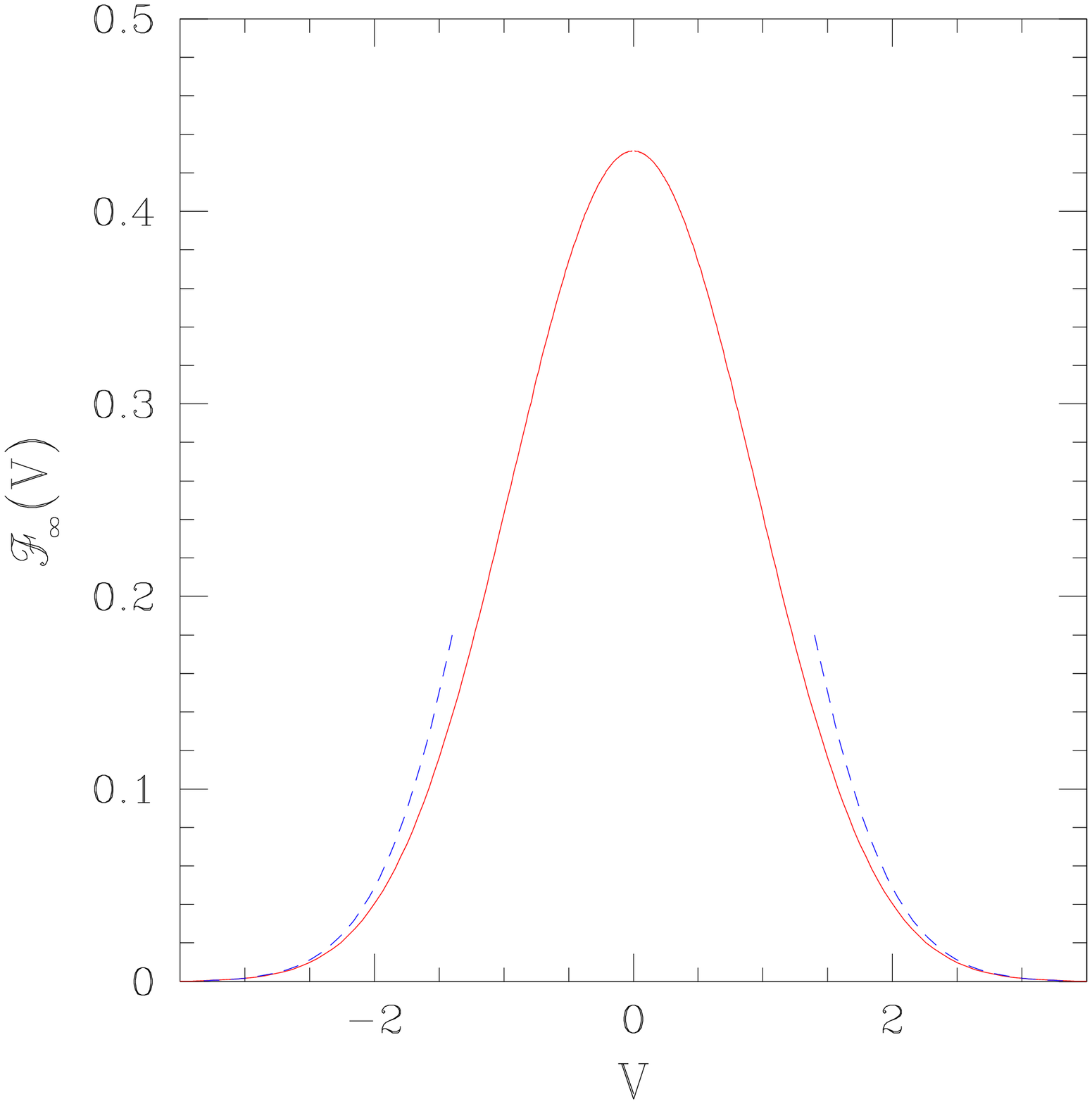}}
\epsfxsize=6.3 cm \epsfysize=5 cm {\epsfbox{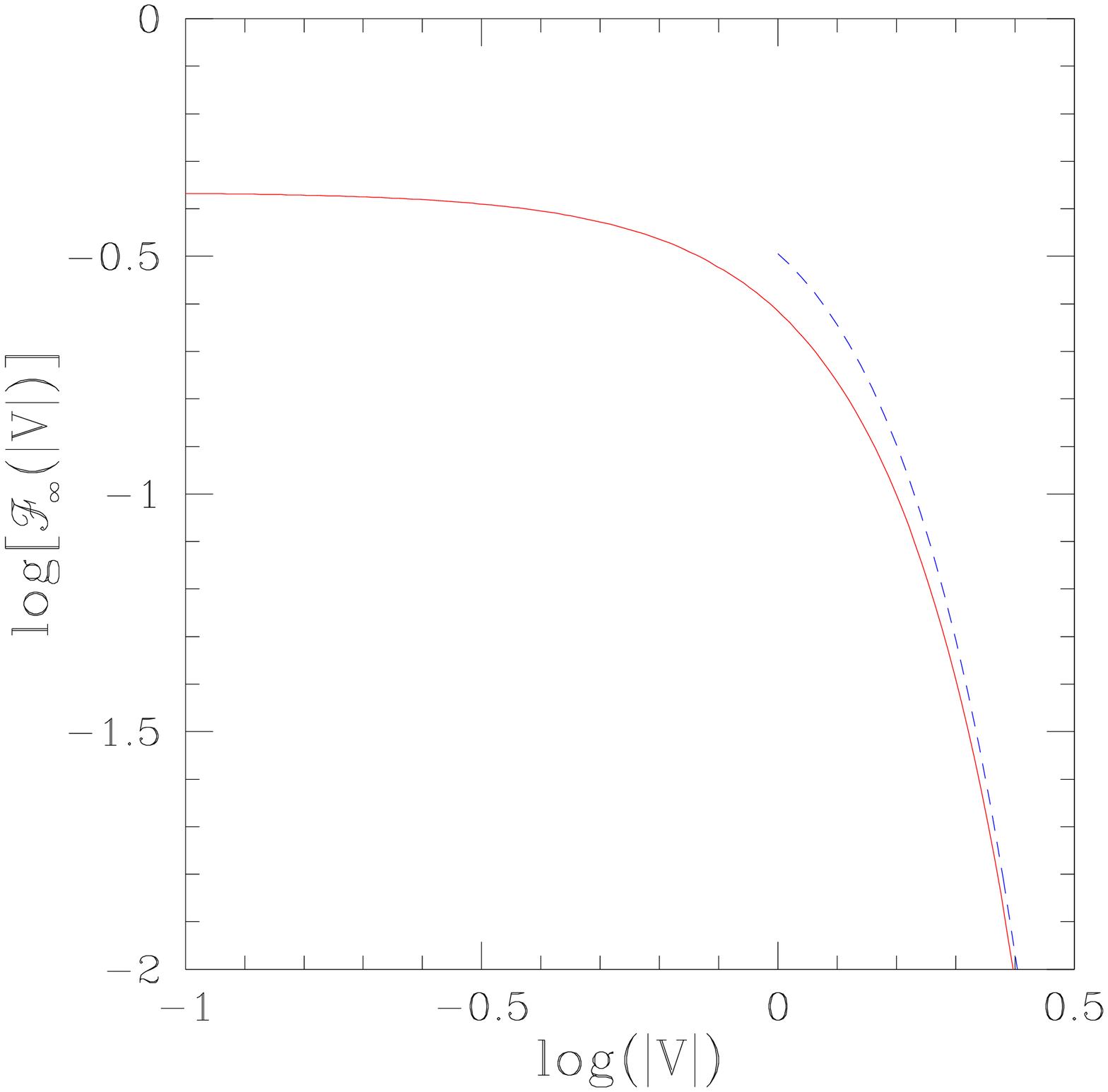}}
\end{center}
\caption{(Color online) {\it Left panel:} The asymptotic distribution $\cFi(V)$
of the velocity increment $V$, reached in the limit of large Eulerian distance $X\gg 1$, 
from Eq.(\ref{hFidef}). 
The dashed lines show the asymptotic behavior (\ref{Fiasymp}).
{\it Right panel:} Same as left panel but on a logarithmic scale.}
\label{figFiV}
\end{figure}

Of course, the scaling function $\cFi(V)$ does not capture the low-$Q$
power-law tail (\ref{PXQasympXl1}). However, using the property
\beq
\inta \frac{\dd s}{2\pi\ii} \frac{1}{\Ai(s)^2} = 1 ,
\label{int1}
\eeq
we can see that the distribution $\cFi(V)$ is normalized to unity as it should,
since the weights of the low-$Q$ tail (\ref{PXQasympXl1}) and of the Dirac
term (\ref{PX0}) vanish in the limit $X\rightarrow \infty$. We show the
scaling function $\cFi(V)$ in Fig.~\ref{figFiV}.

From Eqs.(\ref{PxFi})-(\ref{hFidef}), the moments of the velocity increment 
are given in this limit by
\beq
X \rightarrow +\infty : \;\;\; \lag V^{2n+1}\rag=0 , \;\;\; \lag V^{2n}\rag= 
(-1)^n \frac{\dd^{2n}\hcFi}{\dd k^{2n}}(0) ,
\label{VnXl}
\eeq
whence
\beq
\lag V\rag=0  \;\;\; \mbox{and} \;\;\; \lag Q\rag =X , \;\;\;\;\; 
\lag V^2\rag =  \lag Q^2\rag_c = - \frac{2}{3}\inta\frac{\dd s}{2\pi\ii}
\frac{s}{\Ai(s)^2} \simeq 0.837
\label{V2Xl}
\eeq
We can check that we recover $\lag Q\rag=X$, as implied by the conservation of 
matter. We can see that there is negligible power on large scales since
$\lag Q^2\rag_c$ goes to a constant for $X \rightarrow +\infty$.
This holds for cumulants of all orders, as the scaling function $\cFi(V)$
does not depend on $X$ (see also the left panel of Fig.~\ref{figqnSn} below, where
we can see that both $\lag Q^2\rag_c$ and $\lag Q^4\rag_c$ have a finite nonzero
large-scale limit whereas $\lag Q^3\rag_c$ vanishes by symmetry of $\cFi$).

Equations (\ref{PxFi})-(\ref{Fiasymp}) and (\ref{VnXl}) show that, because of the 
lack of power at large scales in the initial velocity field, at any time $t>0$ 
the system observed at any scale $x$, whatever large, is governed by nonlinear
effects and exhibits strongly non-Gaussian statistics, even though the initial
conditions are Gaussian. Indeed, the redistribution of matter within a series of
discrete shocks has regularized the initially singular white-noise velocity field,
through the balance between the infinite different sign velocities of neighboring
particles, over lengths of order $(2Dt^2)^{1/3}$, that have merged in a single
shock. Moreover, the velocity field in the voids is governed by the motion of
the boundary shocks, since from Eq.(\ref{psinu0}) it has a constant slope
$1/t$ in-between shocks, and the velocity of a shock satisfies
$v^{\rm shock}=(v(x^-)+v(x^+))/2$, see \cite{Burgersbook}. These processes are
clearly non-perturbative and give rise to the non-Gaussian statistical properties
described above in the large-scale limit.
This would not be the case for initial conditions with significant initial power 
on large scales. Then, even though shocks may have formed as soon as $t>0$,
one still recovers the initial Gaussian statistics on large scales, as explicitly
checked in \cite{Valageas2008} for the case of a Brownian initial velocity field
(where the initial energy spectrum is $E_0(k) \propto k^{-2}$ instead of
the constant spectrum associated with the white-noise initial condition studied
in the present article).

Again, these exact results provide a useful confirmation of the results obtained
by approximate methods, such as the steepest-descent approach
of \cite{Valageas2009}. Indeed, there it is found that for $n>D-3$, which includes the
case $\{n=0,D=1\}$ studied in this article, the relevant saddle-points always give
rise to shocks, which is not the case for initial conditions with less initial power
at high wavenumbers (such as $\{n=-2,D=1\}$, i.e. Brownian 1-D initial velocity).

\subsection{Asymptotic distribution on small scales}
\label{Small-scales}

On small scales, apart from the Dirac contribution (\ref{PX0}), associated
with empty cells, and the very large-$Q$ tail (\ref{PXQasympXs3}), the
distributions $P_X(Q)$ and $P_X(V)$ can be described by the function $\cFs$,
\beq
X \ll 1: \;\; P_X(Q) \sim X \, \cFs(Q) \;\;\;\; \mbox{with} \;\;\;\;
Q>0  \;\;\;\; \mbox{and} \;\;\;\; 
\cFs(Q) = \inta \frac{\dd s}{2\pi\ii} \, e^{s Q} \, \tcFs(s) , 
\label{Fsdef}
\eeq
with
\beq
\tcFs(s) = -2 \inta \frac{\dd s'}{2\pi\ii} \frac{1}{\Ai(s')^2}
\frac{\pl}{\pl s'} \frac{\Aip(s'+s)}{\Ai(s'+s)} 
= -4 \inta \frac{\dd s'}{2\pi\ii} \frac{\Aip(s')\Aip(s'+s)}{\Ai(s')^3\Ai(s'+s)} . 
\label{tFsdef}
\eeq
The expressions (\ref{Fsdef})-(\ref{tFsdef}) are obtained from Eq.(\ref{PXQs1})
by taking the Gaussian integration over $s_2$ (since $|s_2-s_1| \sim \sqrt{X}$
we can set at leading order $s_2=s_1$ in (\ref{PXQs1}), apart from the Gaussian 
factor $e^{(s_1-s_2)^2/(4X)}$), and next setting $X=0$, which allows to perform
the integral over $r$.
Note that a change of variable and an integration by parts allow to write
$\cFs(Q)$ as
\beq
\cFs(Q) = 2 Q \, \left( \inta\frac{\dd s'}{2\pi\ii} \, \frac{e^{-s'Q}}{\Ai(s')^2} 
\right) \, \left( \inta\frac{\dd s}{2\pi\ii} \, e^{sQ} \, 
\frac{\Aip(s)}{\Ai(s)} \right) .
\label{Fs1}
\eeq
It is clear that the integral over $Q$ of the distribution (\ref{Fsdef}) is not
normalized to unity since it decreases as $X$ at small Eulerian distance.
Indeed, in this limit almost all Eulerian cells have a zero Lagrangian increment
(whence a zero matter density), associated with the Dirac contribution (\ref{PX0})
(see the first limit in (\ref{PX0asymp})), whereas non-empty cells occur with
a probability proportional to $X$, see Eqs.(\ref{PXQasympXs1})-(\ref{PXQasympXs3}).
This can be directly understood from the fact that all the matter is
condensed into discrete shocks that occur in a finite number per unit length
\cite{She1992,AvellanedaE1995},
so that the probability for an Eulerian interval to contain at least one shock
(which is equal to $1-P_X^0$) scales as $X$ for small cell size $X$.

In fact, the comparison of Eq.(\ref{Fs1}) with results obtained in 
\cite{Frachebourg2000} shows that $\cFs(Q)$ is also the mass function of shocks,
as we shall check in section~\ref{Mass-function} below through a different method.
Therefore, the expression (\ref{Fsdef}) actually means that on small scales the
probability distribution $P_X(Q)$ is asymptotically equal to the probability 
to encounter one shock of strength $Q$ in the interval of size $X$.
Indeed, since shocks are isolated it is clear that in the limit of small size $X$
the probability to have two or more shocks within $X$ goes to zero faster than
$X$, so that $P_X(Q)$ is governed by the probability to encounter one shock
over the length $X$, which directly gives the scaling (\ref{Fsdef}) where
$\cFs(Q)$ would be defined as the shock mass function.
Thus, how results explicitly show how the scaling (\ref{Fsdef}) and the shock
mass function $\cFs(Q)$ arise from the full distribution (\ref{PXQs1}) of the
Lagrangian increment $Q$.

This property is well-known to hold for any Burgers system without dense shocks
\cite{Frisch2001,Tribe2000}. Then, as pointed out in \cite{Tribe2000}, who studied
the case of compactly supported white-noise initial velocity, the statistics of the
velocity field (whence of the Lagrangian increment $Q$) at scales much smaller
than the average distance between shocks are fully determined by the one-point
distribution of shock strength $n(m)$ (which in our case is equal to $\cFs(Q)$
as seen in Eq.(\ref{NF}) below). As noticed in \cite{Tribe2000}, the case of
compact initial conditions is in a different universality class than the system studied
here, where the white-noise initial velocity (\ref{v0def}) extends to the whole
real line, in the sense that the scalings (\ref{QXdef}) no longer hold.
Indeed, the size $L$ of the initially non-zero velocity field introduces a new
scale and at late times there are only two shocks left, which allows \cite{Tribe2000}
to compute both single- and multiple time-velocity structure functions.
Nevertheless, on small scales, much below the typical distance between shocks,
both systems show the same scalings (see Eqs.(\ref{Qns}) and (\ref{Vns}) below),
governed by the one-shock contribution.

\begin{figure}
\begin{center}
\epsfxsize=6.3 cm \epsfysize=5 cm {\epsfbox{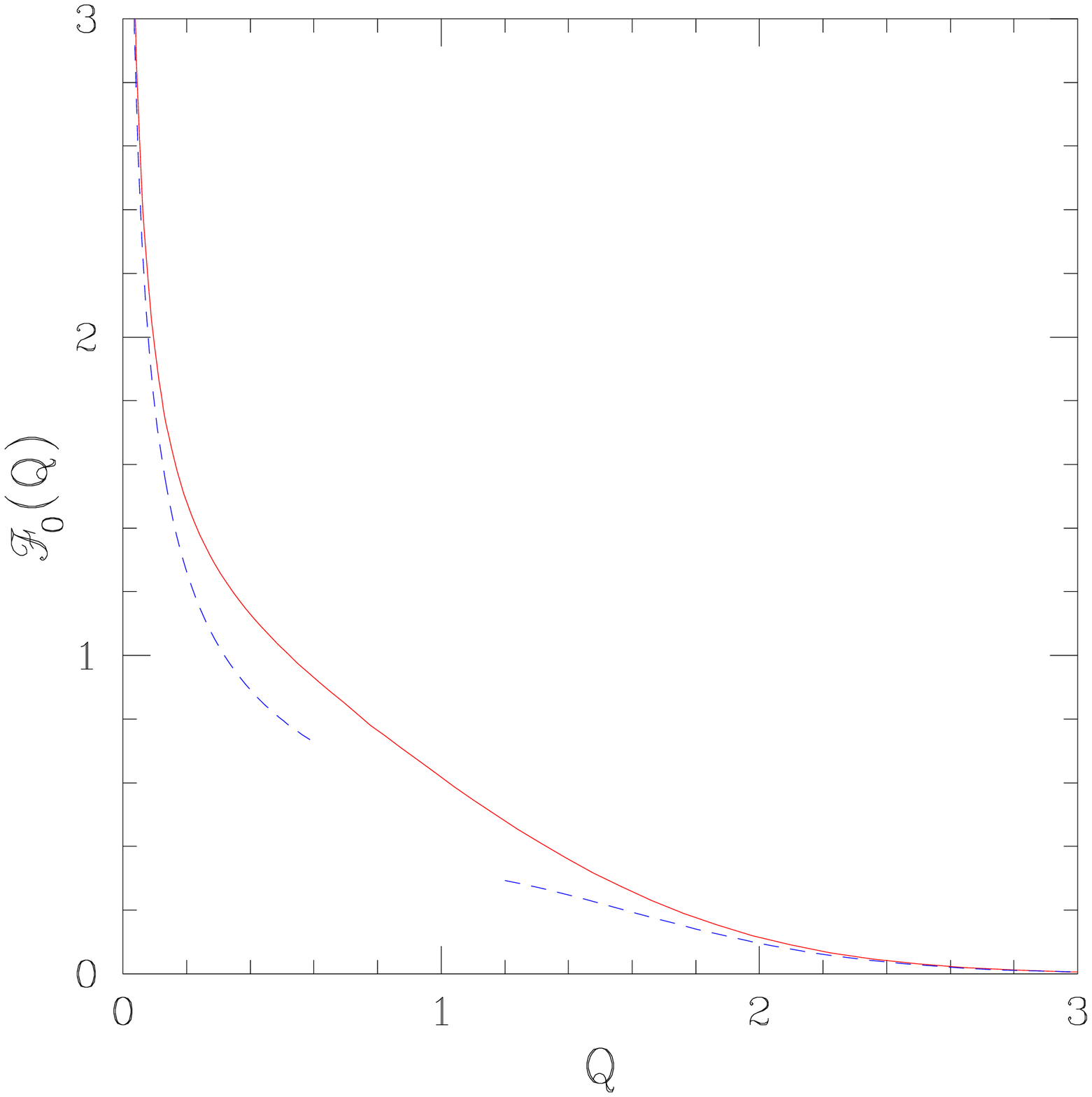}}
\epsfxsize=6.3 cm \epsfysize=5 cm {\epsfbox{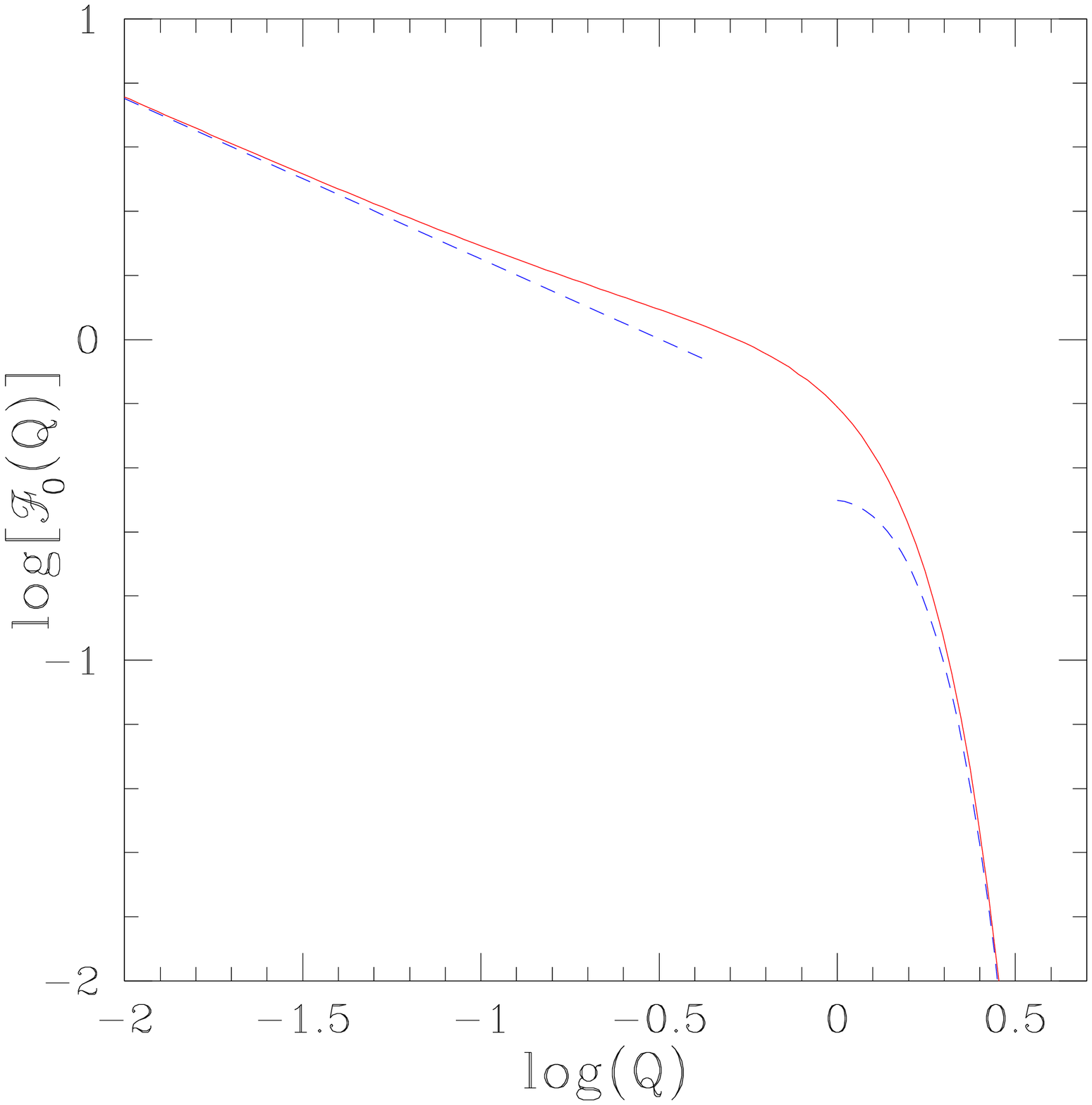}}
\end{center}
\caption{(Color online) {\it Left panel:} The scaling function $\cFs(Q)$ that
describes the distribution of the Lagrangian increment $Q$ in the limit $X \ll 1$, 
from Eqs.(\ref{Fsdef})-(\ref{tFsdef}). This is also the mass function of shocks,
as checked in Eq.(\ref{NF}) below.
The dashed lines are the asymptotic behaviors (\ref{Fsasymp}).
{\it Right panel:} Same as left panel but on a logarithmic scale.}
\label{figF0Q}
\end{figure}

Again the asymptotic behaviors can be read from 
Eqs.(\ref{PXQasympXs1})-(\ref{PXQasympXs3}):
\beq
Q \ll 1 : \;\; \cFs(Q) \sim \frac{1}{\sqrt{\pi Q}} , \;\;\;\;\;
Q \gg 1 : \;\; \cFs(Q) \sim  2\sqrt{\pi} \, Q^{5/2} \, e^{-\om_1 Q - Q^3/12} .
\label{Fsasymp}
\eeq
We show the function $\cFs(Q)$ and its asymptotic tails (\ref{Fsasymp})
in Fig.~\ref{figF0Q}, see also \cite{Frachebourg2000}.
Note that the scaling function $\cFs(Q)$ does not describe the very far tail
$Q\gg X^{-1/2}$ of Eq.(\ref{PXQasympXs3}), which is repelled to infinity in
the limit $X\rightarrow 0$. This very high-$Q$ tail is related to the
behavior of $P_X(Q)$ at large scales, as seen from the comparison with 
Eq.(\ref{PXQasympXs3}). Indeed, it corresponds to the limit of very rare events,
where the tail of the distribution is governed by specific initial conditions,
independently of the scale $X$. These are the saddle-points obtained in
\cite{Valageas2009}, which set the cubic exponential falloffs of both
Eq.(\ref{PXQasympXl2}) and Eq.(\ref{PXQasympXs3}). Thus, for any finite
$X$ the very far tail (\ref{PXQasympXs3}) of the distributions $P_X(Q)$
and $P_X(V)$ is not captured by the shock mass function, but this regime
is repelled to infinity as $X\rightarrow 0$.

For any $\nu>0$, where the contribution from the Dirac term (\ref{PX0}) vanishes,
we obtain for the moments of the Lagrangian increment
\beq
\nu > 0 : \;\;\; \lag Q^{\nu} \rag \sim X \, \Gamma[\nu+1] 
\inta\frac{\dd s}{2\pi\ii} \, (-s)^{-\nu-1} \, \tcFs(s) ,
\label{Qnu}
\eeq
where the integration contour runs to the left of the origin, $\Re(s)<0$.
Thus, we recover the fractality of the inverse Lagrangian map, 
$\lag Q^{\nu} \rag \propto X$, which is well known to be due to the contribution 
from shocks as discussed above \cite{Frisch2001,Tribe2000}.
Indeed, if we have a shock of finite Lagrangian 
length $\delta Q_s$ at position $X_s$, it gives a contribution 
$[Q(X_s+X/2)-Q(X_s-X/2)]^{\nu} \sim (\delta Q_s)^{\nu}$ which remains of order 
unity for $X \rightarrow 0^+$ for any $\nu>0$. Next, the probability to have a 
shock of a given finite strength $\delta Q_s$ in a small Eulerian interval $X$ 
scales as $X$ at small distances, which gives rise to the factor $X$ in 
Eq.(\ref{Qnu}). 
Therefore, the scaling (\ref{Qnu}) is actually quite general and applies
as soon as shocks have formed with a finite probability
\cite{Frisch2001,Tribe2000}, above a critical exponent $\nu_c$ that depends
on the initial conditions (here $\nu_c=0$).
We can note that the moments diverge for $\nu<0$ because of the Dirac contribution
(\ref{PX0}), whereas for other initial conditions such as a Brownian initial
velocity they can remain well-defined and obey a second scaling law below $\nu_c$
\cite{Valageas2008,Aurell1997}.
For integer $\nu$ we obtain:
\beq
n \geq 1: \;\;\; \lag Q^n \rag \sim X \, (-1)^n \, \frac{\dd^n \tcFs}{\dd s^n}(0) ,
\;\;\; \mbox{whence} \;\;\; \lag Q \rag \sim X , \;\;\;\; 
\lag Q^2\rag \sim X \frac{16}{15} \inta\frac{\dd s}{2\pi\ii} \frac{s^2}{\Ai(s)^2} 
\simeq 1.136 \, X .
\label{Qns}
\eeq
Again, we can check that $\lag Q\rag=X$, in agreement with the conservation of 
matter. The scaling (\ref{Qns}) also implies for the cumulants 
$\lag Q^n\rag_c \propto X$ in the small-scale limit $X\ll 1$, as can be checked in 
the left panel of Fig.~\ref{figqnSn}.
This gives for the moments of the velocity increment
\beq
\lag V\rag=0 \;\;\; \mbox{and for}\;\; n\geq 2: \;\; \lag V^n\rag \sim X \, 
\frac{\dd^n \tcFs}{\dd s^n}(0) , 
\;\;\; \lag [v(x_2,t)-v(x_1,t)]^n\rag \sim 
\tcFs^{(n)}(0) 
\left(\frac{2D}{t}\right)^{n/3} \frac{x_2-x_1}{(2Dt^2)^{1/3}} . 
\label{Vns}
\eeq
Thus, we recover the usual anomalous scaling of the structure functions, 
$\lag [v(x+\ell)-v(x)]^n\rag \propto \ell$ at small distance $\ell$, that was also
observed in numerical simulations \cite{She1992}. As explained above, this is
due to the contribution from shocks \cite{Frisch2001,Tribe2000}. We further discuss
these small-scale scalings in section~\ref{General-case-q} below, on a more
general setting.

\section{Density field}
\label{Density}

\subsection{Distribution of the overdensity $\eta$ as a function of scale $x$}
\label{Distribution-of-the-overdensity}

\begin{figure}
\begin{center}
\epsfxsize=6.3 cm \epsfysize=5 cm {\epsfbox{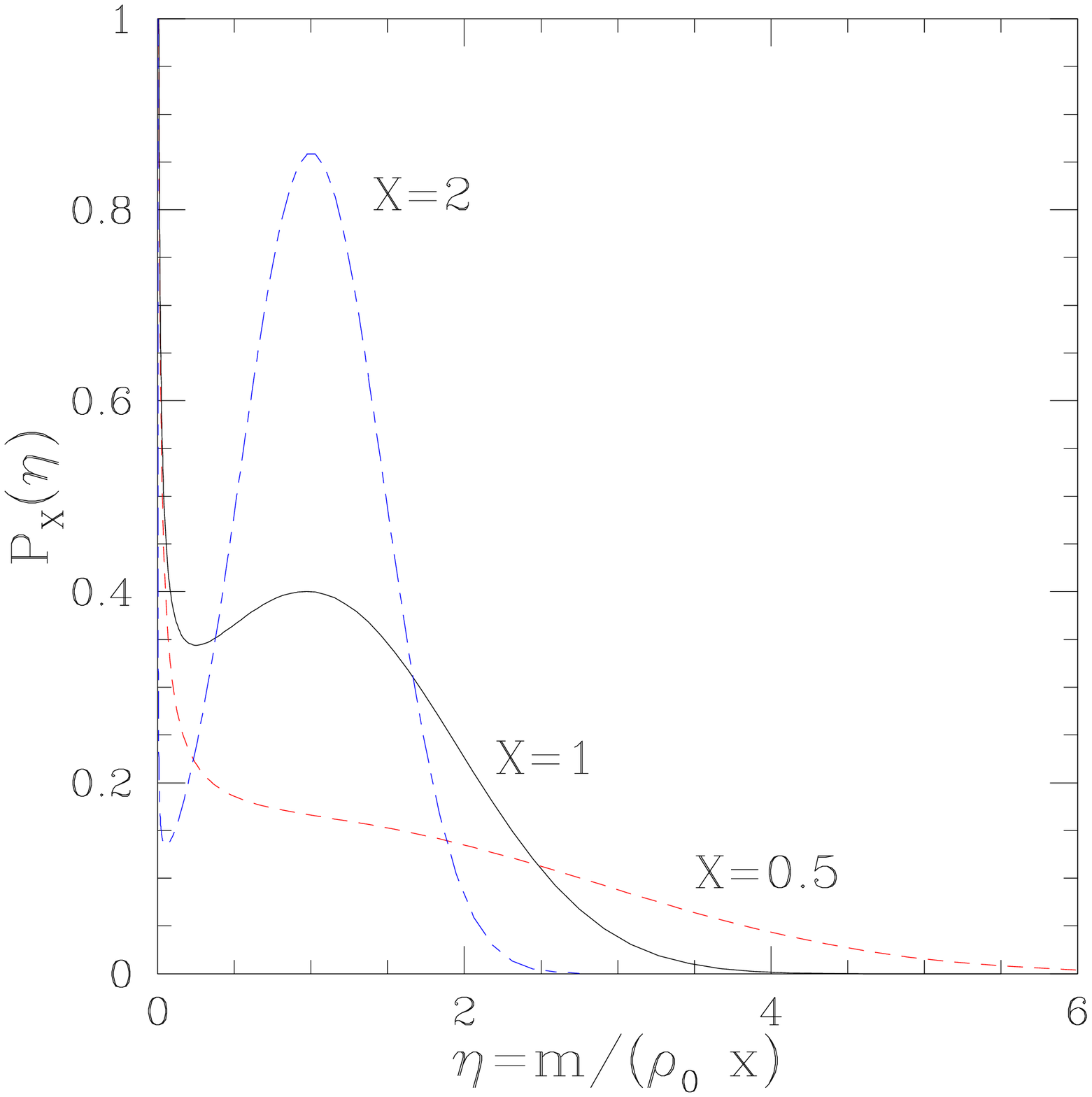}}
\epsfxsize=6.3 cm \epsfysize=5 cm {\epsfbox{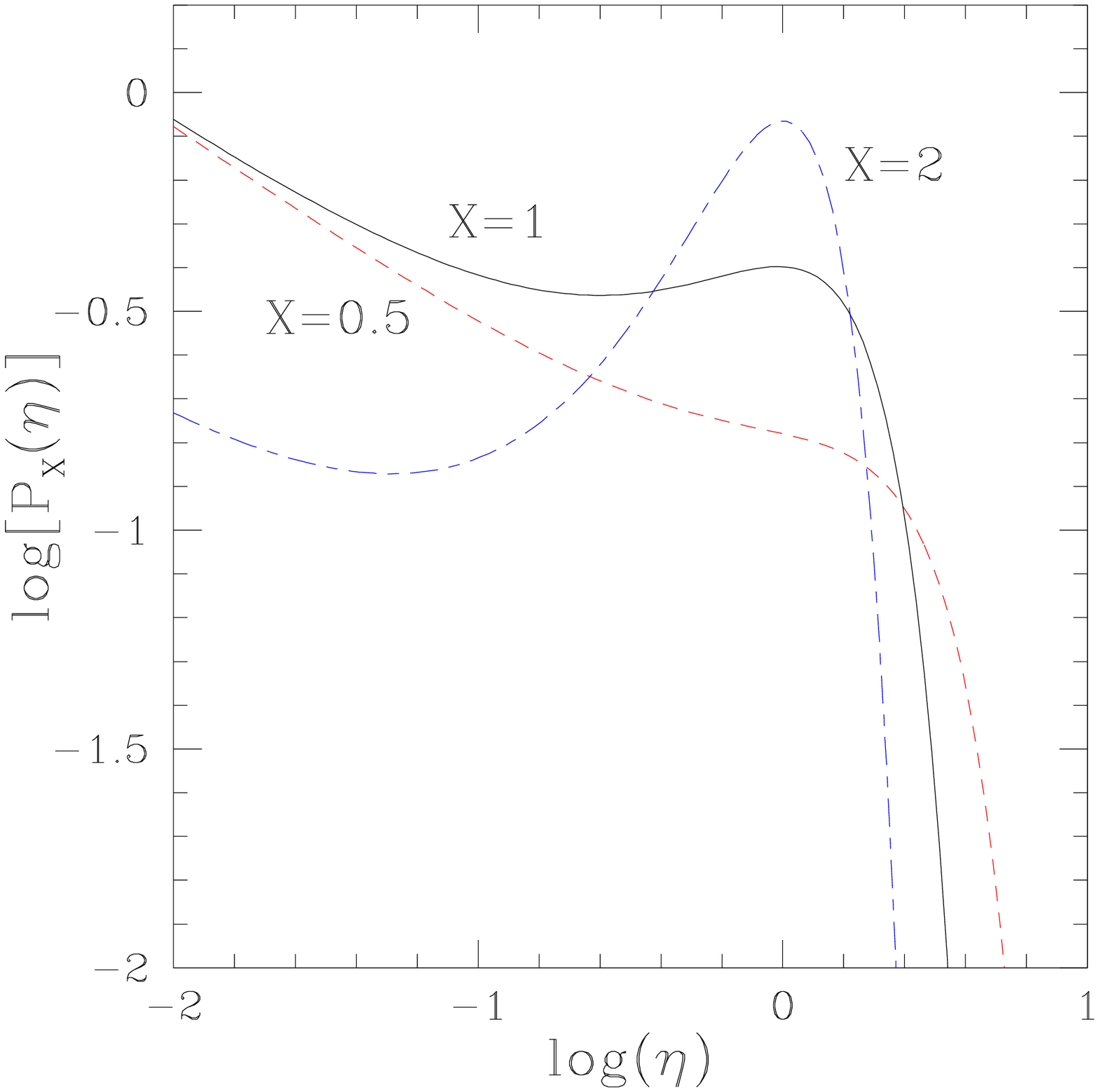}}
\end{center}
\caption{(Color online) {\it Left panel:} The probability distribution $P_X(\eta)$
of the overdensity $\eta=m/(\rho_0 x)=Q/X$, for three Eulerian sizes, $X=0.5,1$
and $2$, from Eqs.(\ref{PXQs1}), (\ref{PXeta}).
All curves display an inverse square root singularity
$\propto 1/\sqrt{\eta}$ at small densities $\eta\rightarrow 0^+$. 
In addition, there is a Dirac contribution, $P_X^0 \delta(\eta)$, with the weight 
$P_X^0$ displayed in Fig.~\ref{figPvoid}, associated with empty cells.
At large scales $X$ we recover a distribution that is sharply peaked around the
mean density, $\lag\eta\rag =1$ (i.e. $\lag \rho\rag=\rho_0$).
{\it Right panel:} Same as left panel but on a logarithmic scale.}
\label{figPXeta}
\end{figure}

We now consider the evolution of the density field, $\rho(x,t)$, that is
generated by the Burgers velocity field, starting at $t=0$ with a uniform
density $\rho_0$. Thus, the density field obeys the usual continuity
equation,
\beq
\frac{\pl\rho}{\pl t} + \frac{\pl}{\pl x}(\rho v) = 0 , \;\;\;\;\;
\mbox{with} \;\;\;\;\; \rho(x,0)= \rho_0 ,
\label{continuity}
\eeq
whereas the velocity field evolves through the Burgers equation (\ref{Burg}).
As recalled in the introduction, in the cosmological context this also provides
an approximation for the formation of large-scale structures (the cosmic web),
known as the ``adhesion model'' \cite{Gurbatove1989,Vergassola1994}.
Then, the density $\rho$ is the comoving density (i.e. measured in comoving
coordinates $x$ that follow the Hubble expansion) and $t$ is linear growing mode
$D_+$.
Thanks to the conservation of matter, the mass $m$ located between the 
Eulerian positions $x_1<x_2$ is $m=\rho_0(q_2-q_1)$, where $q(x,t)$ is the
inverse Lagrangian map.
Then, the overall overdensity, $\eta=m/(\rho_0 x)$, in the interval of size
$x=x_2-x_1$, is given by $\eta=q/x=Q/X$. Thus, the probability distribution,
$P_X(\eta)$, of the overdensity $\eta$, is given by the distribution, $P_X(Q)$, 
of the Lagrangian increment $Q$, through
\beq
\eta=\frac{m}{\rho_0 x} = \frac{Q}{X} \;\;\;\;\; \mbox{whence} \;\;\;\;\;
P_X(\eta) = X P_X(Q) .
\label{PXeta}
\eeq
Explicit expressions are obtained by substituting Eqs.(\ref{PX0}) and (\ref{PXQs1}).
In particular, the singular part (\ref{PX0}) gives the Dirac contribution
$P_X^0 \delta(\eta)$ associated with empty cells.
For the regular part, $\eta>0$, the asymptotic behaviors at large and small 
scales are directly read from Eqs.(\ref{PXQasympXl1})-(\ref{PXQasympXs3}) as
\beqa
X\gg 1 & : & P_X(\eta) \sim \frac{X}{\Aip(-\om_1)^2} \, \eta^{-1/2}  
\, e^{-\om_1 X-X^3/12} \hspace{1.5cm} \mbox{for} \;\; \eta \ll X^{-3} , 
\label{PXetaasympXl1} \\
&& \hspace{-1cm} P_X(\eta) \sim \frac{\sqrt{\pi}X^{5/2}}{\Aip(-\om_1)^2} \, 
|\eta-1|^{3/2} \, e^{-\om_1 X|\eta-1|-X^3|\eta-1|^3/12} \hspace{0.4cm} 
\mbox{for} \;\; |\eta-1|\gg X^{-1} 
\;\; \mbox{and} \;\; \eta\gg X^{-3} ,
\label{PXetaasympXl2}
\eeqa
and
\beqa
X\ll 1 & : & \;\; P_X(\eta) \sim \frac{X^{3/2}}{\sqrt{\pi}} \, \eta^{-1/2} 
\hspace{3.9cm} \mbox{for} \;\; \eta \ll X^{-1} , 
\label{PXetaasympXs1} \\
&& \;\;  P_X(\eta) \sim 2\sqrt{\pi} X^{9/2} \, \eta^{5/2} \, 
e^{-\om_1 X \eta - X^3 \eta^3/12}
\hspace{1cm} \mbox{for} \;\; X^{-1} \ll \eta \ll X^{-3/2} , 
\label{PXetaasympXs2} \\
&& \;\;  P_X(\eta) \sim 2 \pi X^3 \, \eta^{3/2} \, 
e^{-\om_1 X\eta - X^3\eta^3/12} \hspace{1.6cm} \mbox{for} \;\; \eta \gg X^{-3/2} .
\label{PXetaasympXs3}
\eeqa
Thus, at all scales we have an inverse square root tail at low densities,
$\propto 1/\sqrt{\eta}$, that follows from the low-$Q$ tail obtained in 
section~\ref{Lagrangian-increment}. Again, its weight shows the same cubic 
exponential decay at large scales as the weight $P_X^0$ of empty cells 
(\ref{PX0asymp}), whereas at small scales it describes the full low-density
regime. On large scales, the density distribution is centered on the mean
$\lag\eta\rag=1$, with cubic exponential tails on both sides, until it reaches
the very low-density tail $\propto 1/\sqrt{\eta}$ at $\eta \ll 1/X^3$.
On small scales the density distribution shows a monotonous decline, with
again a cubic exponential tail at large densities, $\eta\gg 1/X$.
We display the density distribution $P_X(\eta)$ in Fig.~\ref{figPXeta}, 
for three Eulerian sizes as in Fig.~\ref{figPXQ}, to show its evolution with 
scale. 
Again, the cubic exponential tails (\ref{PXetaasympXl2}) and (\ref{PXetaasympXs3})
can be obtained from a simple and general steepest-descent method \cite{Valageas2009}.
However, the exponent $-1/2$ of the power-law regime that appears at small scales
in Eq.(\ref{PXetaasympXs1}) is beyond the reach of such methods. It would be
interesting to build general approaches that would be able to describe this
highly nonlinear regime, for generic initial conditions and dimensions. Then,
the result (\ref{PXetaasympXs1}) would allow one to check the accuracy of 
such a method for a non-trivial case.

From the results of section~\ref{Large-scales}, we can see that on large scales
the distribution of the overdensity is described by the asymptotic distribution
\beq
X \gg 1: \;\;  P_X(\eta) \sim X \, \cFi(X(\eta-1)) ,
\label{PXetal}
\eeq
which is increasingly peaked around $\eta=1$ at larger scales. Thus, we recover
as expected the uniform density $\rho_0$ on large scales, with a distribution
that falls off faster than a Gaussian, as $e^{-X^3|\eta-1|^3/12}$.
Since $\eta=Q/X=1-V/X$, we obtain from Eqs.(\ref{VnXl})-(\ref{V2Xl}) for the
moments of the density in this large-$X$ limit:
\beq
X \gg 1: \;\;  \lag(\eta-1)^{2n+1}\rag=0 , \;\; \lag(\eta-1)^{2n}\rag= 
\frac{(-1)^n}{X^{2n}} \, \hcFi^{(2n)}(0) ,  \;\; \mbox{whence} \;\;
\lag\eta\rag=1 , \;\;\; \lag\eta^2\rag_c \simeq \frac{0.837}{X^2} .
\label{etanXl}
\eeq
In agreement with the conservation of matter we can check that $\lag\eta\rag=1$.

\begin{figure}
\begin{center}
\epsfxsize=6.3 cm \epsfysize=5 cm {\epsfbox{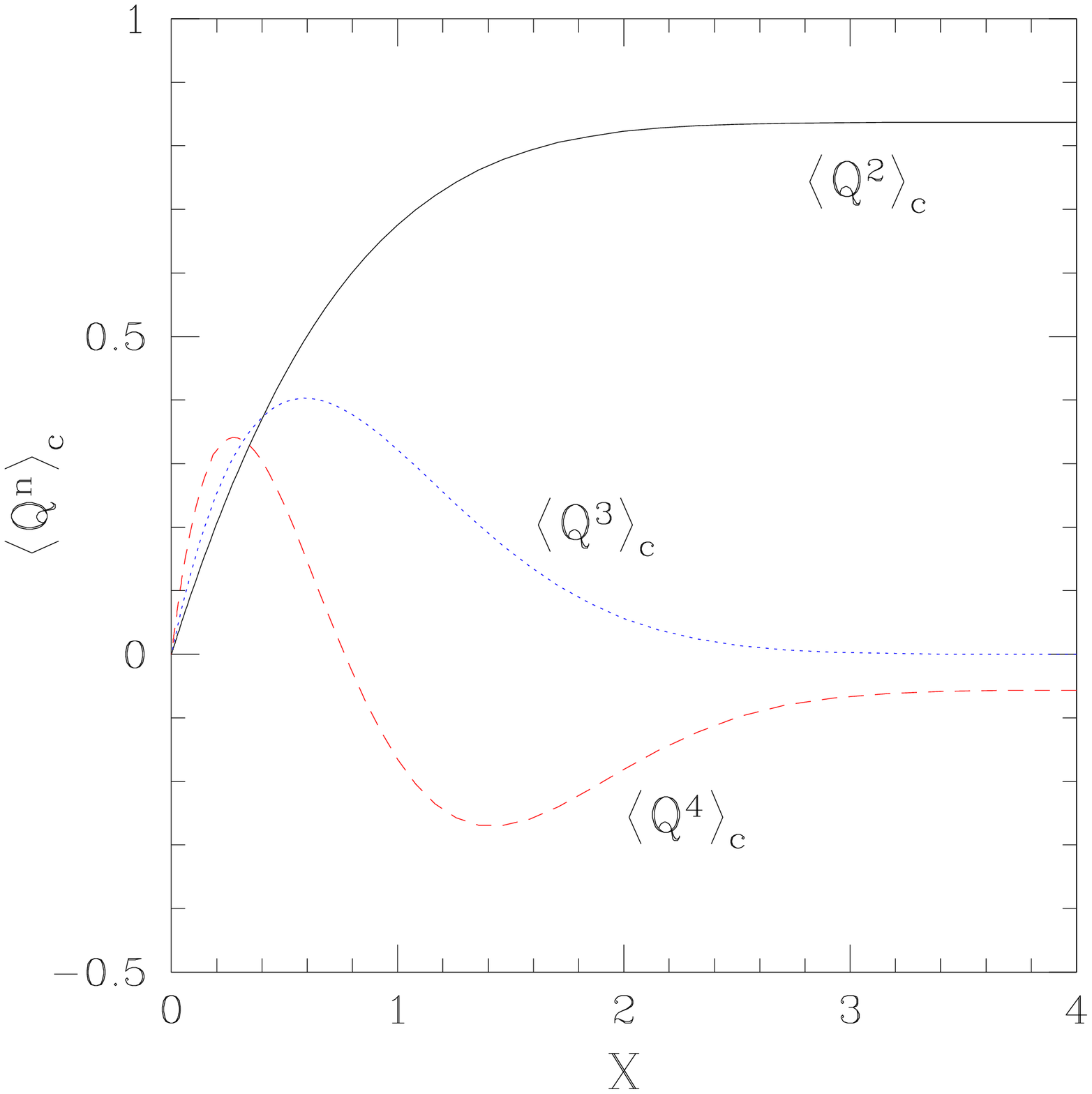}}
\epsfxsize=6.3 cm \epsfysize=5 cm {\epsfbox{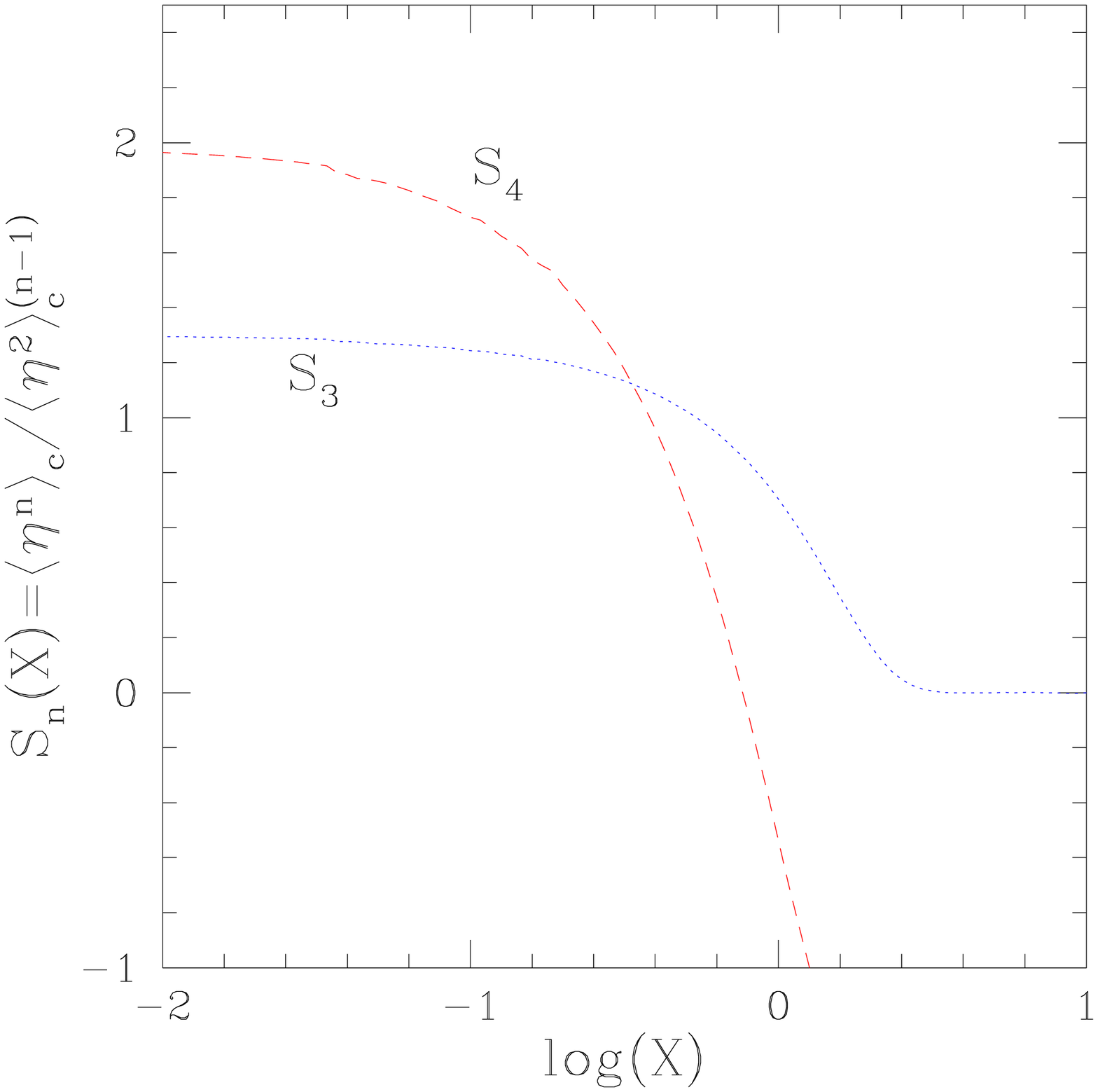}}
\end{center}
\caption{(Color online) {\it Left panel:} The first few cumulants $\lag Q^n\rag_c$
of the Lagrangian increment $Q$, as a function of the dimensionless scale
$X=x/(2Dt^2)^{1/3}$.
{\it Right panel:} The ratios $S_n$, defined by the first equality in (\ref{Sndef}),
as a function of $X$ on a semi-logarithmic scale.}
\label{figqnSn}
\end{figure}

On small scales we obtain from section~\ref{Small-scales}
\beq
X \ll 1: \;\;  P_X(\eta) \sim X^2 \cFs(X\eta) , \;\;\; \mbox{and for} \;\;\;
\nu > 0 : \;\; \lag \eta^{\nu} \rag \sim X^{-\nu+1} \, \Gamma[\nu+1] 
\inta\frac{\dd s}{2\pi\ii} \, (-s)^{-\nu-1} \, \tcFs(s) .
\label{PXetas}
\eeq
As seen in section~\ref{Small-scales}, the scaling function $\cFs(Q)$ in 
(\ref{Fsdef}) describes the probability distribution of the Lagrangian increment
down to $Q=0^+$, hence it also describes the probability distribution of the 
overdensity down to $\eta=0^+$ in (\ref{PXetas}).
Next, the moments of integer order are given by
\beq
X \ll 1 , \;\;  n \geq 1: \;\;\; \lag \eta^n \rag \sim X^{-n+1} \, (-1)^n \, 
\tcFs^{(n)}(0) \;\;\; \mbox{whence} \;\;\; \lag\eta\rag=1 , \;\; 
\lag\eta^2\rag \simeq \frac{1.136}{X} ,
\label{etanXs}
\eeq
which gives the cumulant hierarchy
\beq
X \ll 1 , \;\;  n \geq 1: \;\;\;\; S_n(X)=\frac{\lag\eta^n\rag_c}
{\lag\eta^2\rag_c^{n-1}} \sim (-1)^n \frac{\tcFs^{(n)}(0)}{\tcFs''(0)^{n-1}} .
\label{Sndef}
\eeq
Thus, the ratios $S_n(X)$ have a finite limit for $X\rightarrow 0$ and the
cumulant generating function $\varphi_X(y)$ can be written as
\beq
X \ll 1 : \;\;\; \varphi_X(y)= \sum_{n=1}^{\infty} (-1)^{n-1} \, S_n(X) \,
\frac{y^n}{n!} \sim - \tcFs''(0) 
\left[ \tcFs\left(\frac{y}{\tcFs''(0)}\right)-\tcFs(0) \right] .
\label{phiSn}
\eeq
As explained in section~\ref{Small-scales}, the scalings 
(\ref{PXetas})-(\ref{phiSn}) are due to the presence of shocks and are therefore
quite general: they apply as soon as shocks have formed with a finite probability,
for any initial conditions.
We discuss these small-scale scalings in sections~\ref{General-case-eta}
and \ref{Multifractal-formalism} below, for more general initial conditions and
higher dimensions.

We can note that on large scales the ratios $S_n(X)$ go to zero for odd $n$ and
diverge for even $n$, as seen from (\ref{etanXl}). This is due to the fact that,
even though we start with Gaussian initial conditions at $t=0$, the initial
energy spectrum is so ``blue'' (which also leads to a singular white-noise initial
velocity) that at any time $t>0$ the system is strongly affected by nonlinear
effects (associated with the building of isolated shocks amid empty regions).
This regularizes the density distribution, $p_x(\eta)$, but the latter remains
non-Gaussian in the large-scale limit $x\rightarrow\infty$, as seen in 
Eqs.(\ref{PXetal})-(\ref{etanXl}) or the explicit expression (\ref{PXetaasympXl2}).
Thus, we have $\lag\eta^{2n}\rag_c \sim \lag\eta^2\rag_c^n$ and 
$S_{2n}(X) \sim \lag\eta^2\rag_c^{1-n}\sim X^{2(n-1)}$, for $n\geq 1$ and 
$X\rightarrow\infty$, whereas odd cumulants are exponentially small.
By contrast, for initial conditions with a sufficiently ``red'' spectrum, such as 
the Brownian case, the density distribution becomes Gaussian on large scales
(it remains governed by the initial field and linear theory)
and the ratios $S_n(X)$ have a finite large-scale limit that can be computed 
through perturbative means, see \cite{Valageas2008,Valageas2009}.

We show in Fig.~\ref{figqnSn} the first few cumulants $\lag Q^n\rag_c$ and ratios
$S_n$. As explained above and in section~\ref{Large-scales}, at large scales 
both $\lag Q^2\rag_c$ and $\lag Q^4\rag_c$ reach a nonzero value, as the 
large-scale distribution is non-Gaussian, whereas the odd cumulant $\lag Q^3\rag_c$
shows a cubic exponential decay, since the limiting scaling function $\cFi$ is
even. At small scale all cumulants show a linear dependence on $X$, in agreement
with (\ref{Qns}). Then, the ratio $S_3$ also shows a cubic exponential decay on large
scales whereas $S_4$ goes to $-\infty$; next on small scales both coefficients reach
a nonzero asymptotic value.

\subsection{Density two-point correlation and power spectrum}
\label{Density-correlation}

We now consider the two-point correlation, $\xi(x,t)$, of the density field 
$\rho(x,t)$ itself:
\beq
\lag\rho(x_1,t)\rho(x_2,t)\rag_c = \rho_0^2 \, \xi(x_2-x_1,t) , \;\;\;
\mbox{whence} \;\;\; \lag\eta^2\rag_c = \int_0^x \frac{\dd x_1 \dd x_2}{x^2}
\, \xi(x_2-x_1) .
\label{xidef}
\eeq
In terms of the dimensionless variables (\ref{QXdef}), using $\eta=Q/X$,
we obtain
\beq
\lag Q^2\rag - X^2 = \int_0^X \dd X_1 \dd X_2 \; \xi(X_2-X_1) , \;\;\;
\mbox{whence} \;\;\; \xi(X)= \frac{1}{2}\frac{\dd^2}{\dd X^2}\lag Q^2\rag -1 .
\label{xiQ2}
\eeq
Then, the small-distance behavior (\ref{Qns}), that was associated with shocks,
gives rise to a Dirac contribution
\beq
\xi^0(X) =  \tcFs''(0) \, \delta(X) , \;\;\;
\mbox{whence} \;\;\; \xi^0(x) = (2Dt^2)^{1/3} \, \tcFs''(0) \, \delta(x) \simeq
1.136 \, (2Dt^2)^{1/3} \,\delta(x) .
\label{xi0}
\eeq
In terms of the density power spectrum, $\cP(k,t)$, defined by
\beq
\rho(x,t) - \rho_0 = \int_{-\infty}^{\infty} \frac{\dd k}{2\pi} \, 
e^{\ii k x}\, \rhoh(k,t) , 
\;\;\;\; \lag\rhoh(k_1,t)\rhoh(k_2,t)\rag= \delta(k_1+k_2) \, 2\pi \, 
\cP(k_1,t) , 
\label{Pkdef}
\eeq
this gives the asymptotic behavior at high $k$,
\beq
k\rightarrow\infty :   \;\;\; \cP(k,t) \rightarrow 
\tcFs''(0) \, (2Dt^2)^{1/3} \simeq 1.136 \, (2Dt^2)^{1/3} .
\label{Pkinf}
\eeq
As expected, shocks, that form a series of Dirac peaks in the density field,
give rise to a white-noise power spectrum in the limit of high wavenumbers.
In addition, there are also non-zero correlations at finite distances,
that can be obtained from the
second cumulant of the Lagrangian increment $Q$ through (\ref{xiQ2}). Using
the expression (\ref{PXQs1}), making the change of variable $s_i\rightarrow
s_i+s$ and integrating over $Q$ and $s$, we obtain
\beqa
\lag Q^2\rag = \left. \frac{\pl^2}{\pl s^2} \right|_{0} && 2\sqrt{\pi X} 
e^{-X^3/12} \inta \frac{\dd s_1 \dd s_2}{(2\pi\ii)^2} \,
\frac{e^{(s_1+s_2)X/2+(s_1-s_2)^2/(4X)}}{\Ai(s_1)\Ai(s_2)\Ai(s_1+s)\Ai(s_2+s)}
\nonumber \\
&& \times \int_0^{\infty} \dd r \, e^{Xr} \, \Ai(r+s_1+s) \Ai(r+s_2+s) .
\label{Q2_1}
\eeqa
Next, going back to $s_i\rightarrow s_i-s$, using the expression (\ref{gPhih}),
which allows to remove the asymptotic $X^2$ behavior of $\lag Q^2\rag$, 
and differentiating twice with respect to $X$, the density two-point correlation
reads as
\beqa
X> 0 : \;\;\; \xi(X) & = & \sqrt{\pi}  \inta \frac{\dd s_1 \dd s_2}{(2\pi\ii)^2} 
\, \frac{1}{\Ai(s_1)\Ai(s_2)} \, \biggl \lbrace - \Ax''(X) \int_X^{\infty}\dd y 
\, e^{-\Phix(y)} \, \hx(y) \nonumber \\
&&  \hspace{-2cm} + \, e^{-\Phix(X)}  \, \left[ 2 \Ax'(X) \hx(X) 
+ \Ax(X) ( \hx'(X) - \Phix'(X) \hx(X) ) \right] \biggl \rbrace ,
\label{xiPhih}
\eeqa
where the functions $\Phix(X)$ and $\hx(X)$ are defined in 
Eqs.(\ref{Phidef})-(\ref{hdef}) and we introduced the function $\Ax(X)$ given by
\beq
\Ax(X) =  \left. \frac{\pl^2}{\pl s^2} \right|_{0} 
\frac{e^{-sX}}{\Ai(s_1-s)\Ai(s_2-s)} .
\label{AXdef}
\eeq
This yields the asymptotic behaviors
\beq
\xi(0) \simeq -0.56 , \;\;\; \mbox{and for} \;\; X \rightarrow \infty: 
\;\;\; \xi(X) \sim \frac{-\sqrt{\pi}}{16 \Aip(-\om_1)^2} \, X^{11/2} \,
e^{ -\om_1 X -X^3/12} .
\label{xiasymp}
\eeq

\begin{figure}
\begin{center}
\epsfxsize=6.3 cm \epsfysize=5 cm {\epsfbox{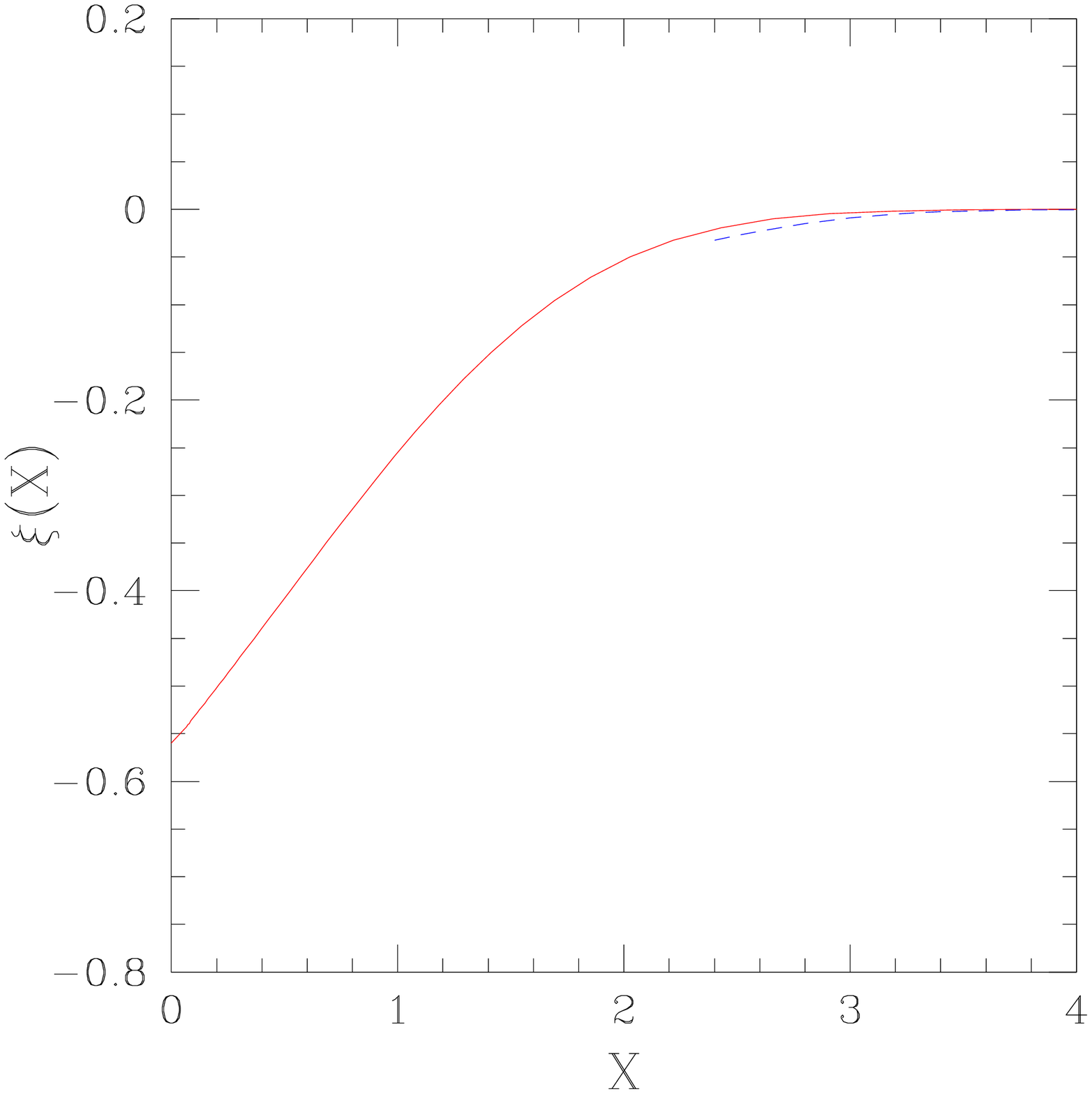}}
\epsfxsize=6.3 cm \epsfysize=5 cm {\epsfbox{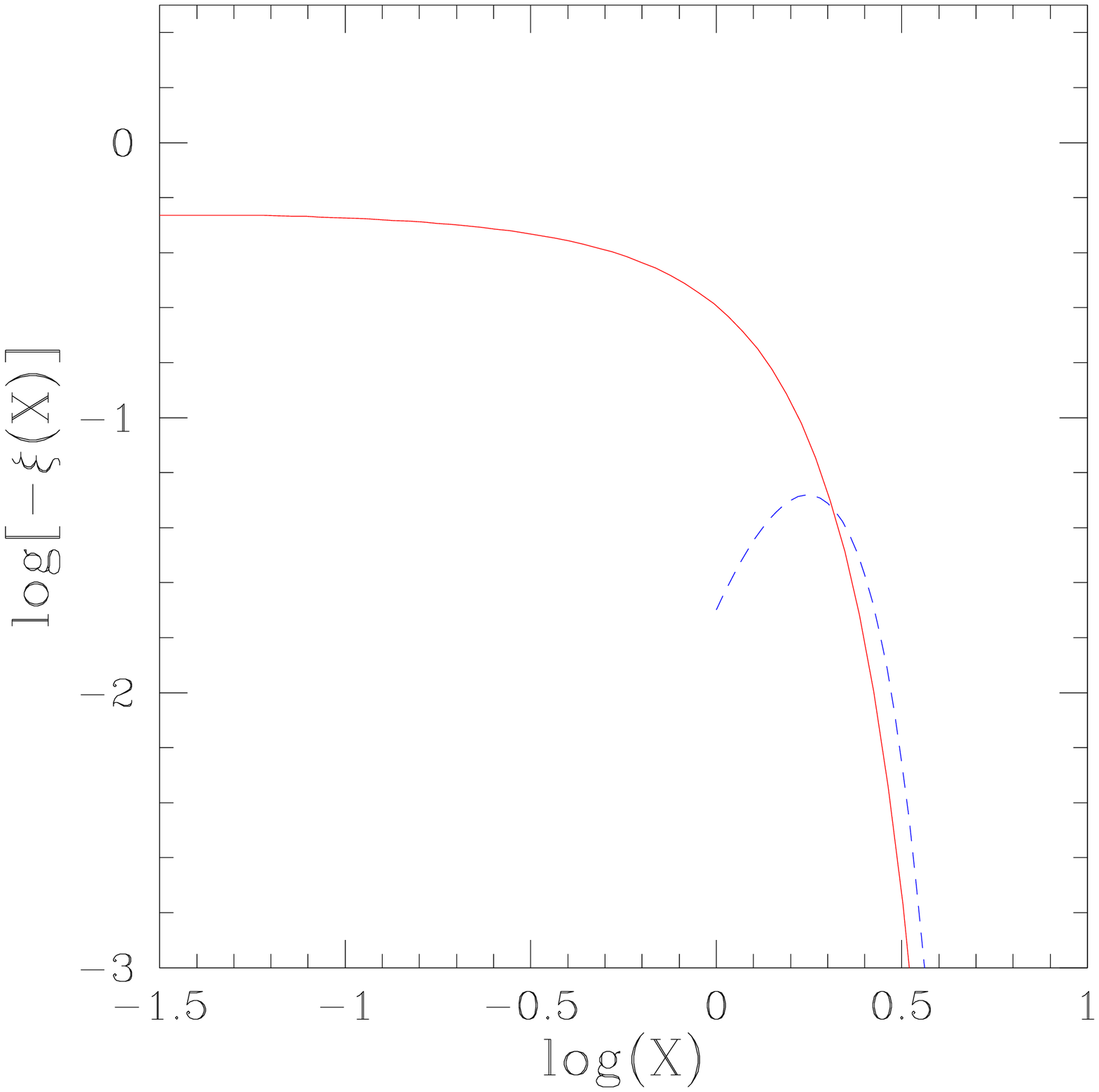}}
\end{center}
\caption{(Color online) {\it Left panel:} The density two-point correlation, 
$\xi(x)=\lag \rho(x_1) \rho(x_1+x)\rag_c/\rho_0^2$, as a function of the
dimensionless scale $X=x/(2Dt^2)^{1/3}$, from Eq.(\ref{xiPhih}). 
It is negative over $x>0$ but there is an additional Dirac contribution at the 
origin, given by Eq.(\ref{xi0}). The dashed line is the large-$X$ asymptotic 
behavior (\ref{xiasymp}).
{\it Right panel:} Same as left panel but on a logarithmic scale.}
\label{figxi}
\end{figure}

Thus, as we can check in Fig.\ref{figxi}, the density correlation is negative for
$x>0$. This may be understood from the fact that, since the matter
collapses within isolated zero-thickness objects (shocks), close to a shock
there is a relative underdensity as matter has already fallen into that shock.
In terms of particles of infinitesimal mass, the massive aggregate associated with
the shock has swept matter from its neighborhood along its motion at previous times
as particles stick together after collisions.
Thus, starting with a white-noise initial velocity which shows no correlations
over finite distance $x>0$, some (anti-)correlations appear as soon as $t>0$ 
over scales of order $x \sim (2Dt^2)^{1/3}$ (i.e. $X\sim 1$), but they
remain very weak as they decay even faster than a Gaussian at larger scales.
Again, the cubic exponential falloff (\ref{xiasymp}) can be understood from simple
arguments. Following the discussion above, correlations at scale $x$ arise from
the motion of shocks over distances of order $x$ and the building of voids of size
$X$. Then, as discussed below Eq.(\ref{PX0asymp}), this can be associated to
an initial mean velocity $\bv_0(x)\sim x/t$ over the interval $x$ and to a 
probabilistic weight $\sim e^{-(x/t)^2/(D/x)}\sim e^{-x^3/(Dt^2)}$, which gives
back the cubic exponential tail (\ref{xiasymp}). In agreement with this discussion,
we can check that the tail (\ref{xiasymp}) is the same (apart from the power-law
prefactor) as the one obtained in Eq.(\ref{PX0asymp}) for the probability of voids.

\begin{figure}
\begin{center}
\epsfxsize=6.3 cm \epsfysize=5 cm {\epsfbox{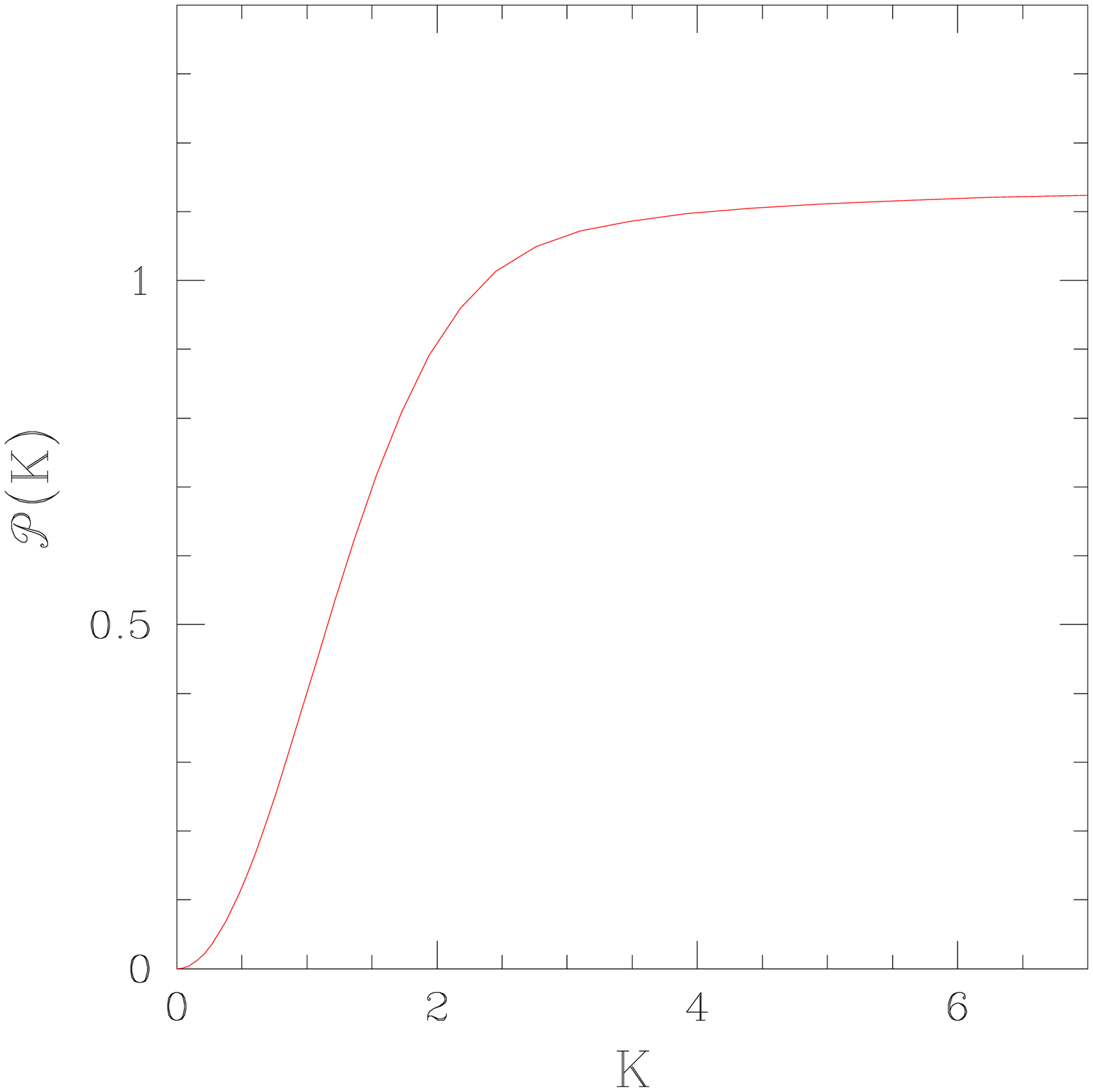}}
\epsfxsize=6.3 cm \epsfysize=5 cm {\epsfbox{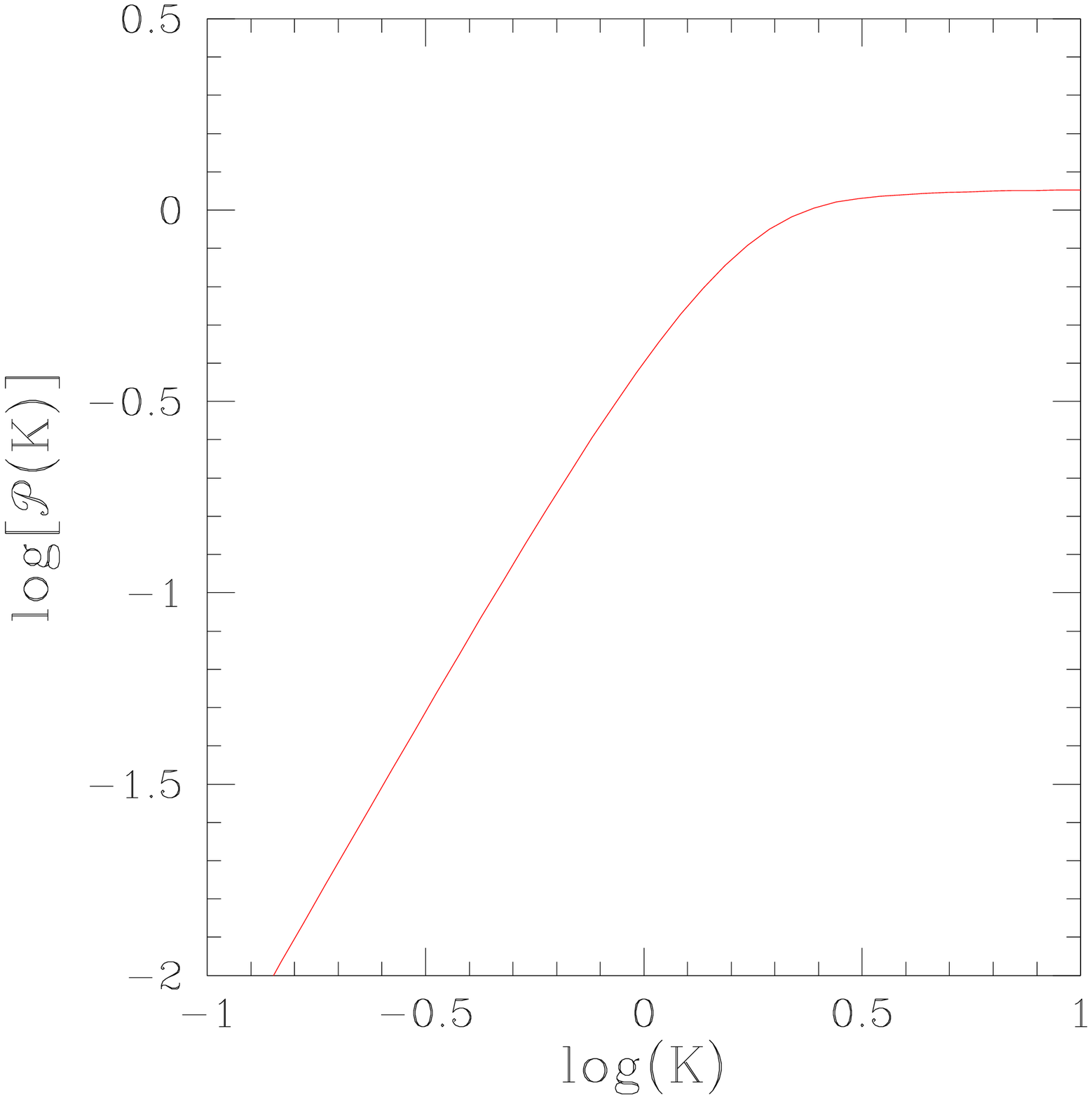}}
\end{center}
\caption{(Color online) {\it Left panel:} The dimensionless density power
spectrum $\cP(K)$, as a function of the dimensionless wavenumber $K=(2Dt^2)^{1/3}k$,
from Eq.(\ref{PkXi}). It goes to zero as $K^2$ at small $K$ and it goes to a constant
at large $K$.
{\it Right panel:} Same as left panel but on a logarithmic scale.}
\label{figPk}
\end{figure}

From Eq.(\ref{Pkdef}) the density power spectrum can be written, in terms of
dimensionless variables, as
\beq
\cP(K)= 2 \int_0^{\infty} \dd X \, \cos(KX) \, \xi(X) 
+ \tcFs''(0) = 2 \int_0^{\infty} \dd X \, 
\left[ \cos(KX)-1 \right] \xi(X) ,
\label{PkXi}
\eeq
where $K=\gam k$ (and $\gam$ was defined in Eq.(\ref{QXdef})).
In the first equality we explicitly separated the Dirac contribution (\ref{xi0}) 
from the integral over $X>0$. The second equality follows by noticing that 
$\cP(0)=0$. Indeed, from Eq.(\ref{xiQ2}) we have
\beq
\cP(0)= \int_{-\infty}^{\infty} \dd X \, \xi(X) =  \lim_{X\rightarrow\infty} 
\left[ \frac{\dd}{\dd X} \lag Q^2\rag - 2 X \right] = 0 ,
\label{Pk0}
\eeq
as the term in brackets decays as $e^{-X^3/12}$ at large $X$, as seen from
Section~\ref{Large-scales} and Appendix~\ref{Laplace}.
We show in Fig.~\ref{figPk} our results for the power spectrum, using 
Eqs.(\ref{xiPhih}) and (\ref{PkXi}). We clearly see the quadratic behavior at
low $K$, that can be obtained by expanding the cosine in Eq.(\ref{PkXi}),
and the saturation at high wavenumbers to the white-noise spectrum (\ref{Pkinf})
due to shocks. Since the correlation function decays faster than a Gaussian at
large distances the power spectrum is actually regular at $k=0$.
The high-wavenumber behavior (\ref{Pkinf}) is universal and appears
as soon as shocks have formed, along with the scalings (\ref{Qns})-(\ref{Vns})
for the Lagrangian and velocity increments observed on small scales.
The quadratic low-$k$ behavior applies from Eq.(\ref{PkXi}) to initial conditions
such that the linear power on large scales decays faster than $k^2$ (by contrast,
if the initial velocity field is given by a Brownian motion, which shows significant
power on large scales, the density power spectrum is exactly a white-noise 
spectrum over all $k$, that is $\cP(k)$ is constant down to $k=0$).

Again, the results (\ref{xiPhih}) and (\ref{PkXi}), shown in Figs.~\ref{figxi}
and \ref{figPk}, may be useful to test general approximation schemes.
In particular, in the cosmological context, the matter two-point correlation
and power spectrum are among the main observables used to
constrain cosmological scenarios (both the global cosmological history,
through the linear growth factor of density fluctuations, and the primordial
initial conditions, generated by an hypothetical inflationary stage, through
the shape of the power spectrum). 
For gravitational systems of this sort, no good approximation scheme has
been obtained yet that is able to estimate the density power spectrum in
both linear and nonlinear regimes (i.e. from large down to small scales),
so that one needs to use numerical simulations. The case studied in this
article provides a rare hydrodynamical example, closely related to 1-D
gravitational dynamics as recalled above, where a complete exact solution
can be derived. In this respect, the present case of white-noise initial velocity
is somewhat more interesting than the case of Brownian initial velocity,
where the power spectrum is simply a constant over all scales, as it shows
a transition between different low- and high-wavenumber regimes.

A key difference between Burgers dynamics and gravitational systems
(and real turbulence) is that the high-$k$ regime is quite simple and universal,
since it is governed by shocks and shows a constant white-noise asymptote as
in Eq.(\ref{Pkinf}). By contrast, in 3-D gravitational (or Navier-Stokes) systems, 
small-scale structures may show a broader variety (extended halos, vortices, ...)
\cite{Vogelsberger2008,Kraichnan1968}
and it is not known whether universal exponents exist and for which class of
initial conditions they hold (in the cosmological context numerical simulations
suggest that there is no such universality as the high-$k$ slope seems to depend
on the initial slope \cite{Peacock1996}).

\section{Lagrangian displacement field}
\label{Lagrangian-displacement}

\subsection{One-point distribution}
\label{One-point-Lag}

We now consider the Burgers dynamics from a Lagrangian point of view, as opposed
to the Eulerian point of view described in the previous sections. Thus, labelling
the particles by their initial position $q$ at the initial time $t=0$, we follow
their trajectory $x(q,t)$. Since particles do not cross each other they remain
well-ordered. Then, the probability, $p_q(\geq x)$, for the particle $q$ to be
located to the right of the position $x$, is equal to the probability,
$p_x(\leq q)$, for the Lagrangian coordinate $q(x)$ associated with position
$x$ to be smaller than or equal to $q$. This yields
\beq
P_Q(X) = - \frac{\pl}{\pl X} P_X(\leq Q) = - \frac{\pl}{\pl X} \int_{-\infty}^Q
\dd Q'P(Q'-X) = P(Q-X)  ,
\;\;\; \mbox{whence} \;\;\; P_Q(X) =P_X(Q) ,
\eeq
where we used from Eq.(\ref{PXQJJ}) the property that the Eulerian distribution
$P_X(Q)$ only depends on the relative distance $Q-X$ as $P_X(Q)=P(Q-X)$ with
$P(V)$ given by Eq.(\ref{PXQJJ}). Thus, the one-point Eulerian and Lagrangian
distributions are identical. This applies to any initial conditions which are 
statistically homogeneous and isotropic, so that $P_X(Q)$ only depends on $|Q-X|$.

\subsection{Two-point distribution and relative distance}
\label{Two-point-Lag}

We now investigate the two-point distribution of the Lagrangian displacement field.
In a fashion similar to the one-point distribution, we can relate the Eulerian
and Lagrangian distributions by
\beq
P_{Q_1,Q_2}(\geq X_1,\leq X_2) = P_{X_1,X_2}(\leq Q_1,\geq Q_2) .
\label{P2EulLag}
\eeq
Then, for $Q_1<Q_2$ the Dirac part (\ref{PQ1Q2ii}) does not contribute and
we obtain from Eq.(\ref{PQ1Q2i}), with $X_1\leq X_2$,
\beq
P_{Q_1,Q_2}(\geq X_1,\leq X_2) = \int_{-\infty}^{Q_1}\dd Q_1' 
\int_{Q_2}^{\infty} \dd Q_2' \, \cJ(Q_1'-X_1) \, \cJ(X_2-Q_2') \, 
\cH_{X_1,X_2}(Q_1',Q_2') .
\label{PX1X2_1}
\eeq
Using Eqs.(\ref{Jdef}) and (\ref{Hdef}), this yields
\beqa
P_{Q_1,Q_2}(X_1,X_2) & = & - \frac{\pl^2}{\pl X_1\pl X_2} 2 \sqrt{\pi(X_2-X_1)} 
\, e^{-(X_2-X_1)^3/12} \inta \frac{\dd s_1 \dd s_2 \dd s_1' \dd s_2'}{(2\pi\ii)^4} 
\, \frac{e^{(s_1-s_2)Q_1 + (s_2'-s_1')Q_2}}{(s_1-s_2)(s_1'-s_2')} \nonumber \\
&& \hspace{-2cm} \times \frac{e^{- s_1 X_1+s_1' X_2 + (s_2-s_2')(X_1+X_2)/2
+ (s_2-s_2')^2/(4(X_2-X_1))}}{\Ai(s_1)\Ai(s_2)\Ai(s_1')\Ai(s_2')}
\int_0^{\infty} \dd r \, e^{(X_2-X_1)r} \, \Ai(r+s_2) \Ai(r+s_2') .
\label{PX1X2_2}
\eeqa
We can check that Eq.(\ref{PX1X2_2}) is invariant through uniform spatial 
translations.
Next, from $P_{Q_1,Q_2}(X_1,X_2)$ we can derive the distribution, $P_Q(X)$, of the
relative Eulerian distance, $X=X_2-X_1$. It only depends on the relative
Lagrangian distance, $Q=Q_2-Q_1$, through
\beq
P_Q(X) = \int_{-\infty}^{\infty} \dd X_1 \, P_{Q_1,Q_1+Q}(X_1,X_1+X) .
\label{PQXdef}
\eeq
This gives
\beqa
X>0 : \;\;\; P_Q(X) & = & \frac{\pl^2}{\pl X^2} \, 2\sqrt{\pi X} \, e^{-X^3/12} 
\inta \frac{\dd s \dd s_1 \dd s_2}{(2\pi\ii)^3} \,
\frac{e^{s (Q-X)+(s_1+s_2)X/2+(s_1-s_2)^2/(4X)}}
{s^2\Ai(s_1)\Ai(s_2)\Ai(s_1-s)\Ai(s_2-s)}
\nonumber \\
&& \hspace{1cm} \times \int_0^{\infty} \dd r \, e^{Xr} \, \Ai(r+s_1) \Ai(r+s_2) ,
 \;\;\; \mbox{with} \;\;\; \Re(s)<0 ,
\label{PQXs1}
\eeqa
where the integration contour over $s$ runs to the left of the pole at $s=0$.
Note that the expression (\ref{PQXs1}) is similar to the result (\ref{PXQs1})
obtained for the distribution of the Lagrangian increment $Q$ over a fixed 
Eulerian interval $X$, except for the double derivative with respect to $X$ and
the factor $1/s^2$. This leads to the relationship between the distributions of the
Eulerian and Lagrangian increments:
\beq
\frac{\pl^2}{\pl Q^2} \, P_Q(X) = \frac{\pl^2}{\pl X^2} P_X(Q) .
\label{d2PP}
\eeq
On the other hand, using the expression (\ref{gPhih}) and the comparison with
Eq.(\ref{PXQs1}), taking the derivatives with respect to $X$ in Eq.(\ref{PQXs1})
gives the relationship 
\beqa
P_Q(X) & = & P_X(Q) + 2\sqrt{\pi} \, \inta \frac{\dd s \dd s_1 \dd s_2}
{(2\pi\ii)^3} \, \frac{e^{s (Q-X)} \; e^{-\Phix(X)}}
{s^2\Ai(s_1)\Ai(s_2)\Ai(s_1-s)\Ai(s_2-s)} \nonumber \\
&& \hspace{2cm} \times \left[ -2s \, \hx(X) - \Phix'(X) \hx(X) + \hx'(X) \right] .
\label{PQXs2}
\eeqa

\begin{figure}
\begin{center}
\epsfxsize=6.3 cm \epsfysize=5 cm {\epsfbox{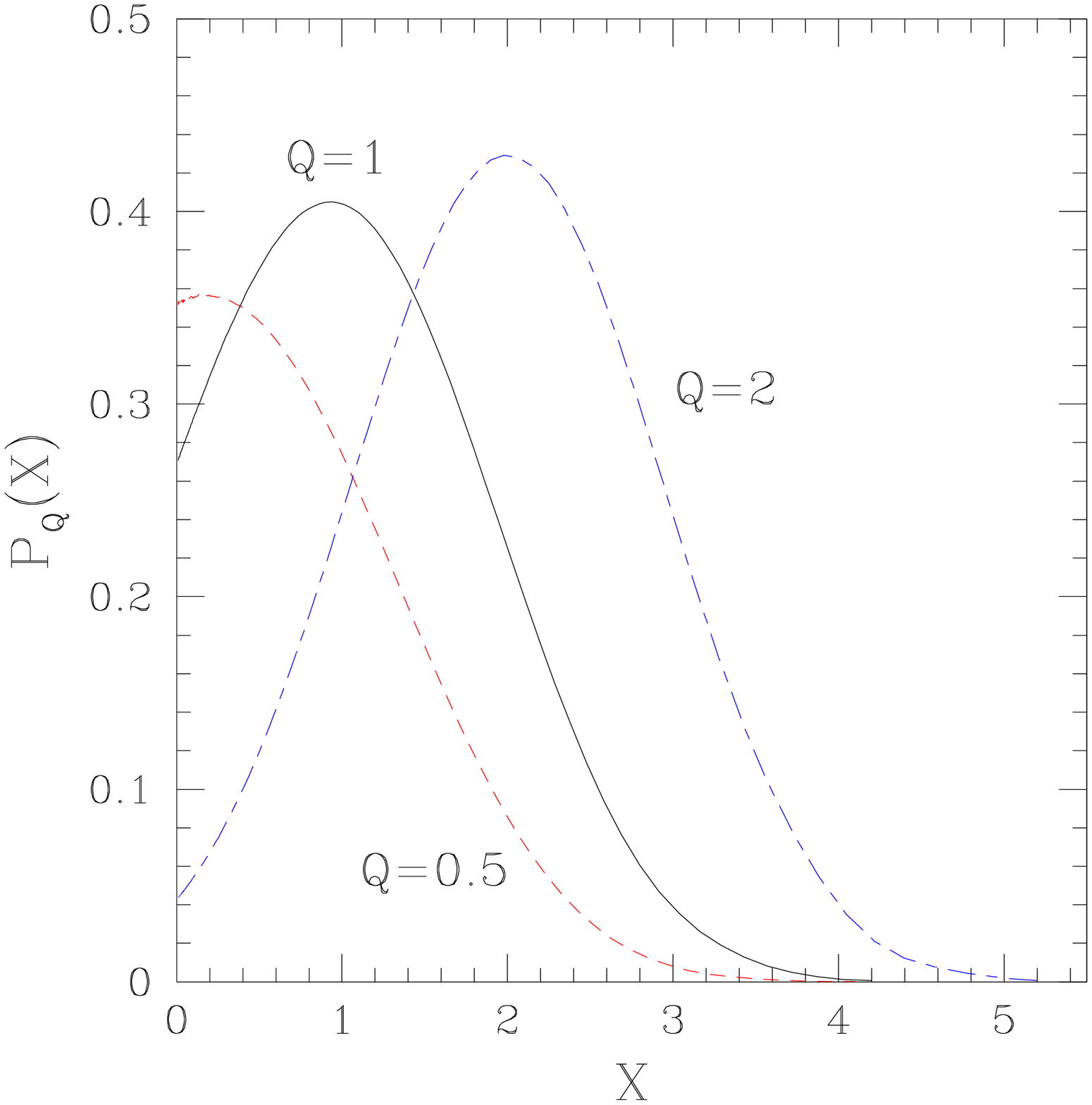}}
\epsfxsize=6.3 cm \epsfysize=5 cm {\epsfbox{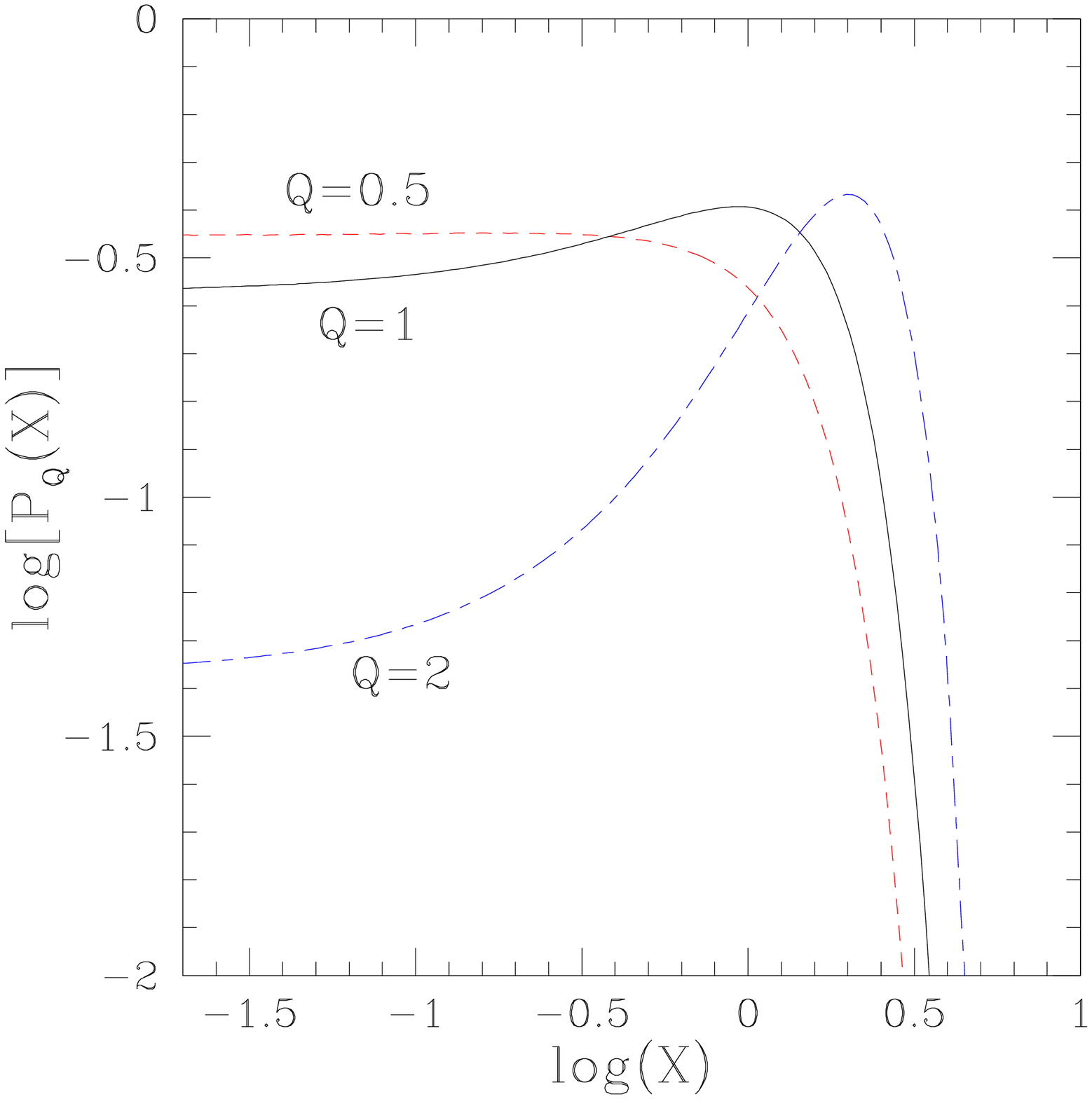}}
\end{center}
\caption{(Color online) {\it Left panel:} The probability distribution $P_Q(X)$
of the Eulerian increment $X$, for three Lagrangian lengths, $Q=0.5,1$ and $2$,
from Eq.(\ref{PQXs2}). The integral over $X>0$ is smaller than unity as
there is an additional Dirac contribution, $\Pshock_Q\delta(X)$, associated 
with shocks, with the weight $\Pshock_Q$ displayed in Fig.~\ref{figPshock}.
{\it Right panel:} Same as left panel but on a logarithmic scale.}
\label{figPQX}
\end{figure}

We show in Fig.~\ref{figPQX} the distribution $P_Q(X)$ obtained for three
Lagrangian intervals $Q$. In a fashion similar to the Eulerian distribution
$P_X(Q)$ shown in Fig.~\ref{figPXQ}, on large scales, $Q\gg 1$, the Lagrangian 
distribution $P_Q(X)$ is centered on $Q$, with cubic exponential tails on both 
sides as seen in (\ref{PQFi}) below, whereas on small scales, $Q \ll 1$, it shows 
a monotonous decline. However, contrary to the Eulerian distribution, the 
Lagrangian distribution $P_Q(X)$ does not show an inverse square-root tail at 
low $X$ as $P_Q(0)$ is finite.
As can be seen in Fig.~\ref{figPQX}, the distribution $P_Q(X)$ given by 
Eqs.(\ref{PQXs1})-(\ref{PQXs2}) over $X>0$ is not normalized to unity as its
weight decreases for smaller $Q$. Indeed, there is an additional Dirac contribution
associated with shocks, of the form $\Pshock_Q \delta(X)$, where the Eulerian
increment $X$ is zero (all particles in the initial range $[Q_1,Q_2]$, of length
$Q=Q_2-Q_1$, have merged into a single shock). Since the weight of this contribution
grows at smaller $Q$, as can be checked in Fig.~\ref{figPshock} below, the 
normalization of the regular contribution (\ref{PQXs1}) decreases at smaller $Q$.

\begin{figure}
\begin{center}
\epsfxsize=6.3 cm \epsfysize=5 cm {\epsfbox{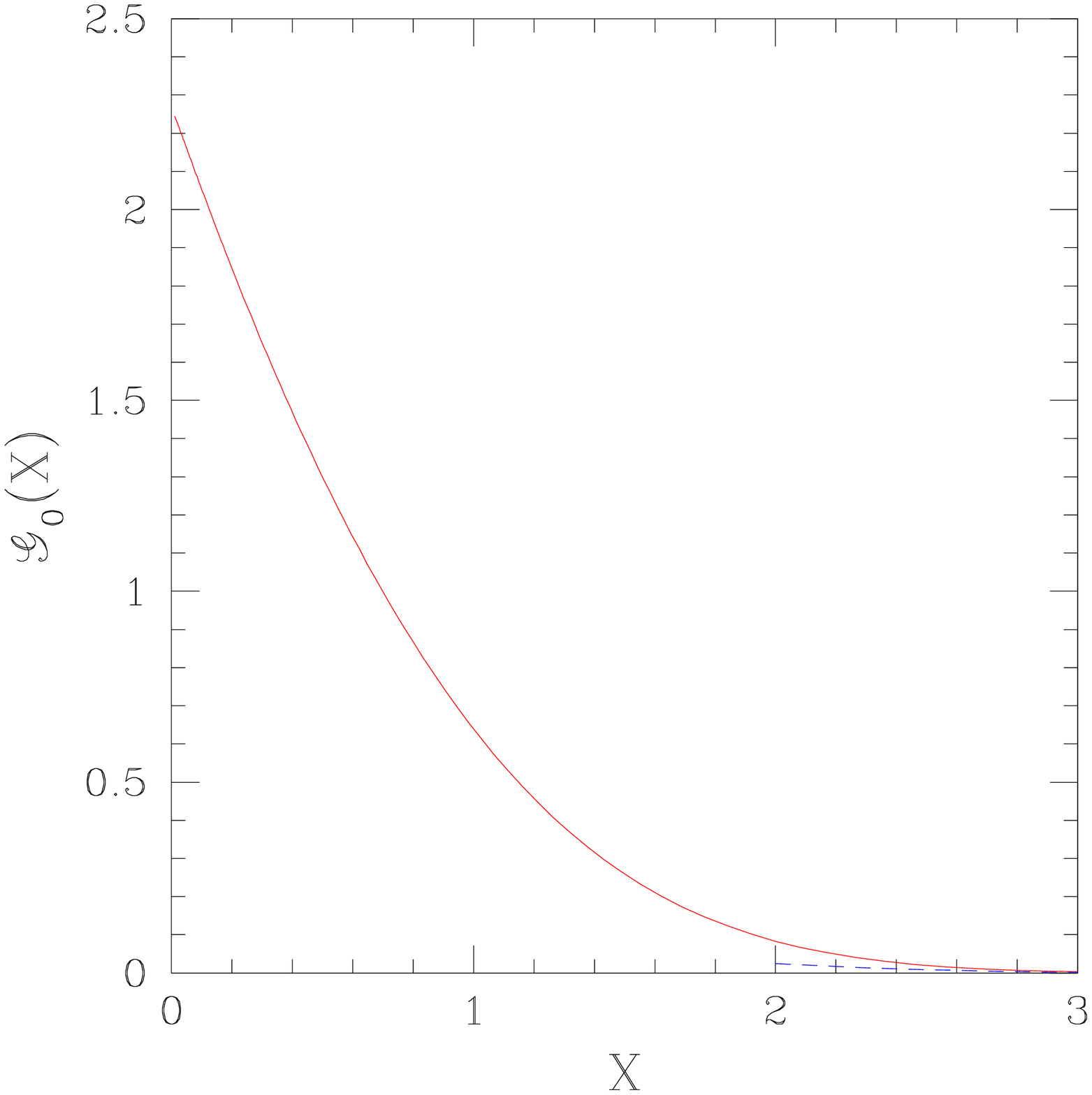}}
\epsfxsize=6.3 cm \epsfysize=5 cm {\epsfbox{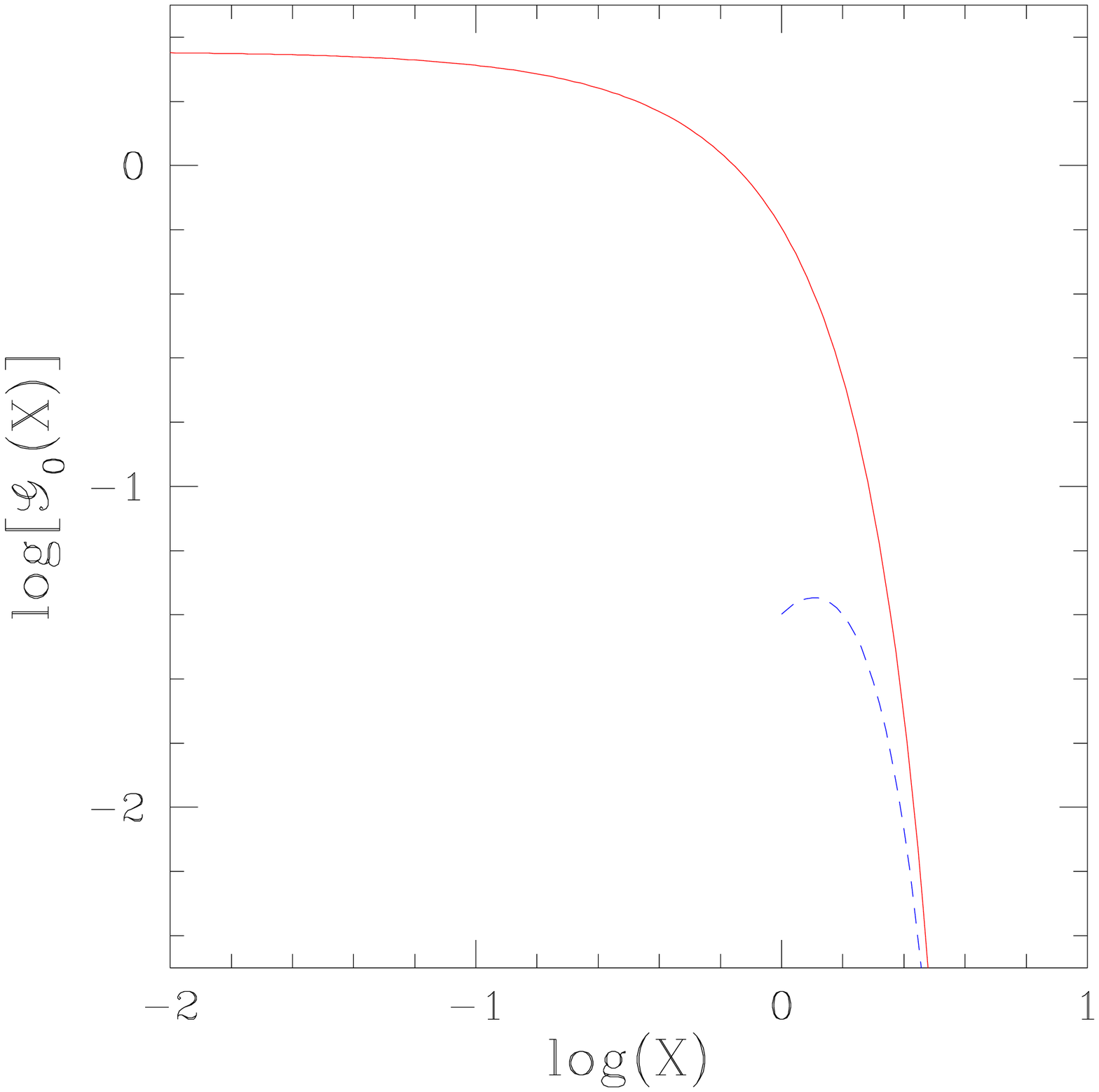}}
\end{center}
\caption{(Color online) {\it Left panel:} The scaling function $\cGs(X)$ that
describes the distribution of the Eulerian increment $X$ in the limit $Q \ll 1$, 
from Eqs.(\ref{PQsX})-(\ref{Gsdef}).
The dashed line is the asymptotic behavior (\ref{GsXl}).
{\it Right panel:} Same as left panel but on a logarithmic scale.}
\label{figG0X}
\end{figure}

On large scales, $Q\gg 1$, from the expression (\ref{PQXs1}) or the relation 
(\ref{d2PP}), we obtain
\beq
Q \gg 1 : \;\; P_Q(X) \sim  \cFi(X-Q) , \;\;\;\; \mbox{whence} \;\;\; 
P_Q(X) \sim P_X(Q) \;\;\; \mbox{for large} \;\; Q \;\; \mbox{and} \;\; X ,
\label{PQFi}
\eeq
where the function $\cFi$ was given in Eq.(\ref{hFidef}) and shown in 
Fig.~\ref{figFiV}. Thus, at large scales the distribution $P_Q(X)$ is peaked 
around $X=Q$, with fluctuations of order unity that become increasingly small 
as compared with $Q$ for $Q\rightarrow\infty$. Moreover, it becomes identical to
the Eulerian distribution in this limit.

On small scales, $Q \ll 1$, using the expression (\ref{gPhih}), we obtain
a scaling similar to the one seen in Eq.(\ref{Fsdef}) for the Eulerian 
distribution,
\beq
Q \ll 1 : \;\;\; P_Q(X) \sim Q \, \cGs(X) , \;\;\; \mbox{with} \;\; X>0 ,
\label{PQsX}
\eeq
and
\beq
\cGs(X)= 2\sqrt{\pi} \inta\frac{\dd s_1\dd s_2}
{(2\pi\ii)^2)} \, \frac{e^{-\Phix(X)}}{\Ai(s_1)^2\Ai(s_2)^2} \,
\left[ \Phix'(X) \hx(X) - \hx'(X) \right] .
\label{Gsdef}
\eeq
In particular, Eq.(\ref{Gsdef}) gives the asymptotic behavior at large $X$:
\beq
X \gg 1 : \;\;\; \cGs(X) \sim \frac{\sqrt{\pi}}{8\Aip(-\om_1)^2}
\, X^{7/2} \, e^{-\om_1 X -X^3/12} .
\label{GsXl}
\eeq
We show the scaling function $\cGs(X)$ in Fig.~\ref{figG0X}.
We can see that it is finite at $X=0$ and is monotonically decreasing.  

The scaling of Eq.(\ref{PQsX}) is related to the fact that the system
is described by a finite number of shocks per unit length, with masses of order
unity (in terms of the dimensionless variables $Q$ and $X$). Then, in the limit 
$Q\rightarrow 0$, the probability $P_Q^{\rm shock}$ that all particles in the
interval $[Q_1,Q_2]$, of length $Q=Q_2-Q_1$, belong to the same shock goes to 
unity, and there remains a probability of order $Q$ that the particles $Q_1$ and 
$Q_2$ belong to different shocks, in which case their Eulerian distance is of order 
unity (i.e. $X\sim 1$). This gives rise to the scaling (\ref{PQsX}) for this 
contribution associated with $X>0$.
This discussion shows that the scaling (\ref{PQsX}) is less general than the
scaling (\ref{Fsdef}) obtained for the small-scale Eulerian distribution,
since it relies on the fact that shocks are well separated by distances of order
unity. For instance, in the case of Brownian initial velocity, the scaling
(\ref{Fsdef}) is still satisfied but the property (\ref{PQsX}) is no longer valid.
Indeed, in this case shocks are dense in Eulerian space and the typical Eulerian
distance $X$ between particles initially separated by the Lagrangian distance $Q$
scales as $X\sim \sqrt{Q}$, as can be seen in \cite{Valageas2008}.

As discussed above, in addition to the contribution (\ref{PQXs1}) associated with
$X>0$, there is a second contribution, of the form $\Pshock_Q\delta(X)$, associated
with the case where both particles $Q_1$ and $Q_2$ belong to the same shock,
whence $X_2=X_1$. Its weight can be derived by computing the weight $P_Q(X>0)$
of the contribution (\ref{PQXs1}). Integrating Eq.(\ref{PQXs1}), which gives
two boundary terms at $X=0$ and $X=+\infty$, yields
\beq
P_Q(X>0)= 1 - \Pshock_Q , \;\; \mbox{with} \;\;
\Pshock_Q =  -2 \inta \frac{\dd s \dd s'}{(2\pi\ii)^2} \frac{e^{sQ}}{s^2\Ai(s')^2}
\frac{\pl}{\pl s'} \frac{\Aip(s'+s)}{\Ai(s'+s)} , \;\; \Re(s)<0 ,
\label{Pshock1}
\eeq
where the integration contour over $s$ runs to the left of the pole at $s=0$.
The comparison with Eqs.(\ref{Fsdef})-(\ref{tFsdef}) gives the relation
\beq
\frac{\dd^2}{\dd Q^2}\Pshock_Q = \cFs(Q) , \;\;\; \mbox{whence} \;\;\;
\Pshock_Q = \int_Q^{\infty} \dd Q' \, (Q'-Q) \, \cFs(Q') .
\label{PshockF}
\eeq
In particular, using the results of section~\ref{Small-scales}, Eq.(\ref{PshockF})
gives at once the asymptotic behaviors
\beq
Q \rightarrow 0 : \;\; \Pshock_Q \sim 1 - \tcFs(0) \, Q \simeq 1 - 1.674 \, Q ,
\;\;\;\;\;\; Q \rightarrow \infty : \;\; \Pshock_Q \sim 32 \sqrt{\pi} \,
Q^{-3/2} \, e^{-\om_1 Q -Q^3/12} .
\label{Pshockasymp}
\eeq
Thus, we recover the fact that the probability, $\Pshock_Q$, for two particles
of initial Lagrangian distance $Q$, to belong to the same shock, goes to unity
for $Q\rightarrow 0$, with a linear deviation so that $P_Q(X>0)\sim \tcFs(0) Q$,
in agreement with the scaling (\ref{PQsX}).

\begin{figure}
\begin{center}
\epsfxsize=6.3 cm \epsfysize=5 cm {\epsfbox{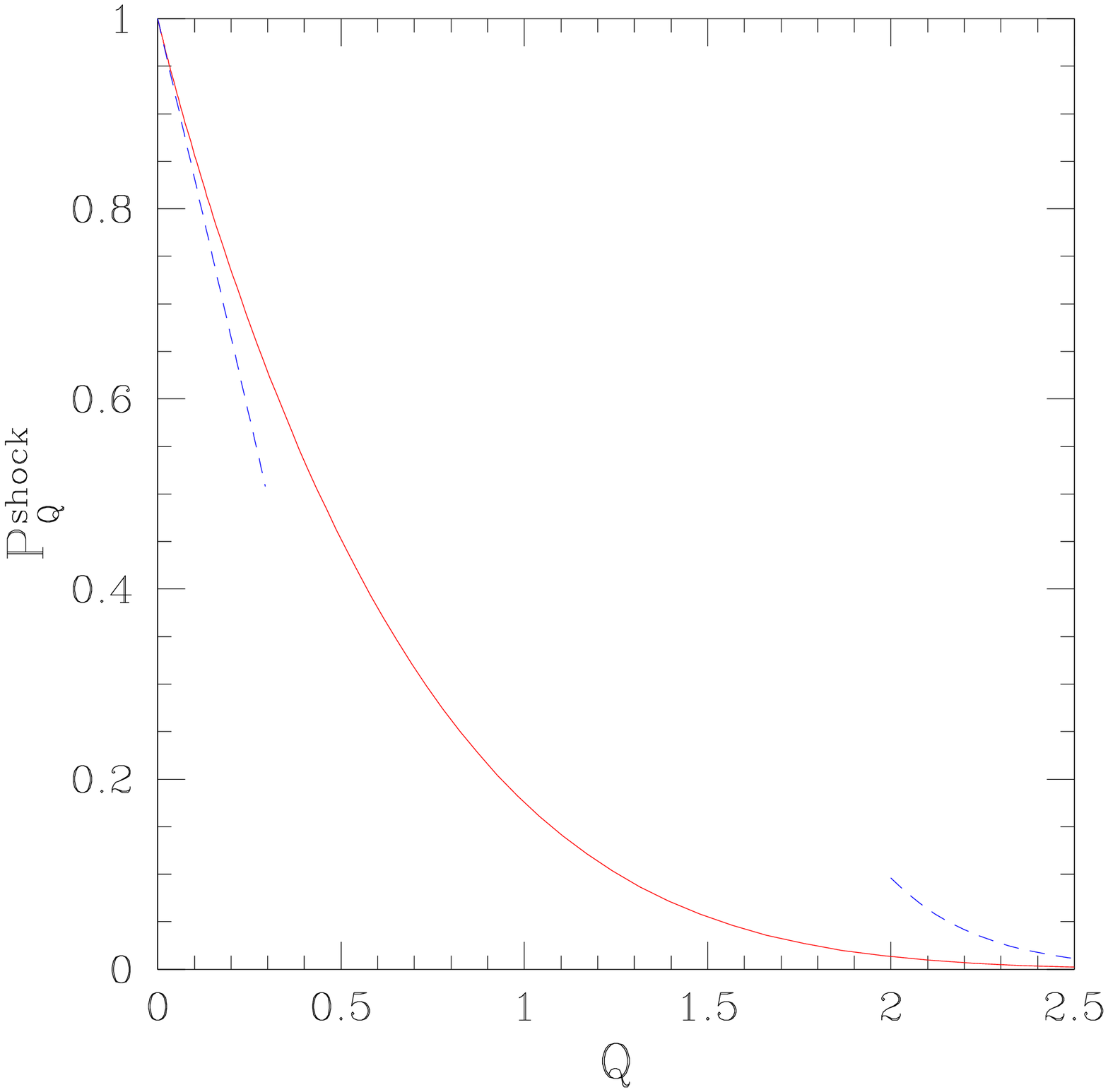}}
\epsfxsize=6.3 cm \epsfysize=5 cm {\epsfbox{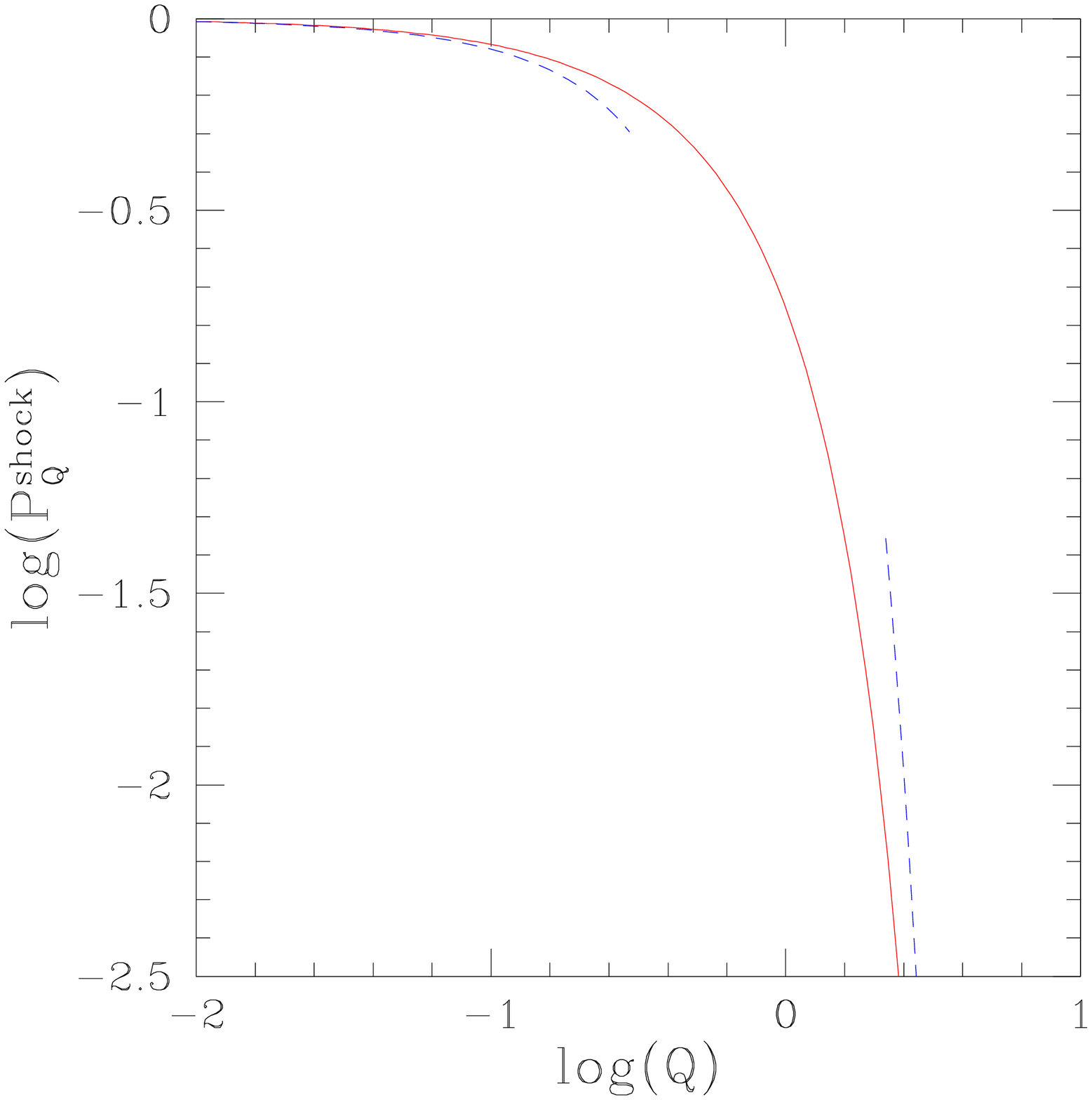}}
\end{center}
\caption{(Color online) {\it Left panel:} The probability $\Pshock_Q$ that a
Lagrangian interval of size $Q$ has collapsed into a single shock, from
Eq.(\ref{PshockF}). The dashed line is the asymptotic behavior (\ref{Pshockasymp}).
{\it Right panel:} Same as left panel but on a logarithmic scale.}
\label{figPshock}
\end{figure}

We display in Fig.~\ref{figPshock}
this probability $\Pshock_Q$, which clearly shows its steep falloff at large $Q$.
Again, the cubic exponential decay can be understood from the same arguments,
as those used for the tails (\ref{PVinf}) or (\ref{PX0asymp}) of Eulerian 
distributions. 
Note that the formation of a single shock of strength $q=q_2-q_1$ 
would be associated with a velocity difference $v_2-v_1=-q/t$, rather than
with the mean velocity $\bv_0$ over the length $q$ as was the case for these 
Eulerian distributions. However, this difference does not give well-defined results
for the initial white-noise velocity field. Again, this is due to the fact that
the dynamics is governed by non-local processes, that is, one cannot obtain
behaviors such as (\ref{Pshockasymp}) from a local analysis (i.e. a local Taylor
expansion) of the initial velocity field. This clearly follows from the fact that,
at any time $t>0$, matter has gathered in a series of discrete shocks, which has
strongly modified the velocity field: the latter has been regularized by the
balance between the (infinite) different sign velocities of neighboring particles
over lengths of order $(2Dt^2)^{1/3}$ and become strongly non-Gaussian, as seen
in sections~\ref{One-point} and \ref{Large-scales}. However, to recover the
tail (\ref{Pshockasymp}), we can split the interval $q$ into two equal parts,
and note that at least one of the two mean initial velocities $\bar{v}_1$ and 
$\bar{v}_2$ of both intervals must be of order $q/t$, which leads back to the
cubic exponential tail (\ref{Pshockasymp}).

\subsection{Mass function of shocks}
\label{Mass-function}

We briefly note here that the mass function of shocks, $n(m)$, can be derived
from the shock probability $p^{\rm shock}_q$ studied in the previous section.
Here we define $n(m)\dd m$ as the mean number of shocks, per unit Eulerian or 
Lagrangian length (both functions are identical since on large scales $X=Q$ up
to fluctuations of order unity, as seen in sections~\ref{Large-scales} and
\ref{Two-point-Lag}), with a mass in the range 
$[m,m+\dd m]$. As in section~\ref{Density}, since we consider a uniform initial
density $\rho_0$, we have $m=\rho_0 q$ for the mass associated with a Lagrangian
interval $q$. Then, the probability, $p^{\rm shock}_q$, that two particles of 
initial Lagrangian distance $q$ belong to the same shock, can be obtained by 
counting the number of shocks of mass $m\geq \rho_0 q$, each shock giving rise
to a factor $(m/\rho_0-q)$ as $q_1$ may be located within the distance 
$(m/\rho_0-q)$ from its left boundary. In terms of dimensionless variables this 
reads as
\beq
\Pshock_Q = \int_Q^{\infty} \dd M \, (M-Q) \, N(M) , \;\;\;\; \mbox{with}
\;\;\; n(m)= \frac{1}{\rho_0\gam^2} N(M) \;\;\; \mbox{and} \;\;\; 
M= \frac{m}{\rho_0 \gamma} .
\label{Ndef}
\eeq
Then, using Eq.(\ref{PshockF}) we obtain at once
\beq
N(M) = \left. \frac{\dd^2 \Pshock_Q}{\dd Q^2} \right|_{Q=M} = \cFs(M) .
\label{NF}
\eeq
From the expression (\ref{Fs1}) we can check that we recover the result of
\cite{Frachebourg2000}, who directly derived the shock mass function from the
geometrical construction (\ref{paraboladef}) without considering the Lagrangian
displacement field. This provides a useful check of the computations performed
in section~\ref{Two-point-Lag} within a Lagrangian framework.
The asymptotic properties of the shock mass function can also be read from 
Eq.(\ref{Fsasymp}), see also
\cite{AvellanedaE1995,Avellaneda1995,Frachebourg2000}.
Moreover, the integral properties
\beq
\int_0^{\infty} \dd M \, N(M) = \tcFs(0) \simeq 1.674 , \;\;\;\;
\int_0^{\infty} \dd M \, M \, N(M) = -\tcFs'(0) = 1 ,
\label{Nint}
\eeq
ensure that mass is conserved and that there is a finite mean number of shocks per
unit length ($\simeq 1.674$ shocks in the mean, in units of $X$ and $Q$).
Many more properties of shocks, such as their $n$-point multiplicity functions
as a function of their mass and velocity, can be found in \cite{Frachebourg2000}.

\section{Small-scale heuristic approach for general initial conditions}
\label{Heuristic}

The results obtained in the previous sections were derived from exact
computations, based on Eqs.(\ref{PQ1Q2i}) and (\ref{PQ1Q2ii}).
In this section, using a heuristic approach that assumes that small-scale
properties are governed by shocks, or point-masses in higher dimensions,
we discuss how small-scale scalings obtained for 1-D white-noise initial
conditions would extend to generic initial conditions and higher dimensions.

\subsection{General 1-D case for the distributions of Lagrangian and velocity 
increments on small scales}
\label{General-case-q}

As explained in section~\ref{Small-scales}, the scalings $P_X(Q) \sim X \cFs(Q)$
and $\lag Q^{\nu} \rag \propto X$ are due to the presence of shocks and as such
they apply to a large class of initial conditions \cite{Frisch2001,Tribe2000}.
Then, the distributions of the Lagrangian increment $q$ and of the velocity
increment $v$ over the distance $x$ are determined by the one-point
distribution of shock strength \cite{Tribe2000}, and they factorize as
\beq
x \rightarrow 0 , \;\; q>0 , \;\; v< x \;\; : \;\; p_x(q) \sim x \, n(q) , \;\;\;
p_x(v) \sim x t \, n(x-v) \sim  x t \, n(-v) ,
\label{pqpvn}
\eeq
where $n(q)$ is the mass function of shocks, that is $n(q)\dd q$ is the number of
shocks of strength $q$ per unit length.
Here we used the dimensional variables $x$ and $q$ because the power 
of time that appears in the relevant scaling variables $X$ and $Q$ depends on the 
initial conditions. For instance, for a power-law initial energy spectrum, 
$E_0(k) \propto k^n$, with $-3<n<1$, we would have $X \propto x/t^{2/(3+n)}$ 
\cite{Gurbatov1997}.
For initial energy spectra that are not a power law there may not exist scaling 
variables such as (\ref{QXdef}) (for instance, for a smooth spectrum one may
expect a time-dependent effective exponent $n(t)$) but as soon as shocks are 
present one still has scalings of the form $\lag q^{\nu}\rag \sim x$ at small 
distance for $\nu\geq 1$. Note that for $\nu=1$ we always have the exact relation
$\lag q\rag=x$,
because of the conservation of matter, but the linear scaling over $x$ does not
always extend to powers $0<\nu<1$, as for the white-noise case studied in the 
present paper, see Eq.(\ref{Qnu}). For instance, for Brownian initial velocity it 
only extends down to $\nu=1/2$ \cite{Aurell1997,Valageas2008}. Indeed, as $\nu$
decreases the moment $\lag q^{\nu}\rag$ becomes increasingly sensitive to 
low-density regions, characterized by low Lagrangian increment $q$, so that shocks
are no longer dominant (for the white-noise case they remain dominant down to
$\nu=0^+$ because shocks are separated by voids where the Lagrangian increment
$q$ is exactly zero). It can be useful to introduce the moment generating function 
$\Psi_x(s)$, defined by
\beq
\Psi_x(s) = \sum_{n=1}^{\infty} \frac{(-s)^n}{n!} \lag q^n\rag = 
\int_0^{\infty} \dd q \, \left(e^{-s q}-1\right) \, p_x(q) , 
\label{Psixdef}
\eeq
whence
\beq
p_x(q) = \inta \frac{\dd s}{2\pi\ii} \, e^{s q} \, \Psi_x(s) \;\;\;\;
\mbox{for} \;\;\;\; q>0 ,
\label{Pxdef}
\eeq
where we assumed that all moments are finite and uniquely determine the function
$\Psi_x(s)$ (note that adding a constant to $\Psi_x(s)$, so that $\Psi_x(0)\neq 0$,
does not contribute to $p_x(q)$ for $q>0$ as it only yields a Dirac $\delta(q)$).
From the small-scale scaling (\ref{pqpvn}) we obtain
\beq
x \rightarrow 0 : \;\; \Psi_x(s) \sim x \, \bPsi(s) \;\; \mbox{and} \;\;
\lag q^n\rag \sim x \, (-1)^n \, \bPsi^{(n)}(0) , \;\;\; \mbox{with} \;\;\;
\bPsi(s) = \int_0^{\infty} \dd q \, \left(e^{-s q}-1\right) \, n(q) .
\label{bPsi}
\eeq
For the case of white-noise initial velocity studied in this article,
using the dimensionless variables $X$ and $Q$, we have $N(Q)=\cFs(Q)$
and $\bPsi(s)=\tcFs(s)-\tcFs(0)$, as shown by Eqs.(\ref{Fsdef})-(\ref{tFsdef})
and Eq.(\ref{NF}).
For the case of Brownian initial velocity, using results from 
\cite{Valageas2008}, we have
\beq
N(Q) = \frac{1}{\sqrt{\pi}} \, Q^{-3/2} \, e^{-Q} , \;\;\; 
\bPsi(s) = 2(1-\sqrt{1+s}) ,
 \;\;\;\; \mbox{for Brownian initial velocity},
\label{Brown1}
\eeq
where we used the relevant scaling variables of the form $Q\propto q/t^2$
(here $E_0(k) \propto k^{-2}$).
We can check in \cite{Valageas2008} that $N(Q)$, given in Eq.(\ref{Brown1}) as the
shock mass function, also describes the probability distribution $P_X(Q)$
through $P_X(Q)\sim X N(Q)$ on small scales, as explained above in (\ref{pqpvn}).
Note that this remains valid even though shocks are no longer isolated but dense
in Eulerian space. Indeed, if we select shocks above a small finite mass threshold
$m_*$, the latter are again isolated so that the previous arguments apply, and
smaller shocks only modify the low-$q$ tail of the distribution. 
Indeed, the previous arguments hold for the limit $x\rightarrow 0$ at 
fixed $q$, or more precisely above a cutoff $q_-(x)$ that goes to zero faster
than $x$, so that the scalings $\lag q^{\nu} \rag \propto x$ for $\nu\geq 1$
do not depend on the behavior of the distribution $p_x(q)$ over this low-$q$ domain.
Thus, the functions $n(q)$ and $\bPsi(s)$ do not necessarily describe the actual 
distribution $p_x(q)$ down to $q=0$ for a finite $x$.
As seen above, for the case of white-noise initial velocity, the scaling function
$\cFs(Q)$ of Eqs.(\ref{Fsdef})-(\ref{tFsdef}) actually applies down to $Q=0^+$,
as it only misses the Dirac contribution (\ref{PX0}) (and the very high-$Q$ tail
(\ref{PXQasympXs3}) which is repelled to infinity). 
However, for the case of Brownian initial velocity for instance, the scaling 
functions (\ref{Brown1}) only apply to $Q\gg X^2$ and they miss a low-$Q$ cutoff 
of the form $e^{-X^2/Q}$ \cite{Valageas2008}.

\subsection{General 1-D case for the distribution of overdensities on small scales}
\label{General-case-eta}

As seen in sect.~\ref{Distribution-of-the-overdensity}, since the overdensity
$\eta$ is also given by the ratio 
$\eta=q/x$, its probability distribution is related to the distribution of the 
Lagrangian increment $q$ through $p_x(\eta)=x p_x(q)$. Then, from 
Eqs.(\ref{pqpvn})-(\ref{bPsi}) we have
\beq
x\rightarrow 0 , \;\; \eta>0 : \;\;\; p_x(\eta) \sim x^2 \, n(x\eta) = x^2 
\inta\frac{\dd s}{2\pi\ii} \, e^{sx\eta} \, \bPsi(s) .
\label{pxetapb}
\eeq
If we define the cumulant ratios $S_n$ and their generating function $\varphi_x(y)$
as in Eqs.(\ref{Sndef})-(\ref{phiSn}),
\beq
S_n(x)= \frac{\lag\eta^n\rag_c}{\lag\eta^2\rag_c^{n-1}} \;\;\; \mbox{and}
\;\;\; \varphi_x(y)= \sum_{n=1}^{\infty} (-1)^{n-1} \, S_n(x) \, \frac{y^n}{n!} ,
\label{Snxdef}
\eeq
using the property that in the small-scale limit, $x\rightarrow 0$, we have
for $n\geq 1$ the asymptotic relationships 
$\lag\eta^n\rag_c \sim \lag q^n\rag/x^n$, we obtain
\beq
x\rightarrow 0 : \;\;\; \varphi_x(y) \sim \bar{\varphi}(y)
\;\;\; \mbox{with} \;\;\; \bar{\varphi}(y) = - \bPsi''(0) \; 
\bPsi\left(\frac{y}{\bPsi''(0)}\right) ,
\label{bvarphidef}
\eeq
and
\beq
x\rightarrow 0 : \;\;\; p_x(\eta) \sim - \inta\frac{\dd y}{2\pi\ii\lag\eta^2\rag_c}
\, e^{y\eta/\lag\eta^2\rag_c} \, \bar{\varphi}(y) \;\;\; \mbox{with} \;\;\; 
\lag\eta^2\rag_c = \frac{\bPsi''(0)}{x} .
\label{pxetavarphi}
\eeq
We can check that this agrees with Eq.(\ref{phiSn}) for the white-noise initial
velocity studied in this article.
For the case of Brownian initial velocity \cite{Valageas2008}, we have, in agreement
with (\ref{Brown1}),
\beq
X \rightarrow 0 : \;\;\; P_X(\eta) \sim \sqrt{\frac{X}{\pi}} \, \eta^{-3/2} 
\, e^{-X\eta} \;\;\; \mbox{and} \;\;\; \bar{\varphi}(y) = \sqrt{1+2y}-1 , 
\;\;\;\; \mbox{for Brownian initial velocity} ,
\label{Brown2}
\eeq
where $X\propto x/t^2$ is again the scaling variable relevant to that initial 
condition.
For the Brownian case it happens that the expression (\ref{Brown2}) for the
density cumulant generating function is actually exact for all scales, but in the 
general case, as discussed below Eq.(\ref{Brown1}), it only applies to the small 
scale limit at fixed density contrast. 
More precisely, although $\cFs$ describes the overdensity 
probability distribution down to $\eta\rightarrow 0^+$ through Eq.(\ref{PXetas})
in the case of white-noise initial velocity, in the general case it only applies
above a cutoff $\eta_-(x)$ that decreases with $x$ (for instance for the
Brownian initial velocity we have $\eta_-(x) \propto x$).

\subsection{Multifractal formalism in $D$ dimensions}
\label{Multifractal-formalism}

These properties can be extended to higher dimensions $D$ through a 
heuristic multifractal
formalism \cite{Balian1989b,Frisch1995,Valageas1999}, without going through 
the inverse Lagrangian map $\bx\mapsto \bq$.
Thus, let us assume that the overdensity within a spherical cell of radius $\ell$ 
centered on $\bx$ scales for $\ell\rightarrow 0$ as 
$\eta_{\ell}(\bx)\sim \ell^{\alpha}$ for points 
$\bx\in\cD_{\alpha}\subset {\mathbb{R}}^{\it D}$, with 
${\rm dim}\cD_{\alpha}=F(\alpha)$. Then, we may write
\beq
\ell\rightarrow 0, \;\; \nu > 0 : \;\;\; \lag \eta_{\ell}^{\nu}\rag = 
\int_0^{\infty} \dd\eta \, \eta^{\nu} p_{\ell}(\eta) \sim \int \dd\alpha \, 
\ell^{\nu\alpha+D-F(\alpha)} p_*(\alpha) ,
\label{palpha}
\eeq
where $p_*(\alpha)$ gives the weight of the various multifractal exponents,
and we used the fact that the probability for a sphere of radius $\ell$ to encounter
an object of dimension $F$ scales as $\ell^{D-F}$ for $\ell\rightarrow 0$.
Using a steepest-descent argument, we obtain the small-scale exponents 
$\gam_{\nu}$,
\beq
\ell \rightarrow 0 : \;\;\; \lag \eta_{\ell}^{\nu}\rag \sim \ell^{-\gam_{\nu}}
\;\;\; \mbox{with} \;\;\; \gam_{\nu} = - \min_{\alpha} [\nu\alpha+D-F(\alpha)] 
= \max_{\alpha} [F(\alpha)-\nu\alpha-D] .
\label{gampalpha}
\eeq
Thus, the exponents $\gam_{\nu}$ and $F(\alpha)$ are related by a Legendre 
transform. The fractal scaling exponent $\alpha_{\nu}$ that is associated to
$\gam_{\nu}$ through (\ref{gampalpha}) is the abscissa of the first-contact point
of the curve $F(\alpha)$ with the family of straight lines, $\nu\alpha+c$, moving
downward from $c=+\infty$, in a fashion similar to the geometrical construction
associated with (\ref{psinu0}). In particular, the exponents $\gam_{\nu}$ only
probe the concave hull of $F(\alpha)$ \cite{Frisch1995}.
Since the matter density is positive the scaling exponents $\alpha_{\nu}$
are restricted to $\alpha\geq -D$. This lower bound corresponds to Dirac
density peaks (i.e. massive points), which have a zero dimension, $F(-D)=0$. 
On the other hand, the constraint associated with the conservation of matter,
$\lag\eta_{\ell}\rag=1$ whence $\gam_1=0$, ensures that the curve $F(\alpha)$
is below the straight line $\alpha+D$, which runs through the point $\{-D,0\}$,
and has at least one contact point with this line. Then, we can see that, as soon
as isolated Dirac density peaks have formed, with a finite probability per unit
volume, the first-contact point between $F(\alpha)$ and the family of straight lines
$\nu\alpha+c$ with $\nu\geq 1$ is the point $\{-D,0\}$ (for $\nu=1$ there can be
other additional contact points), which gives $\gam_{\nu}=(\nu-1)D$ for 
$\nu\geq 1$.
For instance, in three dimensions $D=3$,
we generically expect to first form ``Zeldovich pancakes'' \cite{Zeldovich1970},
that is sheets with a finite surface density, that intersect to form filaments
(i.e. lines of finite line density), which join to form point-like masses (nodes).
This corresponds to objects of fractal exponents and dimensions $\{-1,2\}$,
$\{-2,1\}$, and $\{-3,0\}$, along the line $\{\alpha,F=\alpha+D\}$, that all 
contribute to $\gam_1=0$ while only point-like masses contribute to 
$\gam_{\nu}=3(\nu-1)$ for $\nu>1$.
Then, we obtain
\beq
n \geq 1 : \;\;\; \lag \eta^n\rag \sim\ell^{-(n-1)D} \;\; \mbox{whence} \;\;  
\lag \eta^n\rag_c \sim\ell^{-(n-1)D} , \;\; S_n(\ell) \sim \ell^0 \;\; 
\mbox{and} \;\; \varphi_{\ell}(y) \sim \bar{\varphi}(y) \;\; \mbox{for} \;\;
\ell \rightarrow 0 ,
\label{varphil}
\eeq
where the limiting generating function $\bar{\varphi}(y)$, reached in the 
small-scale limit, no longer depends on $\ell$.
Thus, for stochastic initial conditions, where we generically expect the formation
of isolated Dirac density peaks in finite numbers per unit volume, we obtain
the scaling (\ref{varphil}) and the first expression in (\ref{pxetavarphi}) for
the density probability distribution, where in dimension $D$ we have
$\lag\eta^2\rag_c \sim \ell^{-D}$.
As explained above, within this heuristic multifractal formalism the property
(\ref{varphil}) holds independently of the form of the fractal spectrum $F(\alpha)$
over $\alpha>-D$,
as soon as there is a finite density of point-like masses which gives rise to the
fractal exponent $\{\alpha=-D,F=0\}$. The computation of the complete spectrum
of fractal dimensions, that is necessary for the study of exponents $\nu<1$
in Eq.(\ref{palpha}), is a difficult task and in the general case there can be a
continuous rather than discrete spectrum (as discussed below in a phenomenological
fashion for the case of Brownian initial velocity, where shocks are dense).

Again, the scaling function $\bar{\varphi}(y)$ and the distribution it defines
through (\ref{pxetavarphi}) only apply above a density threshold $\eta_-(\ell)$
that may only show a very slow decrease with $\ell$.
Within the multifractal formalism (\ref{palpha}) the behavior at small densities,
below this threshold $\eta_-(\ell)$, depends on the properties of the curve
$F(\alpha)$ to the right of the point $\{-D,0\}$, see 
\cite{Balian1989b,Valageas1999}.
For the one-dimensional case with white-noise initial velocity studied in this
article, since shocks form isolated density peaks amid empty space, the curve
$F(\alpha)$ is reduced to the single point $\{-1,0\}$ so that the scalings
$\lag\eta^{\nu}\rag \sim X^{1-\nu}$ apply to all $\nu>0$, in agreement with the 
second relation in (\ref{PXetas}) and the fact that the distribution 
$P_X(\eta)\sim X^2 \cFs(X\eta)$ in (\ref{PXetas}) applies downto $\eta=0^+$, 
as seen in the previous sections.

For the one-dimensional case with Brownian 
initial velocity, we have the bifractality $\lag q^{\nu}\rag \sim x$ for 
$\nu \geq 1/2$ and $\lag q^{\nu}\rag \sim x^{2\nu}$ for $\nu \leq 1/2$
\cite{Aurell1997,Valageas2008}. This leads to $\lag \eta^{\nu}\rag \sim x^{1-\nu}$ 
for $\nu \geq 1/2$ and $\lag \eta^{\nu}\rag \sim x^{\nu}$ for $\nu \leq 1/2$.
In terms of the multifractal formalism, this would be interpreted as a second
point $\{1,1\}$. This may be understood as follows. For these initial conditions,
the shock mass function diverges at small masses as $n(m) \propto m^{-3/2}$ and 
shocks are uncorrelated and dense in Eulerian space 
\cite{She1992,Sinai1992,Bertoin1998,Valageas2008}.
Then, choosing a small finite mass threshold $m_*$, the set of shocks of mass 
larger than $m_*$ gives a population of isolated point-like masses that gives 
rise to the fractal exponent $\{-1,0\}$. On the other hand, if we choose a random 
Eulerian interval of size $\ell$, it contains in the mean $m^{-1/2}\ell$ shocks of 
mass in the range $[m,2m]$. Taking $m\sim \ell^{\beta}$ we obtain that i) for any
$\beta>2$ an interval of size $\ell$ contains of the order of $\ell^{1-\beta/2}$
shocks of mass in $[m,2m]$, which leads to an  overdensity larger than 
$\eta_{\ell} \sim \ell^{\beta/2}$, and ii) for any $\beta<2$ an interval of size 
$\ell$ contains with a probability of order $\ell^{1-\beta/2}$ at least one
shock of mass in the range $[m,2m]$, which leads to an  overdensity larger than 
$\eta_{\ell} \sim \ell^{\beta-1}$. The first point i) leads to the multifractal
point $\{1,1\}$, and the second point ii) leads to the points 
$\{\beta-1,\beta/2\}$ for $0<\beta<2$, that is to the segment joining the
points $\{-1,0\}$ and $\{1,1\}$. Therefore, we obtain in this case the
multifractal spectrum $F(\alpha)=\alpha/2+1/2$ with $-1\leq\alpha\leq 1$, see also
\cite{Aurell1997} for more rigorous discussions. More generally, in one dimension
the scalings obtained for $\nu<1$ and the low-density tail are related to the 
low-mass tail of the shock mass function.

We can note that the small-scale limit of finite ratios $S_n$ and generating 
function $\varphi_x(y)$, as in (\ref{bvarphidef}), also corresponds to the 
``stable-clustering ansatz'' introduced in the cosmological context as a 
phenomenological model for the highly nonlinear regime \cite{Peebles1980}.
There, it was derived by assuming that on small physical scales, after nonlinear 
collapse and gravitational relaxation, overdensities decouple from the Hubble 
expansion and keep a constant physical size \cite{Davis1977}.
It can also be associated with a multifractal formalism, where the moments
of the density with $\nu\geq 1$ would be governed by a single fractal exponent
as in (\ref{varphil}), which however would not be associated with point-like
masses but with structures of exponent $\alpha\sim 1.3$ and dimension 
$F=3-\alpha\sim 1.2$ \cite{Balian1989b,Valageas1999}. However, contrary to the
Burgers dynamics, this behavior may not be exactly reached on small scales
for the gravitational dynamics, as the coefficients $S_n$ still appear to show 
a weak dependence with scale \cite{Colombi1996}.
On the other hand, the Burgers dynamics itself is also known as the ``adhesion
model'' in this cosmological context \cite{Gurbatove1989,Vergassola1994},
where it provides a good description of the large-scale filamentary structure of 
the cosmic web \cite{Melott1994}. It is not clear whether the reasonably good
match of the ``stable-clustering ansatz'' could be understood from the
the exact scaling (\ref{varphil}) achieved in the small-scale limit by the
``adhesion model'', since the nonlinear structures are different (point-like
masses as opposed to extended halos) and no detailed comparisons have been performed
yet in terms of the ratios $S_n$ themselves.

\section{Conclusion}
\label{Conclusion}

We have obtained in this article some equal-time properties of the Burgers dynamics,
in the inviscid limit for white-noise initial velocity. In agreement with previous
works, the initially singular distributions are regularized as soon as $t>0$, but
little power is transfered to large scales. Thus, the distributions of the 
fluctuations of the Lagrangian increment, $q$, and of the velocity increment, $v$,
around their means, have a finite limit in the large-scale limit $x\rightarrow 0$.
We recover the characteristic cubic exponential tails associated with white-noise 
initial conditions. Voids lead to an additional Dirac-type contribution to these 
distributions, that also decays as a cubic exponential at large scales and is
preceded by an inverse square root tail with a weight of the same order.
On small scales, where the probability to be within a void goes to unity,
the regular part factorizes as $X\,\cFs(Q)$, which corresponds to the probability
to contain one shock of strength $Q$. In particular, the scaling function $\cFs(Q)$
is also the mass function of shocks. This leads to the standard linear scaling 
with $x$ of the velocity structure functions at small scale, due to shocks.

Next, we have derived the distribution of the density within intervals of size $x$.
It presents similar properties to those obtained for the Lagrangian increment,
and exhibits the corresponding large-scale and small-scale scalings. In particular,
at small scales this gives rise to the scaling hierarchy for the density cumulants
known as the ``stable-clustering ansatz'' in cosmology. Here it is due to the
presence of shocks. We also
obtain the density two-point correlation and power spectrum, with the
high-wavenumber constant asymptote associated with shocks.

Turning to the Lagrangian displacement field, associated with a description
of the dynamics in terms of Lagrangian coordinates, we have obtained the
distribution of the Eulerian increment $x$ for a given mass $\rho_0 q$. 
On large scales the Lagrangian distribution $p_q(x)$ becomes identical, at leading
order, to the Eulerian distribution $p_x(q)$. On small scales there is also
a factorization of the form $Q\,\cGs(X)$, but this is less general than for the 
small-scale Eulerian distribution since it only applies to initial conditions such
that shocks are isolated, that is initial energy spectra with $-1<n<1$, whereas the
Eulerian factorization remains valid for the whole range $-3<n<1$.
Contrary to the Eulerian distribution, the Lagrangian distribution $p_q(x)$ does
not show divergent tails as it remains finite for $x\rightarrow 0$, but there is
again an additional Dirac contribution, which is now due to shocks.

Finally, within a heuristic approach we have discussed how these small-scale
properties generalize to other initial conditions and give rise to
a universal scaling for the distribution of the Lagrangian increment (and of the
velocity increment) above a lower cutoff $q_-(x)$, that goes to zero faster than
$x$ in a fashion that depends on the initial conditions.
A heuristic multifractal formalism allows to extend these results to higher
dimensions. It generically leads
to a universal scaling hierarchy for the density cumulants in the small-scale limit,
that is governed by point-like masses. This also corresponds to the 
``stable-clustering ansatz'' introduced in the cosmological context, where the
Burgers dynamics is known as the ``adhesion model'' and is used to describe the
large-scale cosmic web.

The results obtained in this article may prove useful to test approximation schemes 
devised to handle other initial conditions or closely related dynamics, such as
Navier-Stokes turbulence or gravitational dynamics, where no exact results are 
available, as in \cite{Fournier1983,Valageas2007,Valageas2009}. 
In this respect, the case of white-noise initial velocity studied here
would present a severe test for non-perturbative methods. Indeed, the initial
energy spectrum is so ``blue'' that nonlinear effects are dominant up to the largest
scales, $x\rightarrow\infty$, and perturbative expansions already encounter
ultraviolet divergences at leading orders. This implies that alternative approaches
must be able to take into account shocks, as for the steepest-descent methods
presented in \cite{Valageas2009}.  
Another interesting feature of the case of white-noise initial velocity studied in this
article is that it shows a density power spectrum that displays two different large-scale
and small-scale regimes, as for the gravitational dynamics in the cosmological
context, but can still be computed exactly.

\appendix

\section{Transition kernel with parabolic absorbing barrier}
\label{sec:transition}

For the white-noise initial conditions (\ref{v0def}), the process 
$q\mapsto\psi_0$ is Markovian and a key quantity is the conditional
probability density $K_{x,c}(q_1,\psi_1;q_2,\psi_2)$ for the Markov
process $\psi_0(q)$, starting from $\psi_1$ at $q_1$, to end at
$\psi_2$ at $q_2 \geq q_1$, while staying above the parabolic barrier,
$\psi_0(q)>\cP_{x,c}(q)$, for $q_1\leq q\leq q_2$. We briefly recall
here its derivation, obtained in \cite{Frachebourg2000}, using our
notations. It obeys the diffusion equation
\beq
q_2 \geq q_1 : \;\;\;\; \frac{\pl}{\pl q_2} K_{x,c}(q_1,\psi_1;q_2,\psi_2) 
= \frac{D}{2} \frac{\pl^2}{\pl\psi_2^2} K_{x,c}(q_1,\psi_1;q_2,\psi_2)
\label{diffK}
\eeq
over the domain $\psi \geq \cP_{x,c}(q)$, with the initial condition at $q_2=q_1$,
$K_{x,c}(q_1,\psi_1;q_1,\psi_2) = \delta(\psi_2-\psi_1)$,
and the boundary conditions, $K_{x,c}(q_1,\psi_1;q_2,\psi_2) = 0$ 
for $\psi_1=\cP_{x,c}(q_1)$ or $\psi_2=\cP_{x,c}(q_2)$.
The kernel associated with the propagation towards the left side, $q_2 \leq q_1$, 
is obtained from the parity symmetry
\beq
q_2 \leq q_1 : \;\;\;\; K_{x,c}(-q_1,\psi_1;-q_2,\psi_2) = 
K_{-x,c}(q_1,\psi_1;q_2,\psi_2) .
\label{Kmq}
\eeq
In terms of the dimensionless coordinates (\ref{QXdef}) the kernel $K_{x,c}$
can be written as
\beq
K_{x,c}(q_1,\psi_1;q_2,\psi_2) \, \dd \psi_2 = e^{(Q_2-X)r_2-(Q_1-X)r_1
-(Q_2-X)^3/3+(Q_1-X)^3/3} \; G(\tau;r_1,r_2) \, \dd r_2 ,
\label{KxcG}
\eeq
where we defined
\beq
\tau= Q_2-Q_1, \;\;\; r_i= 2\left[ \Psi_i+\frac{(Q_i-X)^2}{2}-C \right] ,
\label{taudef}
\eeq
and the propagator $G$ obeys the Schrodinger-like equation
\beq
\frac{\pl G}{\pl\tau} = - r_2 \, G + \frac{\pl^2 G}{\pl r_2^2}  
\;\;\; \mbox{over} \;\;\;\; \tau\geq 0 , \;\;\; r\geq 0 ,
\label{Gforward}
\eeq
with the initial condition $G(0;r_1,r_2) = \delta(r_2-r_1)$ and the boundary
conditions $G(\tau;r_1,r_2)=0$ for $r_1=0$ or $r_2=0$.
This reduced propagator 
$G$ can be solved as \cite{Groeneboom1989,Frachebourg2000}
\beq
G(\tau;r_1,r_2)= \sum_{k=1}^{\infty} e^{-\om_k \tau} \, 
\frac{\Ai(r_1-\om_k)\Ai(r_2-\om_k)}{\Aip(-\om_k)^2} ,
\label{GAiAi}
\eeq
where $-\om_k$ are the zeros of the Airy function $\Ai(x)$ (in particular,
$\om_1 \simeq 2.338$). Thus, $G(\tau;r_1,r_2)$ is symmetric over $\{r_1,r_2\}$,
and it also obeys the backward equation (compare with Eq.(\ref{Gforward}))
\beq
\frac{\pl G}{\pl\tau} = - r_1 \, G + \frac{\pl^2 G}{\pl r_1^2}  
\;\;\;\; \mbox{over} \;\;\;\; \tau\geq 0 , \;\;\; r\geq 0 .
\label{Gbackward}
\eeq
Next, it is convenient to introduce the probability density, 
$E_{x,c}(q_1,\psi_1;q_2,\psi_2;q)\dd q \dd c \dd\psi_2$, for the curve 
$\psi_0(q)$, starting from $\psi_1$ at $q_1$, to end at $\psi_2$ at 
$q_2\geq q_1$, while staying above the parabolic barrier $\cP_{x,c}$,
and with a last excursion below $\cP_{x,c+\dd c}$ in the range $[q,q+\dd q]$.
From the definition of the kernel $K_{x,c}$, it reads as
\beq
E_{x,c}(q_1,\psi_1;q_2,\psi_2;q) = \frac{\pl}{\pl q} 
\lim_{\delta c\rightarrow 0} \frac{1}{\delta c} \int \dd\psi 
\left[ K_{x,c}(q_1,\psi_1;q,\psi)-K_{x,c+\delta c}(q_1,\psi_1;q,\psi) \right]
K_{x,c}(q,\psi;q_2,\psi_2) .
\label{E1}
\eeq
Using a Taylor expansion and integrations by parts, this yields
\cite{Frachebourg2000}
\beq
E_{x,c}(q_1,\psi_1;q_2,\psi_2;q) = \frac{D}{2} \, \left. 
\frac{\pl K_{x,c}}{\pl\psi_2}(q_1,\psi_1;q,\psi) \, 
\frac{\pl K_{x,c}}{\pl\psi_1}(q,\psi;q_2,\psi_2) \right|_{\psi=\cP_{x,c}(q)} ,
\label{E2}
\eeq
which gives in terms of the reduced propagator $G$ introduced in Eq.(\ref{KxcG})
\beqa
E_{x,c}(q_1,\psi_1;q_2,\psi_2;q) & = & \frac{8Dt^4}{\gam^8} \,  
e^{(Q_2-X)r_2-(Q_1-X)r_1-(Q_2-X)^3/3+(Q_1-X)^3/3}  \nonumber \\
&& \times \; \frac{\pl G}{\pl r_2}(Q-Q_1;r_1,0) \,
\frac{\pl G}{\pl r_1}(Q_2-Q;0,r_2) .
\label{ExcG}
\eeqa
As could be expected, the expression (\ref{ExcG}) shows that the probability 
density $E_{x,c}$ depends on the behavior of the propagator $G$ close to
the boundary $r=0$ at one end. This corresponds to the contact point of abscissa
$q$ between the curve $\psi_0$ and the parabola $\cP_{x,c}$ that is involved in 
the definition of $E_{x,c}$.

\section{Eulerian distributions}
\label{Eulerian-distributions}

\subsection{One-point distributions $p_x(q)$ and $p_x(v)$}
\label{One-point}

Substituting Eqs.(\ref{ExcG}) and (\ref{GAiAi}) into Eq.(\ref{pxq1}), one
obtains in terms of the dimensionless variables (\ref{QXdef}),
\beq
P_X(Q) = \cJ(X-Q) \, \cJ(Q-X) \;\;\; \mbox{and} \;\;\; P(V) = \cJ(V) \cJ(-V)
,\label{PXQJJ}
\eeq
where we used the relation $X=Q+V$, with \cite{Frachebourg2000}
\beq
\cJ(u) = \lim_{\tau\rightarrow\infty} \int_0^{\infty}\dd r \, 
e^{-(\tau-u)^3/3+(\tau-u)r} \sum_{k=1}^{\infty} e^{-\om_k \tau} 
\frac{\Ai(r-\om_k)}{\Aip(-\om_k)} = \inta \frac{\dd s}{2\pi\ii} \,
\frac{e^{s u}}{\Ai(s)} .
\label{Jdef}
\eeq
From the asymptotic behaviors of the function $\cJ(u)$,
\beq
u \rightarrow +\infty: \;\; \cJ(u) \sim \frac{e^{-\om_1 u}}{\Aip(-\om_1)} ,
\;\;\;\;\; \mbox{and for} \;\;\; u \rightarrow -\infty: \;\;\;\; \cJ(u) \sim 
-2 u \, e^{u^3/3} ,
\label{Jasymp}
\eeq
one obtains the asymptotic behavior (\ref{PVinf}) of the distribution
of the velocity $V$ (and of the Lagrangian coordinate $Q=X-V$)
\cite{Frachebourg2000}.

\subsection{Two-point distributions $p_{x_1,x_2}(q_1,q_2)$ and 
$p_{x_1,x_2}(v_1,v_2)$}
\label{Two-point}

We first consider the case i) of section~\ref{Two-point-Eulerian}, when
the two first-contact parabolas $\cP_{x_1,c_1}$ and $\cP_{x_2,c_2}$
have two different contact points $q_1$ and $q_2$ with the curve $\psi_0(q)$
(and there is at least one shock in the interval $[x_1,x_2]$ since the
map $x\mapsto q$ is constant outside of shocks \cite{She1992,Frachebourg2000}). 
Then, noting $q_*$ the abscissa of the intersection between both parabolas, 
in a fashion similar to (\ref{pxq1}) we can write this contribution to 
$p_{x_1,x_2}(q_1,q_2)$ as
\beq
p_{x_1,x_2}^{\neq}(q_1,q_2) = \lim_{q_{\pm}\rightarrow\pm\infty} 
\int \dd c_1\dd c_2 \dd\psi_*\dd\psi_+ \, E_{x_1,c_1}(q_-,0;q_*,\psi_*;q_1)
\, E_{x_2,c_2}(q_*,\psi_*;q_+,\psi_+;q_2) .
\label{pq1q2i}
\eeq
Substituting the expression (\ref{ExcG}) of the kernel $E_{x,c}$ gives,
in agreement with \cite{Frachebourg2000}, the expression (\ref{PQ1Q2i}),
where we introduced the function $\cH$ defined by
\beqa
\cH_{X_1,X_2}(Q_1,Q_2) & = & 2(X_2-X_1) \int_0^{\infty} \dd r_* \int_{Q_1}^{Q_2} 
\dd Q_* \, e^{(X_2-X_1) r_*-(Q_*-X_1)^3/3+(Q_*-X_2)^3/3} \, \nonumber \\
&& \times \frac{\pl G}{\pl r_1}(Q_*-Q_1;0,r_*) \, 
\frac{\pl G}{\pl r_2}(Q_2-Q_*;r_*,0) .
\label{Hdef}
\eeqa
We can check that the function $\cH_{X_1,X_2}(Q_1,Q_2)$, whence the distribution
$P_{X_1,X_2}^{\neq}(Q_1,Q_2)$, are invariant with respect to uniform translations 
of $X_i$ and $Q_i$, in agreement with the statistical homogeneity of the system.

We now consider the second case ii) of section~\ref{Two-point-Eulerian}, when
the two parabolas intersect at the common point $q_1=q_2$,
and we can write this contribution to $p_{x_1,x_2}(q_1,q_2)$ as
\beq
p_{x_1,x_2}^=(q_1,q_2) = \delta(q_2-q_1) \; \lim_{q_{\pm}\rightarrow\pm\infty}
\lim_{q_*\rightarrow q_1^+} \int \dd c_1 \dd\psi_*\dd\psi_+ \, 
E_{x_1,c_1}(q_-,0;q_*,\psi_*;q_1) \, K_{x_2,c_2}(q_*,\psi_*;q_+,\psi_+) .
\label{pq1q2ii}
\eeq
This gives in terms of dimensionless variables \cite{Frachebourg2000}
the expression (\ref{PQ1Q2ii}).

\subsection{Probability $P_X^0$ of empty intervals of size $X$}
\label{Void-distribution}

\begin{figure}
\begin{center}
\epsfxsize=6.3 cm \epsfysize=5 cm {\epsfbox{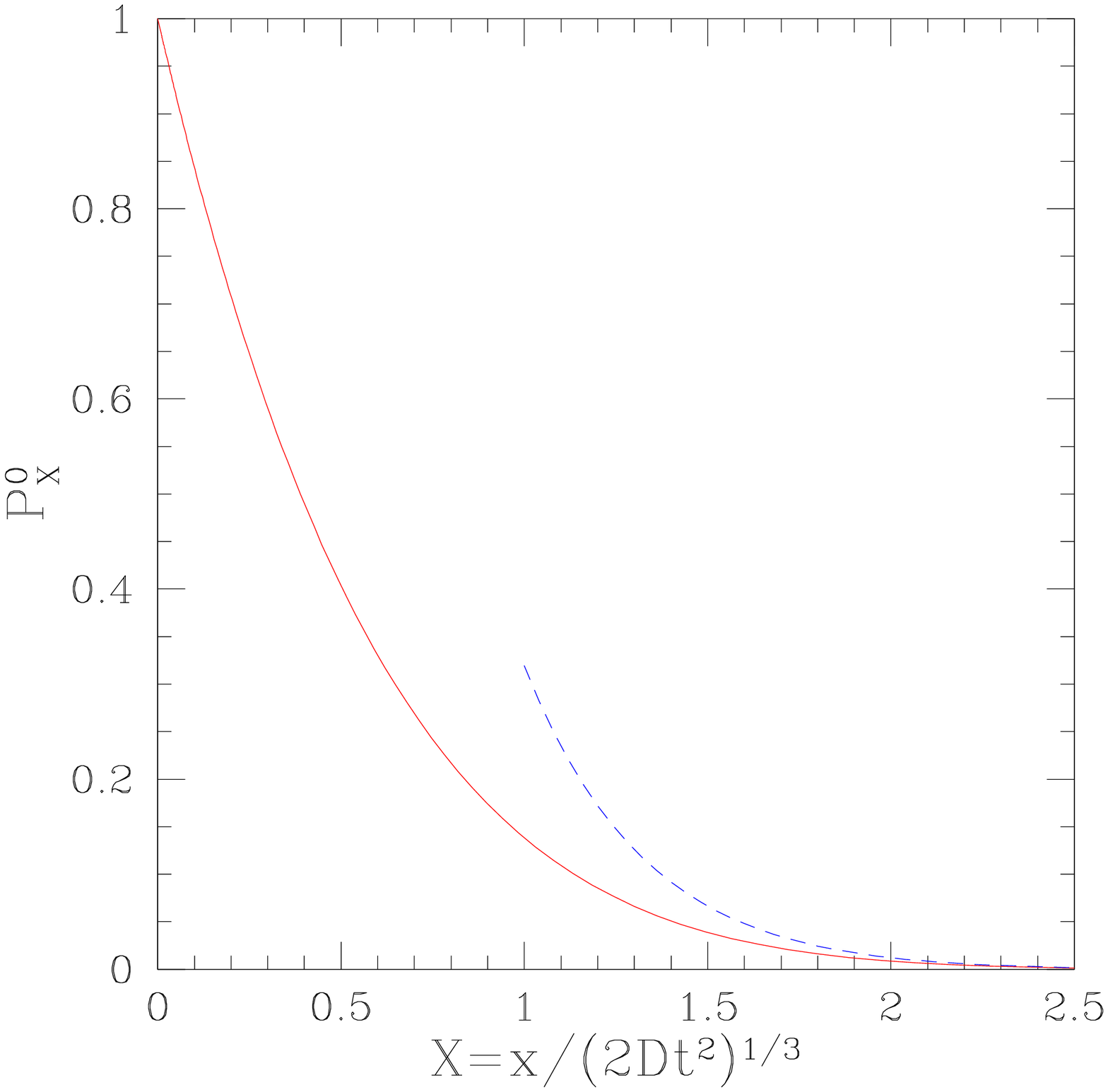}}
\epsfxsize=6.3 cm \epsfysize=5 cm {\epsfbox{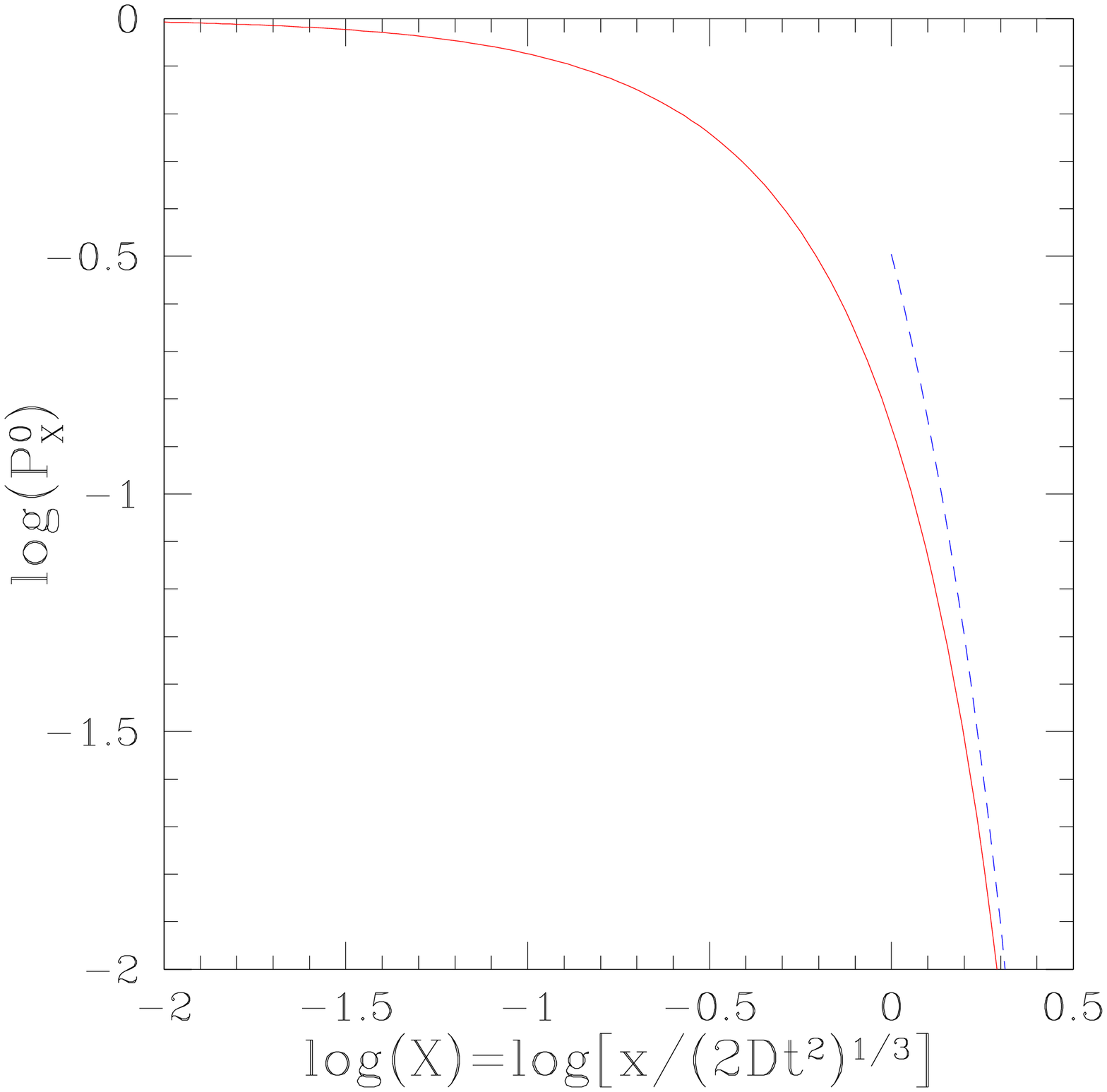}}
\end{center}
\caption{(Color online) {\it Left panel:} The probability $P_X^0$ that an Eulerian
interval of size $X$ is empty, that is, that the Lagrangian increment $Q$ over this interval
is zero, from Eq.(\ref{PX0}). The dashed line is the asymptotic behavior 
(\ref{PX0asymp}).
{\it Right panel:} Same as left panel but on a logarithmic scale.}
\label{figPvoid}
\end{figure}

Using Eq.(\ref{Jdef}),
we can write the second term in (\ref{PXQ1}) as \cite{Frachebourg2000}
\beq
P_X^=(Q) = \delta(Q) \, P_X^0 \;\;\; \mbox{with} \;\;\;
P_X^0 = \sqrt{\frac{\pi}{X}} \, e^{-X^3/12} \inta\frac{\dd s_1\dd s_2}{(2\pi\ii)^2}
\; \frac{e^{(s_1+s_2)X/2+(s_1-s_2)^2/(4X)}}{\Ai(s_1) \, \Ai(s_2)} .
\label{PX0}
\eeq
This yields for the probability $P_X^0$ to have a vanishing Lagrangian increment
the asymptotic behaviors
\beq
X\rightarrow 0 : \;\; P_X^0 \rightarrow 1 , \;\;\;\; \mbox{and for}  \;\;\;   
X\rightarrow\infty : \;\; P_X^0 \sim \frac{\sqrt{\pi}}{\Aip(-\om_1)^2} \, 
X^{-1/2} \, e^{-\om_1 X-X^3/12} .
\label{PX0asymp}
\eeq
Since Eulerian intervals with $Q=0$ have a zero matter content,
$P_X^0$ is also the probability for an interval of size $X$ to be empty,
in agreement with the result obtained in \cite{Frachebourg2000} for this void
probability.
We compare this probability $P_X^0$ with its asymptotic behavior (\ref{PX0asymp})
in Fig.~\ref{figPvoid}, see also \cite{Frachebourg2000}. The cubic exponential 
tail (\ref{PX0asymp}) may be understood using the same arguments as those
used for the tail of the velocity distribution (\ref{PVinf}) discussed above.
Thus, for the Eulerian interval of size $x$ to be empty, its initial matter 
content must have traveled by a distance of order $x$, which requires a mean
velocity over this interval of order $v\sim x/t$. Again, since the initial
Gaussian velocity over scale $x$ is $\bv_0(x)=(\psi_2-\psi_1)/x$, with a variance
$\sigma_{\bv_0}^2(x)=D/x$, this yields the probability 
$\sim e^{-(x/t)^2/\sigma_{\bv_0}^2(x)}\sim e^{-x^3/(Dt^2)}$, which gives back the
cubic exponential tail (\ref{PX0asymp}).

\section{Laplace transform of the product of two Airy functions}
\label{Laplace}

We recall here the results obtained by \cite{Frachebourg2000} for the
integral over two Airy functions that appears in Eq.(\ref{PXQs1}).
Thus, if we define $f(r)$ and $g(x)$ by
\beq
f(r) = \Ai(r+s_1) \Ai(r+s_2) , \;\;\;  
g(x) = \int_0^{\infty} \dd r \, e^{x r} \, f(r) ,
\label{fgdef}
\eeq
the Laplace transform $g(x)$ can be integrated as
\beq
g(x) = \frac{1}{2\sqrt{\pi}} \, e^{\Phix(x)} -  e^{\Phix(x)} 
\, \int_x^{\infty} \dd y \, e^{-\Phix(y)} \, \hx(y) ,
\label{gPhih}
\eeq
with
\beq
\Phix(x) = \frac{x^3}{12} - \frac{s_1+s_2}{2} x - \frac{1}{2} \ln x 
- \frac{(s_1-s_2)^2}{4x} , 
\label{Phidef}
\eeq
and
\beq
\hx(x) = \frac{f(0)}{4} x - \frac{f'(0)}{4} + \frac{f''(0)-2(s_1+s_2)f(0)}{4x}
- \frac{f^{(3)}(0)-2(s_1+s_2)f'(0)-2f(0)}{4x^2} .
\label{hdef}
\eeq
The first term in the right hand side of Eq.(\ref{gPhih}) gives the large-$x$
behavior of $g(x)$, up to terms of relative order $e^{-x^3/12}$.

|
\bibliographystyle{plain}

\bibliography{ref}   

\begin{thebibliography}{10}

\bibitem{Aurell1997}
E.~Aurell, U.~Frisch, A.~Noullez, and M.~Blank.
\newblock Bifractality of the devil's staircase appearing in the burgers
  equation with brownian initial velocity.
\newblock {\em J. Stat. Phys.}, 88:1151--1164, 1997.

\bibitem{Avellaneda1995}
M.~Avellaneda.
\newblock Statistical properties of shocks in burgers turbulence ii: tail
  probabilities for velocities, shock-strengths and rarefaction intervals.
\newblock {\em Commun. Math. Phys.}, 169:45--59, 1995.

\bibitem{AvellanedaE1995}
M.~Avellaneda and Weinan E.
\newblock Statistical properties of shocks in burgers turbulence.
\newblock {\em Commun. Math. Phys.}, 172:13--38, 1995.

\bibitem{Balian1989b}
R.~Balian and R.~Schaeffer.
\newblock Scale-invariant matter distribution in the universe. ii - bifractal
  behaviour.
\newblock {\em Astron. Astrophys.}, 226:373--414, 1989.

\bibitem{Bec2007}
J.~Bec and K.~Khanin.
\newblock Burgers turbulence.
\newblock {\em Phys. Rep.}, 447:1--66, 2007.

\bibitem{Bertoin1998}
J.~Bertoin.
\newblock The inviscid burgers equation with brownian initial velocity.
\newblock {\em Commun. Math. Phys.}, 193:397--406, 1998.

\bibitem{Burgersbook}
J.~M. Burgers.
\newblock {\em The nonlinear diffusion equation}.
\newblock D. Reidel, Dordrecht, 1974.

\bibitem{Cole1951}
J.~D. Cole.
\newblock On a quasi-linear parabolic equation occuring in aerodynamics.
\newblock {\em Quart. Appl. Math.}, 9:225--236, 1951.

\bibitem{Colombi1996}
S.~Colombi, F.~R. Bouchet, and L.~Hernquist.
\newblock Self-similarity and scaling behavior of scale-free gravitational
  clustering.
\newblock {\em Astrophys. J.}, 465:14, 1996.

\bibitem{Davis1977}
M.~Davis and P.~J.~E. Peebles.
\newblock On the integration of the bbgky equations for the development of
  strongly nonlinear clustering in an expanding universe.
\newblock {\em Astrophys. J. Supp. S.}, 34:425--450, 1977.

\bibitem{LeDoussal2008}
P.~Le Doussal.
\newblock Exact results and open questions in first principle functional rg.
\newblock {\em arXiv:0809.1192}, 2008.

\bibitem{Fournier1983}
J.-D. Fournier and U.~Frisch.
\newblock L'equation de burgers deterministe et statistique.
\newblock {\em J. Mec. Theor. Appl.}, 2:699--750, 1983.

\bibitem{Frachebourg2000}
L.~Frachebourg and Ph.~A. Martin.
\newblock Exact statistical properties of the burgers equation.
\newblock {\em J. Fluid Mech.}, 417:323--349, 2000.

\bibitem{Frisch1995}
U.~Frisch.
\newblock {\em "Turbulence"}.
\newblock Cambridge University Press, Cambridge, 1995.

\bibitem{Frisch2001}
U.~Frisch and J.~Bec.
\newblock {\em "Burgulence", Les Houches 2000: New trends in turbulence}.
\newblock M. Lesieur, A. Yaglom \& F. David, Springer EDP-Sciences, 2001.

\bibitem{Groeneboom1989}
P.~Groeneboom.
\newblock Brownian motion with a parabolic drift and airy functions.
\newblock {\em Probab. Theory Related Fields}, 81:79--109, 1989.

\bibitem{Gurbatov1991}
S.~N. Gurbatov, A.~Malakhov, and A.~Saichev.
\newblock {\em Nonlinear random waves and turbulence in nondispersive media:
  waves, rays and particles}.
\newblock Manchester University Press, 1991.

\bibitem{Gurbatove1989}
S.~N. Gurbatov, A.~I. Saichev, and S.~F. Shandarin.
\newblock The large-scale structure of the universe in the frame of the model
  equation of non-linear diffusion.
\newblock {\em Mont. Not. Roy. Astron. Soc.}, 236:385--402, 1989.

\bibitem{Gurbatov1997}
S.~N. Gurbatov, S.~I. Simdyankin, E.~Aurell, U.~Frisch, and G.~Toth.
\newblock On the decay of burgers turbulence.
\newblock {\em J. Fluid Mech.}, 344:339--374, 1997.

\bibitem{Hopf1950}
E.~Hopf.
\newblock The partial differential equation $u_t+u u_x=u_{xx}$.
\newblock {\em Commun. Pure Appl. Math.}, 3:201--230, 1950.

\bibitem{Kida1979}
S.~Kida.
\newblock Asymptotic properties of burgers turbulence.
\newblock {\em J. Fluid Mech.}, 93:337--377, 1979.

\bibitem{Kraichnan1968}
R.~H. Kraichnan.
\newblock Lagrangian-history statistical theory for burgers' equation.
\newblock {\em Phys. Fluids}, 11:265--277, 1968.

\bibitem{Melott1994}
A.~L. Melott, S.~F. Shandarin, and D.~H. Weinberg.
\newblock A test of the adhesion approximation for gravitational clustering.
\newblock {\em Astrophys. J.}, 428:28--34, 1994.

\bibitem{Molchan1997}
G.~M. Molchan.
\newblock Burgers equation with self-similar gaussian initial data: tail
  probabilities.
\newblock {\em J. Stat. Phys.}, 88:1139--1150, 1997.

\bibitem{Noullez2005}
A.~Noullez, S.~N. Gurbatov, E.~Aurell, and S.~I. Simdyankin.
\newblock Global picture of self-similar and non-self-similar decay in burgers
  turbulence.
\newblock {\em Phys. Rev. E}, 71:056305, 2005.

\bibitem{Peacock1996}
J.~A. Peacock and S.~J. Dodds.
\newblock Non-linear evolution of cosmological power spectra.
\newblock {\em MNRAS}, 280:L19--L26, 1996.

\bibitem{Peebles1980}
P.~J.~E. Peebles.
\newblock {\em The large scale structure of the universe}.
\newblock Princeton university press, Princeton, 1980.

\bibitem{She1992}
Z.-S. She, E.~Aurell, and U.~Frisch.
\newblock The inviscid burgers equation with initial data of brownian type.
\newblock {\em Commun. Math. Phys.}, 148:623--641, 1992.

\bibitem{Sinai1992}
Ya.~G. Sinai.
\newblock Statistics of shocks in solutions of inviscid burgers equation.
\newblock {\em Commun. Math. Phys.}, 148:601--621, 1992.

\bibitem{Tribe2000}
R.~Tribe and O.~Zaboronski.
\newblock On the large time asymptotics of decaying burgers turbulence.
\newblock {\em Commun. Math. Phys.}, 212:415--436, 2000.

\bibitem{Valageas1999}
P.~Valageas.
\newblock Non-linear gravitational clustering: smooth halos, substructures and
  scaling exponents.
\newblock {\em Astron. Astrophys.}, 347:757--768, 1999.

\bibitem{Valageas2007}
P.~Valageas.
\newblock Using the zeldovich dynamics to test expansion schemes.
\newblock {\em Astron. Astrophys.}, 476:31--58, 2007.

\bibitem{Valageas2009}
P.~Valageas.
\newblock Quasi-linear regime and rare-event tails of decaying burgers
  turbulence.
\newblock {\em Phys. Rev. E}, 80:016305, 2009.

\bibitem{Valageas2008}
P.~Valageas.
\newblock Statistical properties of the burgers equation with brownian initial
  velocity.
\newblock {\em J. Stat. Phys.}, 134:589, 2009.

\bibitem{Vergassola1994}
M.~Vergassola, B.~Dubrulle, U.~Frisch, and A.~Noullez.
\newblock Burgers' equation, devil's staircases and the mass distribution for
  large-scale structures.
\newblock {\em Astron. Astrophys.}, 289:325--356, 1994.

\bibitem{Vogelsberger2008}
M.~Vogelsberger, S.~D.~M. White, A.~Helmi, and V.~Springel.
\newblock The fine-grained phase-space structure of cold dark matter haloes.
\newblock {\em MNRAS}, 385:236--254, 2008.

\bibitem{Zeldovich1970}
Y.~B. Zeldovich.
\newblock Gravitational instability: An approximate theory for large density
  perturbations.
\newblock {\em Astron. Astrophys.}, 5:84--89, 1970.

\end{thebibliography}

\end{document}